\pdfoutput=1
\documentclass[compsoc, conference, a4paper, 10pt, times]{IEEEtran}
\usepackage{amsmath,amsfonts,mathtools}
\usepackage{adjustbox}
\usepackage{graphicx}
\usepackage{diagbox}
\usepackage{textcomp}
\usepackage[table]{xcolor}
\usepackage[linesnumbered,ruled,vlined]{algorithm2e}
\usepackage{pdfcomment}
\usepackage{xcolor}
\usepackage{soul}
\usepackage{booktabs}
\usepackage{multicol}
\usepackage{algpseudocode}
\usepackage{multirow}
\usepackage[caption=false]{subfig}
\usepackage{svg}
\usepackage{xurl}
\usepackage[utf8]{inputenc}
\usepackage[T2A,T1]{fontenc}
\usepackage{stackengine}
\usepackage{multicol}
\usepackage{fancyhdr}

\usepackage{enumitem}
\usepackage{multirow,rotating}
\usepackage{alphalph}
\usepackage{diagbox}
\usepackage{rotating}
\usepackage{standalone}
\usepackage{preview}

\usepackage{tikz}

\usetikzlibrary{arrows,shapes}
\usetikzlibrary{positioning}
\newcommand{\6}{\raisebox{2pt}{\tikz \draw[black,fill=black] circle(2pt);}}
\newcommand{\3}{\raisebox{2pt}{\tikz \draw[black,fill=black] circle(2pt);} & }
\newcommand{\2}{\raisebox{2pt}{\tikz \draw[line width=1.0pt] circle(1.7pt);} & }

\newcommand{\4}{\raisebox{2pt}{\tikz \draw[line width=1.0pt] circle(1.7pt);}}
\newcommand{\1}{& }

\usepackage{colortbl}
\newcolumntype{L}[1]{>{\raggedright\let\newline\\\arraybackslash\hspace{0pt}}m{#1}}
\newcolumntype{C}[1]{>{\centering\let\newline\\\arraybackslash\hspace{0pt}}m{#1}}
\newcolumntype{R}[1]{>{\raggedleft\let\newline\\\arraybackslash\hspace{0pt}}m{#1}}
\definecolor{myblue}{HTML}{1f77b4}
\definecolor{myorange}{HTML}{ff7f0e}
\definecolor{mygreen}{HTML}{2ca02c}
\definecolor{myred}{HTML}{d62728}
\definecolor{lightred}{rgb}{1,0.5,0.5}
\definecolor{lightgreen}{rgb}{0.5,1,0.5}
\definecolor{lightlime}{rgb}{0.75,1,0.75} 
\definecolor{darkpink}{RGB}{231, 84, 128}
\usepackage{hyperref}
\hypersetup{
  pdfborder = {0 0 0},
  urlcolor=blue,
  linkcolor=green,
  citecolor=red,
}

\def\BibTeX{{\rm B\kern-.05em{\sc i\kern-.025em b}\kern-.08em
    T\kern-.1667em\lower.7ex\hbox{E}\kern-.125emX}}

\begin{document}
\title{\textsc{Gotcha}: Real-Time Video Deepfake Detection via Challenge-Response}

\author{
{\rm Anonymous Euro S\&P 2024 submission}
}

\author{
{\rm Govind Mittal,  Chinmay Hegde, Nasir Memon}\\
Dept. of Computer Science and Engineering, Tandon School of Engineering, New York University, USA\\
\{mittal, chinmay.h, memon\}@nyu.edu
}

\maketitle

\begin{abstract}
With the rise of AI-enabled Real-Time Deepfakes (RTDFs), the integrity of online video interactions has become a growing concern. RTDFs have now made it feasible to replace an imposter's face with their victim in live video interactions. Such advancement in deepfakes also coaxes detection to rise to the same standard. However, existing deepfake detection techniques are asynchronous and hence ill-suited for RTDFs. To bridge this gap, we propose a challenge-response approach that establishes authenticity in live settings. We focus on talking-head style video interaction and present a taxonomy of challenges that specifically target inherent limitations of RTDF generation pipelines. We evaluate representative examples from the taxonomy by collecting a unique dataset comprising eight challenges, which consistently and visibly degrades the quality of state-of-the-art deepfake generators. These results are corroborated both by humans and a new automated scoring function, leading to 88.6\% and 80.1\% AUC, respectively. The findings underscore the promising potential of challenge-response systems for explainable and scalable real-time deepfake detection in practical scenarios. We provide access to data and code at \url{https://github.com/mittalgovind/GOTCHA-Deepfakes}.
\end{abstract}

\section{Introduction}
\label{sec:intro}

In an increasingly interconnected world, the adoption of live, online video interactions has become widespread. Projections for 2023 indicate that 86\% of companies are conducting interviews online, while approximately 83\% of employees estimate spending about a third of their working week in virtual meetings~\cite{videoconf}. This significant uptick in online interactions creates a fertile ground for novel social engineering attacks. 
Specifically, high-quality tools capable of creating video deepfakes, which can convincingly mimic a target's facial appearance, are becoming more accessible and can readily circumvent commercial APIs designed for liveness detection and identity verification~\cite{sensitykyc}. While earlier versions of deepfakes primarily targeted public figures~\cite{deepfake_scam0}, recent advancements in deepfake generation technology have made it possible to impersonate ordinary individuals~\cite{deepfake_scam1}, even with a limited set of training images~\cite{mollickquickfake}, in a real-time setting~\cite{iperov_deepfacelab_2021}. These systems, named \emph{Real-Time Deepfakes (RTDFs)}, are a sophisticated variant of deepfakes involving live, interactive video impersonations of a target, posing an urgent, burgeoning threat to the integrity of online human interactions across the globe~\cite{bloomberg1}. 

RTDFs have already become prevalent to the extent that the FBI has warned of their imminent threat and pervasiveness~\cite{fbiworries}. Known recent incidents include an individual in China transferring \$620,000 to an impersonator of their friend after a video call~\cite{deepfake_scam1}. Also, an impersonation of the former president of Ukraine joined a Kremlin critic during a recorded video call to lure the critic into performing embarrassing acts~\cite{deepfake_scam2}.

Conventional techniques~\cite{marraGANsLeaveArtificial2019},~\cite{guo2022eyes},~\cite{dong2022protecting},~\cite{mazaheri2022detection},~\cite{boccignone2022deepfakes} have considered deepfake detection, but in an offline and non-interactive setting. Despite being technically impressive, such techniques are not explicitly designed for RTDFs and operate under the assumption of \textit{no interaction between an imposter and the detector}. This assumption creates a threat model that favors imposters, who can leverage offline resources to refine their models to target specific individuals. Meanwhile, detection often needs to be performed in real-time without knowledge of the creation process and in potentially constrained environments. 

Contrary to this traditional strategy, we explore an alternate approach where the detector can interactively present non-trivial tasks or \emph{challenges} to the imposter. Under this model, the onus of consistently maintaining high-quality deepfakes in real-time, under challenging situations, is now squarely on the imposter. We leverage this asymmetric advantage to design and validate a \emph{challenge-response} approach for identifying RTDFs.

The fundamental conceptual problem is to develop a suite of practical challenges (i.e., the ``real user'' experience is not significantly altered) yet maximally informative (i.e., ``fake user'' video outputs are statistically and visually abnormal). In this work, we propose and categorize challenges designed to exploit limitations specific to components of an RTDF generation pipeline, such as a facial landmark detector and auto-encoder. As we propose a variety of challenge categories, the detection accuracy of each challenge may vary depending on the deepfake generation model and the conditions under which the deepfake is enacted, such as ambient lighting or video quality. Therefore, more than a single challenge may be required for accurate detection.

A desirable feature of the proposed challenges is that they induce human-visible artifacts in the responses and provide robust signals for downstream automated machine-learning detectors. This feature facilitates easy audit and explainability in their practical implementations.

 We call the proposed technique \textsc{Gotcha}, echoing the ubiquitous \textsc{Captcha} test used to identify online robots masquerading as humans. At its core, \textsc{Gotcha} presents one or more challenges to a suspected RTDF, which could require physical action or involve digital manipulations in the video feed. We validate \textsc{Gotcha} on a novel video dataset of 56,247 videos derived by collecting data from 47 legitimate users. We conduct both human-based and automated evaluations to test its potential in practical scenarios while establishing a security-usability tradeoff.

Our evaluation of \textsc{Gotcha} demonstrates consistent and measurable degradation of deepfake quality across users, highlighting its promise for RTDF detection in real-world settings. Our contributions are as follows:
    \begin{itemize}[nosep,parsep=0pt,leftmargin=*]
    \item We explore a challenge-response approach for authenticating live video interactions and develop a taxonomy of challenges by exploiting vulnerabilities in real-time deepfake generation pipelines.

    \item We collect video data of 47 real users in person, performing eight challenges. The new dataset consists of 56,247 short real and fake videos.

    \item We perform human and automated evaluations of the approach, demonstrating consistent and visible degradation of deepfake quality caused by challenges. We train a new self-supervised ML-based fidelity scoring model for this purpose.  \footnote{\textbf{Reproducibility:} All participants consented to release their data for academic research. Hence, we release originally recorded challenges and corresponding deepfakes for non-commercial research, along with the code for the fidelity score model and survey instruments used for human evaluations.}
\end{itemize}
\vspace{0.5em}
\noindent\textbf{Authentication and Liveness Detection.} It is important to note that a \textsc{Captcha}-like test does not assume the availability of any prior knowledge or identification of the target user. This is fundamentally different from ``what-you-know'' (e.g. passwords) or ``who-you-are'' (e.g., biometrics) type of authentication systems. Such systems require the user to enroll prior to authentication and perform it by using an identity-matching algorithm. 

In this work, the imposter is explicitly considered to be \emph{present} during an interaction, and all impersonations are performed live (consequently, in real-time). The proposed solution trivially addresses presentation attacks, thus differentiating it from methods explicitly designed for liveness detection.

\section{Real-Time Deepfake Generation}
\label{sec:background}

 \noindent\textbf{Definition.}  A \emph{deepfake} refers to a digital impersonation in which an \emph{imposter} mimics audio or visual characteristics to convincingly match a specific \emph{target}'s likeness. When such impersonations can be done live\footnote{Computer graphics literature traditionally considers real-time to be 30 frames/s; however, for video calling, 15 frames/s is often sufficiently "real-time" or live.} with sufficient fidelity, we call them RTDFs.
 
 This work focuses on talking-head videos, which encompass techniques such as facial reenactment or face-swapping. A typical use case for deploying RTDFs are hoax video call interactions.

 \subsection{Dissecting the Generation Pipeline}
 \label{sec:rtdf_pipeline}

  \begin{figure}[t!]
    \centering
    \includegraphics[width=0.9\columnwidth]{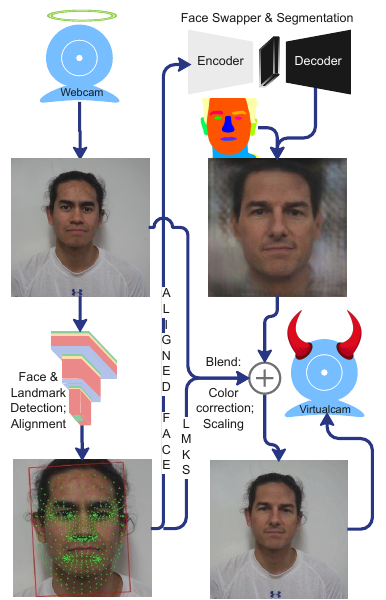}
    \caption{A generic face-swapping RTDF pipeline containing a physical webcam (top), face and landmark detector, face-swapper (auto-encoder), blending operator and a virtual webcam (right). The virtual webcam is piped into a video conferencing software (not shown). Arrows indicate relevant data flows.}
    \label{fig:rtdf_pipeline}
  \end{figure}
  
 Several approaches exist for generating real-time deepfakes~\cite{iperov_deepfacelab_2021,nirkin_fsganv2_2022,wang2022latent}. While their details differ, they all utilize the same key components. The following list briefly describes each of these components, as depicted in Fig.~\ref{fig:rtdf_pipeline}.  We assume that the computing device being used for video conferencing is instrumented to channel the imposter's video stream through a deepfake generation pipeline, and then forward it to the video-conferencing client. 

  \begin{itemize}[nosep,leftmargin=*]

    \item \emph{Face Detector} is a neural network that predicts a bounding box per face in a video frame.
     
        \item \emph{Landmark Detection} is a neural network that detects facial key-points called landmarks. Several reeactment techniques use only landmarks as the driving signal for the target's face image. Landmarks also aid in conforming the target's predicted face to fit the imposter's face shape. 

    \item \emph{Face Alignment} module aligns a given input face with respect to the landmarks, which is vital for a robust prediction by the face-swapper. Subsequently, the applied alignment is reversed and re-applied to the predicted face, in order to match back with the imposter. 

    \item \emph{Segmentation} (a) separates the face region into distinct regions of interest (lips, eyes, nose), (b) derives the convex hull of the face that is visible, and (c) determines facial boundaries around occlusions (e.g., hand).

    \item \emph{Face-Swapper}, typically, involves an autoencoder. The auto-encoder takes the input from the face detector and predicts how the target's face would look under a given set of facial landmarks, occlusions, and lighting.

    \item \emph{Blending Operator} overlays the inner predicted face of target onto the outer head of imposter. This post-processing step tends to vary across RTDF generation pipelines, involving some combination of blurring, degrading, scaling, compression, and fading.

    \item \emph{Color Correction}: In case of swapping only the inner face of the imposter with that of the target, this module can adjust difference in complexion, by sampling color from the outer face region (around forehead and neck) and adjusting the inner swapped face~\cite{reinhard2001color}.
  \end{itemize}

\vspace{0.5em}
\noindent\textbf{Note on choice of generative models.} Although, this work explores Auto-Encoder-based generative models, but we acknowledge the innovations in synthetic media generation, especially based on NERFs~\cite{athar2022rignerf}, and Diffusion Models~\cite{dhariwal2021diffusion} coupled with CLIP~\cite{ramesh2022hierarchical}. However, none of these simultaneously support real-time throughput and high-fidelity deepfake outputs, which are essential ingredients to conceive realistic RTDFs.

\subsection{Hurdles to Generating Realistic Deepfakes}
  \label{sec:constraints}

Launching a convincing deepfake of a specific target involves overcoming multiple impediments. Below we describe some of the \emph{key impediments we leverage while designing our approach}. 
 
\vspace{0.5em}
\noindent \textbf{Data Diversity.} Creating a persuasive deepfake of a specific individual often requires their diverse facial data. Ideally, this collection should include the target's face captured under various lighting conditions and angles, with or without occlusions, preferably drawn from high-resolution videos. For instance, using a face-swapping framework like DeepFaceLab~\cite{iperov_deepfacelab_2021} typically necessitates 4,000 diverse images. In stark contrast, facial reenactment techniques only expect a single image of the target, which is then animated via text~\cite{synthesia}, speech~\cite{thies2020neural}, or facial landmarks~\cite{siarohin2019first}. However, these reenactments are far more fragile when confronted with unexpected scenarios. We delve deeper into the implications of this inherent constraint in \S\ref{sec:insights}.

\begin{figure}[t!]
    \centering
    \resizebox{0.8\columnwidth}{!}{%
        \subfloat{\includegraphics[]{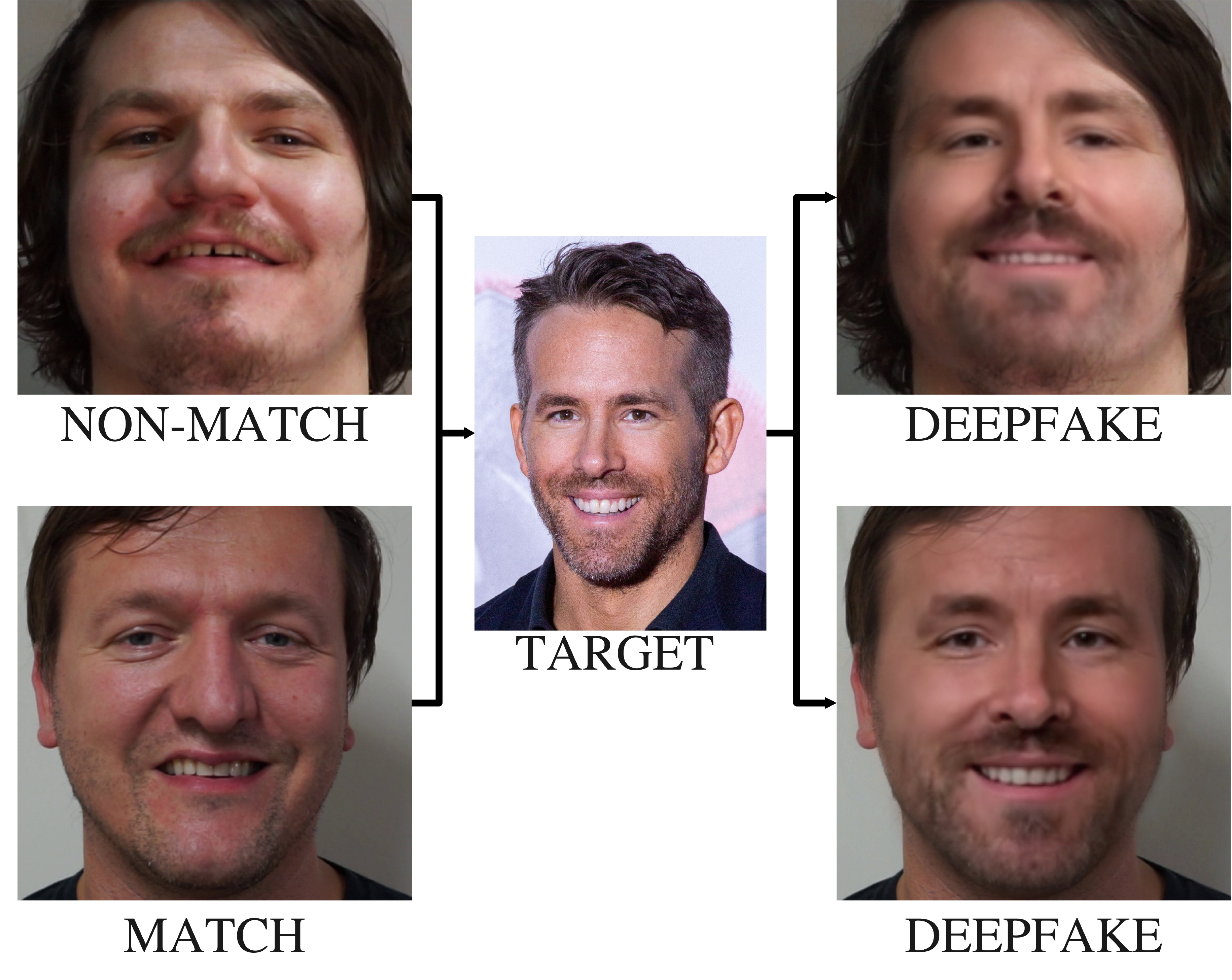}}
    }
    \vspace{-1em}
    \caption{Importance of facial shape similarity. Two imposters with distinct facial shapes result in differing quality outputs, while assuming Ryan Reynolds as target. Qualitative observations imply that better match yields a better fit.}
    \label{fig:face_similarity}
\end{figure}
  
    \vspace{0.5em}
  \noindent \textbf{Face Shape Similarity.}
    Successful impersonation also hinges on the similarity between the facial shapes of an imposter and their target. Any significant mismatch, such as in the spacing between the eyes or the shape of the nose or lips, can result in unrealistic artifacts. These anomalies can manifest as stretching or contraction of the deepfake mask, blending issues, or rigid expressions.
    
    An imposter attempts to project their target's identity, which is often strongly tied to their facial shape and skin color. Hence, having similar face shapes is crucial for a convincing impersonation. However, discrepancies in skin color can often be mitigated using a color transfer technique~\cite{colortransferrct}. Fig.~\ref{fig:face_similarity} compares two individuals with distinct facial shapes; both face-swapped into the same target identity, indicating that a face shape with a `closer' match yields a more convincing impersonation.
    
    \begin{table}[t!]
    \caption{Inference speed of an RTDF. Red and green colors indicate performance below and above the real-time threshold of 15 frames/s.}
    \resizebox{\columnwidth}{!}{\begin{tabular}{C{1.2cm}C{1.5cm}C{1.8cm}C{1.7cm}C{1.8cm}}
      \hline
      \textbf{\diagbox[innerwidth=1.2cm]{Speed}{Config.}}  & \textbf{$\alpha$ : 8-core i7 CPU } & \textbf{$\beta$ : $\alpha$ + RTX 3070 (8 GB)} & \textbf{$\beta$ + WiFi Camera} & \textbf{$\beta$ + LTE Phone Cam.}  \\ \hline
      Frames/s & \cellcolor{lightred} 1.2    & \cellcolor{lightgreen}23.5   & \cellcolor{lightgreen}21.1 & \cellcolor{lightgreen} 20.2   \\ \hline
      \end{tabular}
      }
      \label{tab:df_speed}
  \end{table}

    \vspace{0.5em}
   \noindent \textbf{Computational Resources and Real-Time Constraints.} Producing high-quality deepfakes is computationally expensive, typically requiring specialized hardware such as GPUs or TPUs. Table~\ref{tab:df_speed} showcases the inference speed of DeepFaceLab~\cite{iperov_deepfacelab_2021} with various hardware configurations. 
   
   The computational load escalates when the deepfake generation has to happen in real time. Envision a scenario where the subject needs to perform an activity such as rapidly moving a hand in front of the face. While human observers might see the motion as a blur, it is still discernible. For deepfakes, handling such high-activity scenarios in real time is challenging. Nimble movements can significantly reduce throughput leading to dropped frames or noticeable lags.

    Furthermore, popular cloud service providers such as Google Colab~\cite{googlecolab} have placed restrictions on training deepfakes on their platform. This constraint adds another layer of complexity on easy access to requisite computational resources, necessitating personal hardware investments to train] high-fidelity deepfakes.

    \vspace{0.5em}
     \noindent \textbf{Narrow Portability and Technical Effort}: Each new target identity necessitates training a fresh face-swap model from scratch or retraining an existing pre-trained model using the target's data for a substantial duration, taking up to a week.
    
    While there exist target-agnostic, ready-to-use face-swap methods such as FSGAN~\cite{nirkin_fsganv2_2022}, the resulting deepfakes tend to be more fragile. The fragility is primarily due to the imperfect disentanglement of identity attributes from other aspects like pose and expression. Also, in recent years, the broader community has focused on advancing facial reenactment~\cite{wang2022latent,stypulkowski2023diffused,GAvatar2023,kim2023dcface}, and such methods trivially degrade under our approach (see Fig~\ref{fig:challenges}).

    Hence, an imposter aiming to navigate around this impediment has to enhance the existing face-swapping methods or build a custom one from scratch. Therefore, possessing (a) substantial software development proficiency, (b) an understanding of underlying principles, and (c) the potential to conduct extensive trial-and-error iterations to achieve a viable, effective solution. 

\vspace{0.5em}
Thus, an imposter attempting to successfully deepfake a target needs to manage all the constraints listed above and possibly more.

\section{Problem Description}

Given the design of an RTDF generator and the aspects an imposter needs to consider for using them in practice, we describe our threat model, setup and hypothesis.

\vspace{0.5em}
\noindent\textbf{Threat Model} involves three subjects -- an imposter, a target, and a defender. The defender is an individual on a video call, intending to interact with the target. The imposter is another individual on the same call, seeking to deceive the defender by driving an AI-generated talking head portrayed as the target.

Practical examples of this threat model exists in online \textit{Know-Your-Customer (KYC) verification process}, crucial for companies like Airbnb~\cite{airbnb} and Uber~\cite{uberkyc}, as well as US government agencies utilizing ID.me~\cite{idme}, involves capturing a selfie with a photo ID. RTDFs threaten these face-matching systems by altering faces in both the ID and live video, bypassing liveness detection mechanisms as reported by Sensity~\cite{sensitykyc}.

Additionally, online exams, such as those administered by ETS~\cite{ets_gre}, and interviews are vulnerable against RTDFs. This technology enables impostors to fraudulently take tests or interviews with a fake identity or on behalf of the intended person. Such a tactic comes close to North Korean IT workers, who secured remote employment under fake identities~\cite{koreanfreelancer}.

\vspace{0.5em}
\noindent\textbf{Setup.} When the defender and the potential imposter join an online video call, the defender provides instructions to the potential imposter and requests that they respond to one or more specific actions, which we will call challenges. After receiving the response(s), the defender accepts or rejects the portrayed identity. 
  
  \vspace{0.5em}
  \noindent\emph{Defenders.}  The defender assumes the existence of generic deepfake generation components without knowledge of any exact specification. The defender does not assume any trust in the imposter or their devices and does not keep the nature of the requested challenges secret. Also, no identifiable information, such as biometrics or face, is collected through an extensive enrollment process. Here, we assume a stronger threat model, as having such information can help the defender.

  \vspace{0.5em}
  \noindent\emph{Imposters.} An imposter can train and use deepfake models and have enough computational capacity to perform a successful live impersonation. The imposter is also capable of understanding, interacting, and responding to requests made to them by the defender. The imposter could be aware of the nature of the requested tasks and can deploy adaptive countermeasures.

  Initially, we consider a scenario where the target individual has a limited online footprint, leading to a constrained range of available high-quality facial data. Such cases confine an imposter to using images from specific sources like social media or recorded webinars. However, we later broadened our scope to include more data-diverse contexts featuring potential accomplices or public figures as targets (\S\ref{sec:adaptive}). In such scenarios, the imposter can access a \emph{rich} set of data, either willingly or inadvertently provided by the target. Access to such a dataset can improve the realism of RTDFs, making them harder to detect. This extension is particularly relevant in contexts such as job interviews, where the target might be an accomplice of the imposter.

  \vspace{0.5em}
\noindent\textbf{Hypothesis.} \emph{We posit that a user can perform the specific tasks required by the challenge during a live interaction, which is difficult for a deepfake generation pipeline to model in real-time}.

Each component within such a pipeline has optimal operating conditions, defined by the type of component, its architecture, and the biases learned during its training phase. As a result, an imposter can successfully generate a convincing deepfake only when the live inputs fall within the overlapping range of these optimal conditions. Any divergence from this range can compromise the performance of one or more components in the pipeline, leading to discernible artifacts in the final deepfake video. Grounded on our hypothesis and this insight, we design a series of tasks that impair the performance of the RTDF pipeline.

\section{Challenges}
\label{sec:challenges}
 \noindent \textbf{Definition.} A challenge $\mathcal{C}$ is a task  that has the following properties:
  \begin{itemize}[nosep,parsep=0pt,leftmargin=*]
    \item degrades real-time deepfakes (optionally visibly),
    \item is can be perforned during a live call, and 
    \item can be specified by a set of randomizable parameters.
  \end{itemize}
      \begin{figure}[t!]
        \centering
        \includegraphics[width=0.6\columnwidth]{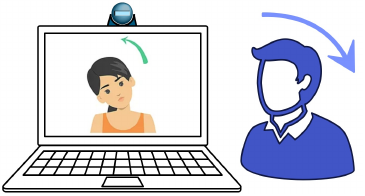}
        \caption{A method to guide a user and randomize the challenge is illustrated. The user follows on-screen instructions to mimic the actions of an avatar, performing required head movements.}
        \label{fig:laptop_pattern}
    \end{figure}

   In order to verify a potential imposter, a defender picks a random challenge (and its specification), and conveys it to the potential imposter to perform. For example, if the challenge picked is head rotation, then a concrete specification could be  `rotate the head to the left by an angle of $45^\circ$ and downwards by $30^\circ$ for a specified time,' creating a unique challenge. In order to enhance usability, a defender can demonstrate the challenge themselves or by using an automated avatar as shown in Fig.~\ref{fig:laptop_pattern}.

    Since current RTDF pipelines essentially replace the imposter's face with that of a target while preserving the remainder of the scene, hence in this work, we only consider challenges that necessitate non-trivial user actions involving faces, such as performing a specific sequence of head movements, occlusions, face deformations, etc.  

  \label{sec:taxonomy}
    \begin{table*}[ht!]
      \caption{Taxonomy of  Challenge Categories Considered. Usability benefits are described in \S\ref{subsec:usability}.}
      \centering
      \resizebox{1.6\columnwidth}{!}{
\begin{preview}
\begin{tabular}{C{1.8cm}|C{3.4cm}|
C{0.2cm}C{0.2cm}C{0.2cm}C{0.2cm}C{0.2cm}C{0.2cm}C{0.2cm}C{0.2cm}|
C{0.2cm}C{0.2cm}C{0.2cm}C{0.2cm}C{0.2cm}}
\toprule
 & & \multicolumn{8}{c}{\textbf{Component Effected}} & \multicolumn{5}{c}{\textbf{Usability}}  \\ 
\textbf{Category} & \textbf{Examples} & 
\begin{turn}{90} \textit{Face Detector} \end{turn} & 
\begin{turn}{90} \textit{Landmark Detection} \end{turn} & 
\begin{turn}{90} \textit{Face Alignment} \end{turn} & 
\begin{turn}{90} \textit{Segmentation} \end{turn} & 
\begin{turn}{90} \textit{Auto-encoder} \end{turn} & 
\begin{turn}{90} \textit{Blending} \end{turn} & 
\begin{turn}{90} \textit{Color Correction} \end{turn} & 
\begin{turn}{90} \textit{Inference Speed} \end{turn} & 
 \begin{turn}{90} \textit{Easy-to-Comprehend}\end{turn} &
 \begin{turn}{90} \textit{Appropriate-to-Request} \end{turn} &
 \begin{turn}{90} \textit{Physically-Effortless}\end{turn} &
 \begin{turn}{90} \textit{No-Equipment-Needed} \end{turn} &
 \begin{turn}{90} \textit{Detected-by-Humans} \end{turn} \\
\midrule
    \multirow{2}{1.8cm}{Face Occlusion}
   & Hand                       & \3\2\2\3\2\3\1\1 \3\3\2\3\6 \\
   & Object                     & \3\1\2\3\2\3\2\1 \3\2\2\1\6 \\
\midrule

\multirow{2}{1.8cm}{Head Movement}
   & Directed Movement       & \2\1\2\3\1\3\1\1 \3\3\2\3\6 \\
   & Whole Body Movement     & \3\1\2\3\1\2\1\1 \3\2\1\3\6 \\
\midrule
\multirow{2}{1.8cm}{Facial Deformation}  
   & Manual Deformation            & \1\3\1\3\2\1\1\2 \3\1\2\3\6 \\
   & Expression Alteration         & \1\1\1\1\3\1\1\2 \3\2\2\3\6 \\
\midrule
\multirow{2}{1.8cm}{Face Illumination}  
   & User-Guided        & \1\1\1\1\2\3\3\2 \3\2\1\1\4 \\
   & Display-Induced    & \1\1\1\1\2\3\3\2 \2\2\3\3 \\
\midrule
\multicolumn{15}{l}{Benefits: \raisebox{1pt}{\tikz \draw[black,fill=black] circle(2.3pt);} = Fully Offered; \raisebox{1pt}{\tikz \draw[line width=1.12pt] circle(2pt); }= Quasi/Partially Offered ; \textbf{Empty} = Not Offered ;}\\ 
\bottomrule
\end{tabular}
\end{preview}
}
      \label{tab:taxonomy}
    \end{table*}
\subsection{A Taxonomy of Facial Challenges}
  This subsection introduces a taxonomy of challenge categories considered in this work. Challenges are tailored to impact facial fidelity and are classified based on specific facial transformations they address. Table~\ref{tab:taxonomy} summarizes the taxonomy and symbolizes the components that get disrupted due to each challenge along with usability characteristics. The categories in the taxonomy are as follows:
  
   \begin{itemize}[nosep,parsep=2pt,leftmargin=*]

    \item\textbf{Head Movement} encompasses challenges that involve physical motions of the head or the entire body, which can modify the point-of-view of the facial appearance. Examples of head movement-based challenges include:
    
    \begin{itemize}[nosep,leftmargin=*] 
        \item \emph{Directed Head Movement}: A user moves their head in specific directions - for example, nodding, shaking, or tilting the head to various angles. These movements can complicate the task for RTDF generators, which typically see more stable and frontal facial views during training.
        
        \item \emph{Whole Body Movement}: More drastic movements, such as turning around, walking, or standing up, could introduce complex temporal and spatial distortions, altering the facial perspective and distance viewed by the camera.
    \end{itemize}
    
    \item\textbf{Face Occlusion}. This category includes challenges involving reduced face visibility due to an occluding object. Ways to occlude a face include:
      
      \begin{itemize}[nosep,leftmargin=*]
       \item \emph{Hand}: A user can partially occlude their face using their hands, to block full facial visibility.

       \item \emph{Objects}: External objects can also be used for occlusion. Readily available items, such as masks, sunglasses, and tissues, work for this purpose.
      \end{itemize}
      
    \item\textbf{Facial Deformations} comprise challenges designed to induce modifications to the facial structure or presentation. These challenges could be:
    
    \begin{itemize}[nosep,leftmargin=*] 
        \item \emph{Manual Deformations}: The user manipulates their facial features using their hands. For example, users press a finger against their cheek or forehead or stretch their lips, creating deformations.
    
        \item \emph{Expression Alterations}: These challenges consist of users altering their facial expressions dramatically. Examples include wide grins, deep-set frowns, elevated eyebrows, and squinting eyes.
    \end{itemize}
    
    \item\textbf{Face Illumination} includes challenges that alter the illumination on the user's face. Examples include:

    \begin{itemize}[nosep,leftmargin=*]
       \item \emph{User-Guided Illumination}: Users change their position relative to light sources, adjust the light source itself, or use handheld devices like a flashlight or camera flash to introduce diverse lighting conditions, producing shadows and highlights. For example, having light illuminate just one side of the face can cause difficulty replicating it accurately.
        
       \item \emph{Display Induced Illumination}: Face illumination can also be affected by causing the imposter's display to emit sudden flashes or structured light patterns to reflect from the user's face. 
    \end{itemize}
    \end{itemize}
    
\begin{figure}[t]
    \centering
    \resizebox{\columnwidth}{!}{%
    \begin{tabular}{C{1.75cm}C{2.2cm}|C{2.2cm}C{2.3cm}C{2.2cm}}
         No Challenge & \begin{tikzpicture}\node[anchor=south west,inner sep=0] (image) at (0,0) { \includegraphics[width=0.3\columnwidth]{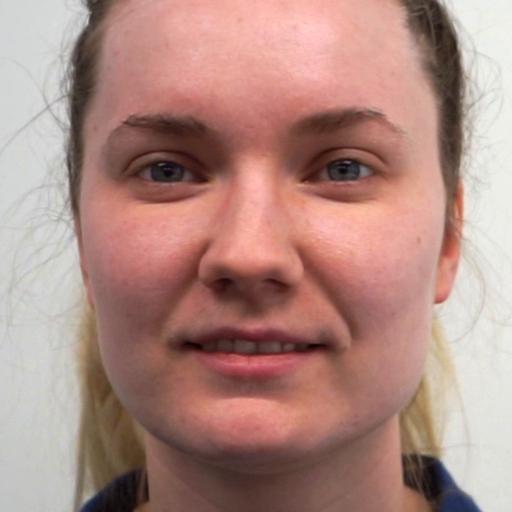}};\end{tikzpicture} &
        \begin{tikzpicture}
            \node[anchor=south west,inner sep=0] (image) at (0,0) { \includegraphics[width=0.3\columnwidth]{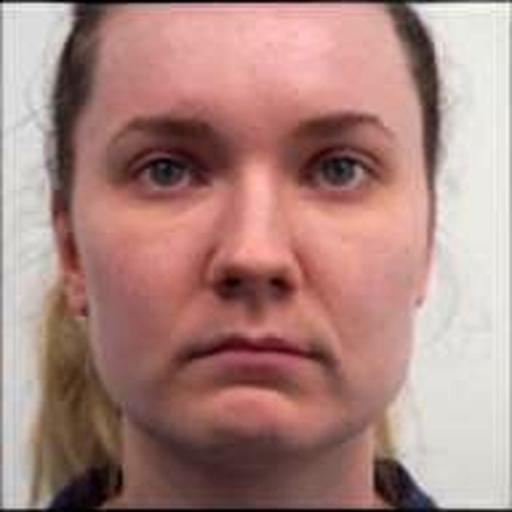}};
            \begin{scope}
                \fill[green] ([xshift=-1.5mm]image.south east) rectangle ([yshift=9.9mm]image.east);
            \end{scope}
        \end{tikzpicture} &
        \begin{tikzpicture}
            \node[anchor=south west,inner sep=0] (image) at (0,0) { \includegraphics[width=0.3\columnwidth]{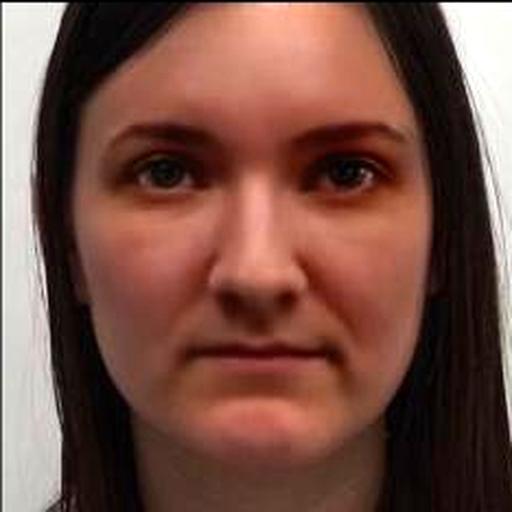}};
            \begin{scope}
                \fill[green] ([xshift=-1.5mm]image.south east) rectangle ([yshift=4.95mm]image.east);
            \end{scope}
        \end{tikzpicture} &
        \begin{tikzpicture}
            \node[anchor=south west,inner sep=0] (image) at (0,0) { \includegraphics[width=0.3\columnwidth]{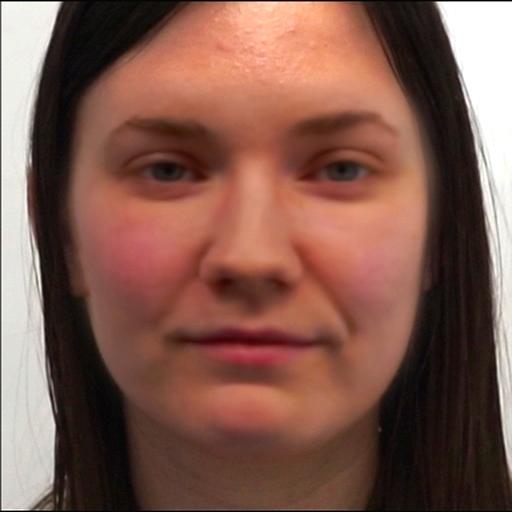}};
            \begin{scope}
                \fill[green] ([xshift=-1.5mm]image.south east) rectangle ([yshift=4.95mm]image.east);
            \end{scope}
        \end{tikzpicture}\\
        \midrule
         Head movement & \begin{tikzpicture}    \node[anchor=south west,inner sep=0] (image) at (0,0) { \includegraphics[width=0.3\columnwidth]{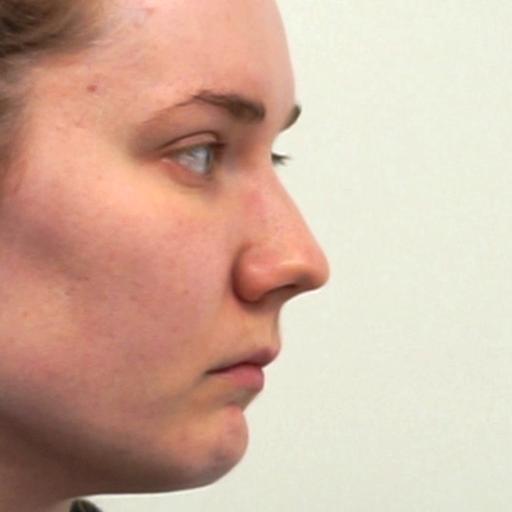}};\end{tikzpicture} &
        \begin{tikzpicture}
            \node[anchor=south west,inner sep=0] (image) at (0,0) { \includegraphics[width=0.3\columnwidth]{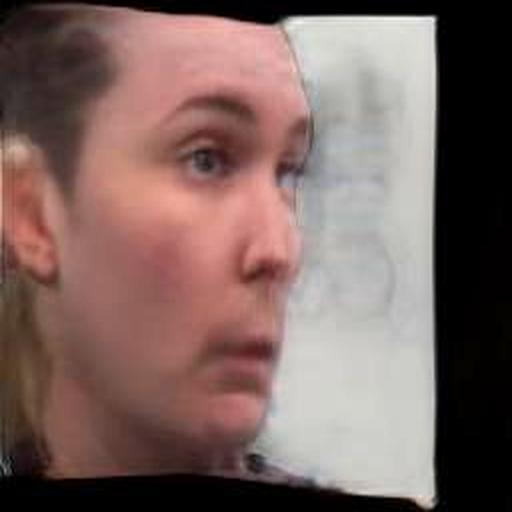}};
            \begin{scope}
                \fill[green] ([xshift=-1.5mm]image.south east) rectangle ([yshift=-3mm]image.east);
            \end{scope}
        \end{tikzpicture} &
        \begin{tikzpicture}
            \node[anchor=south west,inner sep=0] (image) at (0,0) { \includegraphics[width=0.3\columnwidth]{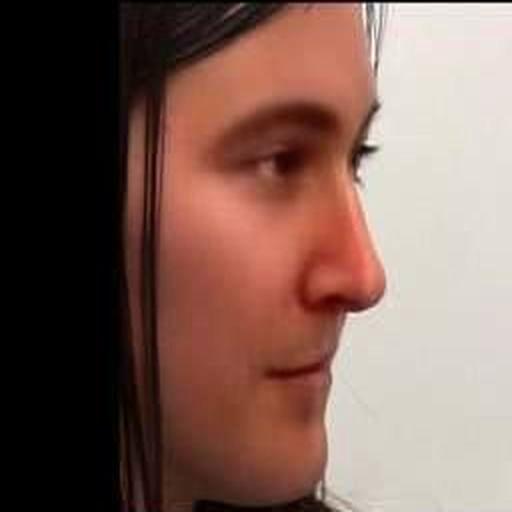}};
            \begin{scope}
                \fill[green] ([xshift=-1.5mm]image.south east) rectangle ([yshift=1mm]image.east);
            \end{scope}
        \end{tikzpicture} &
        \begin{tikzpicture}
            \node[anchor=south west,inner sep=0] (image) at (0,0) { \includegraphics[width=0.3\columnwidth]{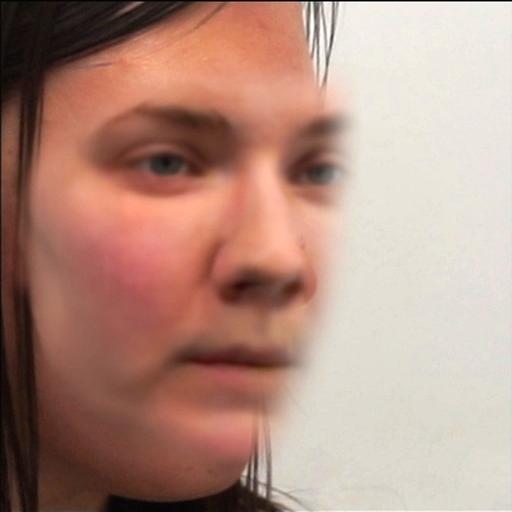}};
            \begin{scope}
                \fill[green] ([xshift=-1.5mm]image.south east) rectangle ([yshift=-1mm]image.east);
            \end{scope}
        \end{tikzpicture} \\
         Occlude with Hand & \begin{tikzpicture}    \node[anchor=south west,inner sep=0] (image) at (0,0) { \includegraphics[width=0.3\columnwidth]{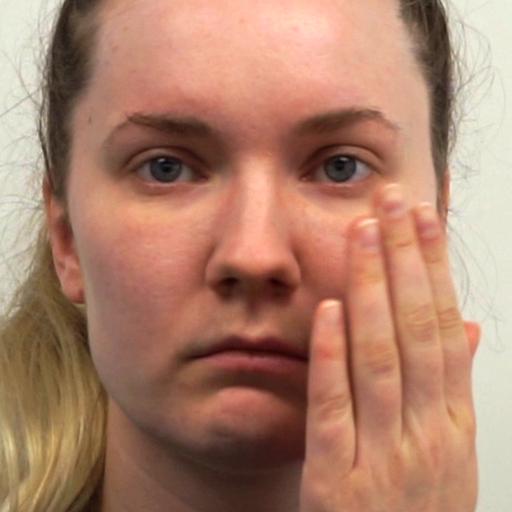}};\end{tikzpicture} &
        \begin{tikzpicture}
            \node[anchor=south west,inner sep=0] (image) at (0,0) { \includegraphics[width=0.3\columnwidth]{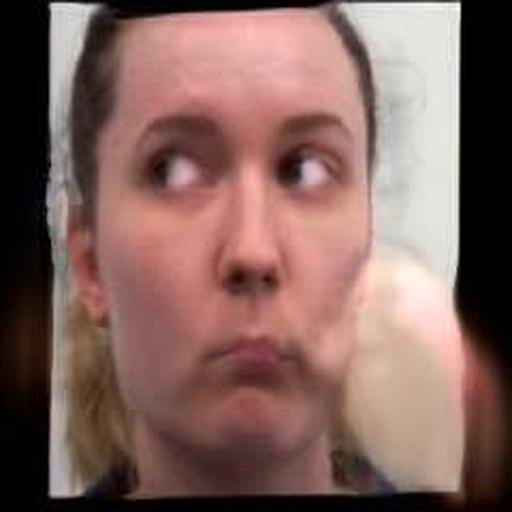}};
            \begin{scope}
            \end{scope}
        \end{tikzpicture} &
        \begin{tikzpicture}
            \node[anchor=south west,inner sep=0] (image) at (0,0) { \includegraphics[width=0.3\columnwidth]{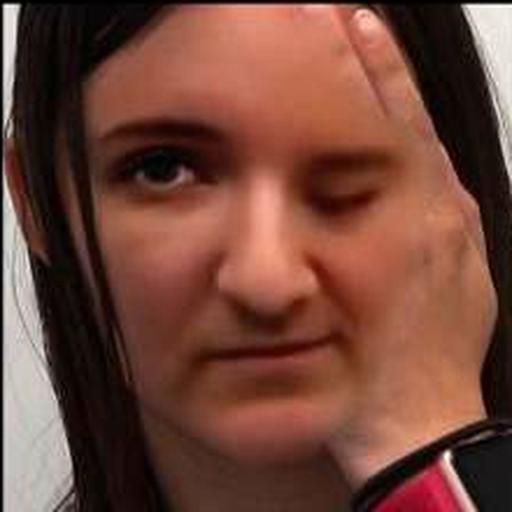}};
            \begin{scope}
            \end{scope}
        \end{tikzpicture} &
        \begin{tikzpicture}
            \node[anchor=south west,inner sep=0] (image) at (0,0) { \includegraphics[width=0.3\columnwidth]{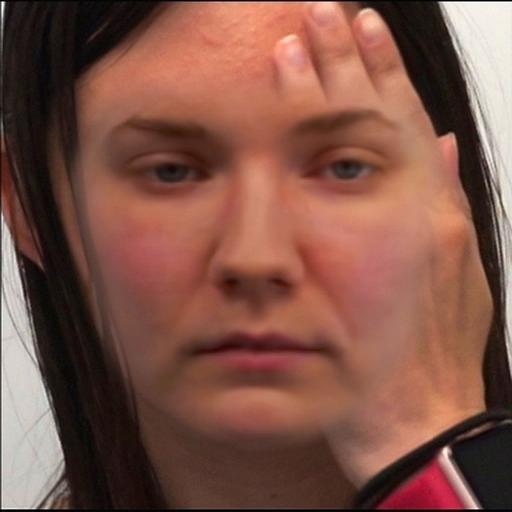}};
            \begin{scope}
            \end{scope}
        \end{tikzpicture}\\
         Manual Facial Deformation & \begin{tikzpicture}    \node[anchor=south west,inner sep=0] (image) at (0,0) { \includegraphics[width=0.3\columnwidth]{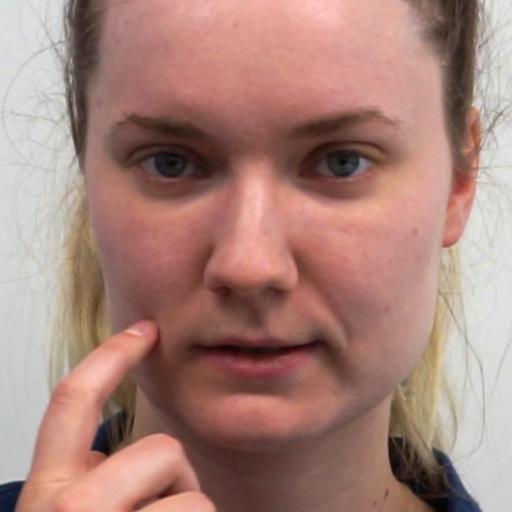}};\end{tikzpicture} &
        \begin{tikzpicture}
            \node[anchor=south west,inner sep=0] (image) at (0,0) { \includegraphics[width=0.3\columnwidth]{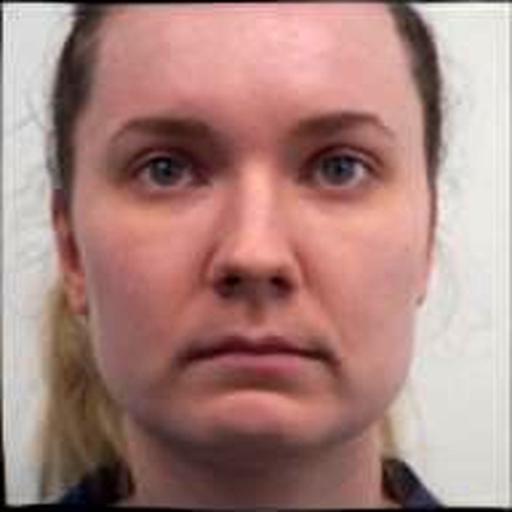}};
            \begin{scope}
            \end{scope}
        \end{tikzpicture} &
        \begin{tikzpicture}
            \node[anchor=south west,inner sep=0] (image) at (0,0) { \includegraphics[width=0.3\columnwidth]{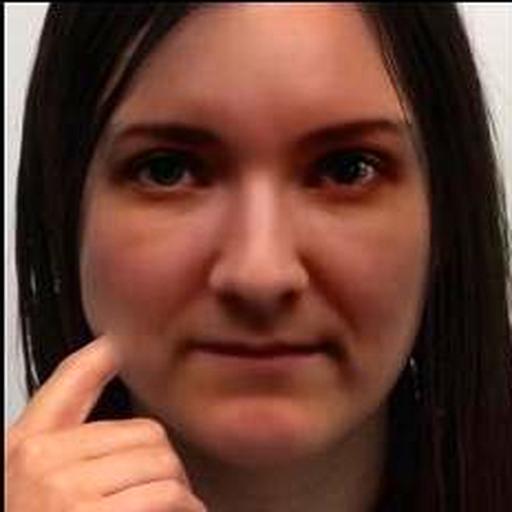}};
            \begin{scope}
            \end{scope}
        \end{tikzpicture} &
        \begin{tikzpicture}
            \node[anchor=south west,inner sep=0] (image) at (0,0) { \includegraphics[width=0.3\columnwidth]{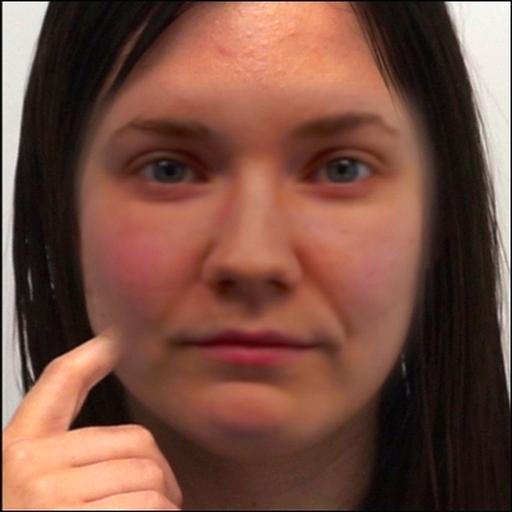}};
            \begin{scope}
            \end{scope}
        \end{tikzpicture} \\
         Face Illumination with Flash & \begin{tikzpicture}    \node[anchor=south west,inner sep=0] (image) at (0,0) { \includegraphics[width=0.3\columnwidth]{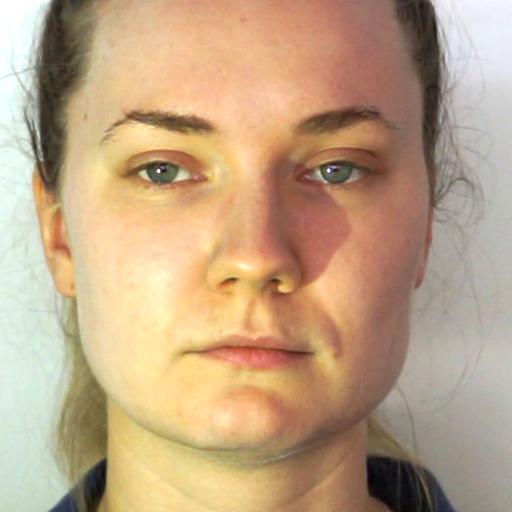}};\end{tikzpicture} &
        \begin{tikzpicture}
            \node[anchor=south west,inner sep=0] (image) at (0,0) { \includegraphics[width=0.3\columnwidth]{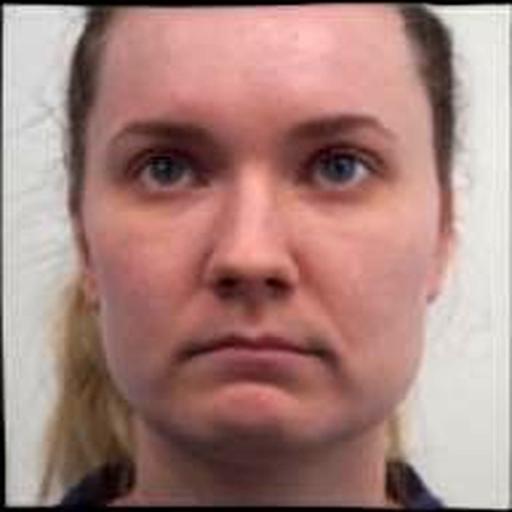}};
            \begin{scope}
            \end{scope}
        \end{tikzpicture} &
        \begin{tikzpicture}
            \node[anchor=south west,inner sep=0] (image) at (0,0) { \includegraphics[width=0.3\columnwidth]{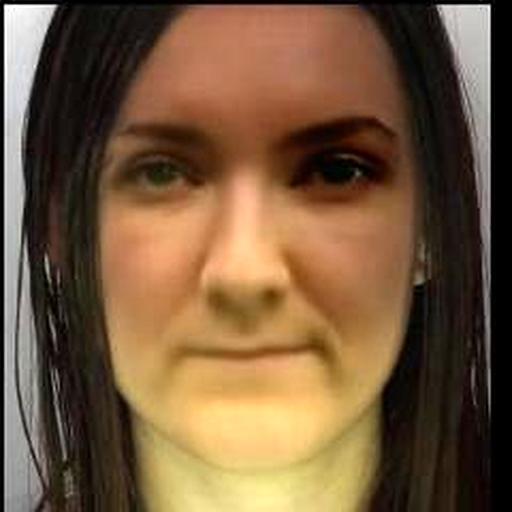}};
            \begin{scope}
                \fill[green] ([xshift=-1.5mm]image.south east) rectangle ([yshift=0mm]image.east);
            \end{scope}
        \end{tikzpicture} &
        \begin{tikzpicture}
            \node[anchor=south west,inner sep=0] (image) at (0,0) { \includegraphics[width=0.3\columnwidth]{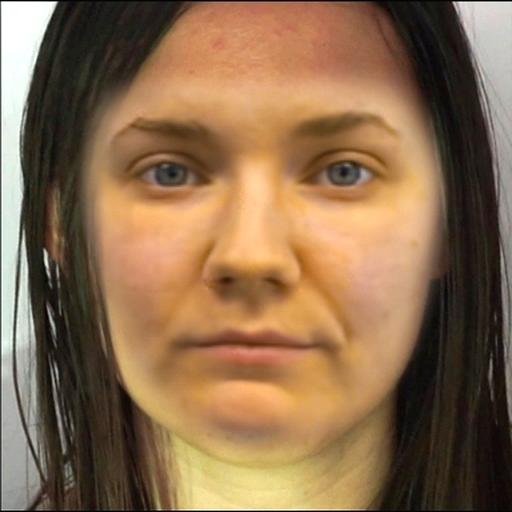}};
            \begin{scope}
                \fill[green] ([xshift=-1.5mm]image.south east) rectangle ([yshift=3mm]image.east);
            \end{scope}
        \end{tikzpicture}
        \\
        & Target & LIA~\cite{wang2022latent} & FSGAN~\cite{nirkin_fsganv2_2022} & DFL~\cite{iperov_deepfacelab_2021} 
    \end{tabular}
}
\caption{Challenge frame of original and deepfake videos. Each row aligns outputs against the same instance of challenge, while each column aligns the same deepfake method. The green bars are a metaphor for the fidelity score, with taller bars suggesting higher fidelity. Missing bars imply the specific deepfake failed to do that specific challenge. Video version at \url{http://govindm.me/gotcha-figures}.}
\label{fig:challenges}
\end{figure}
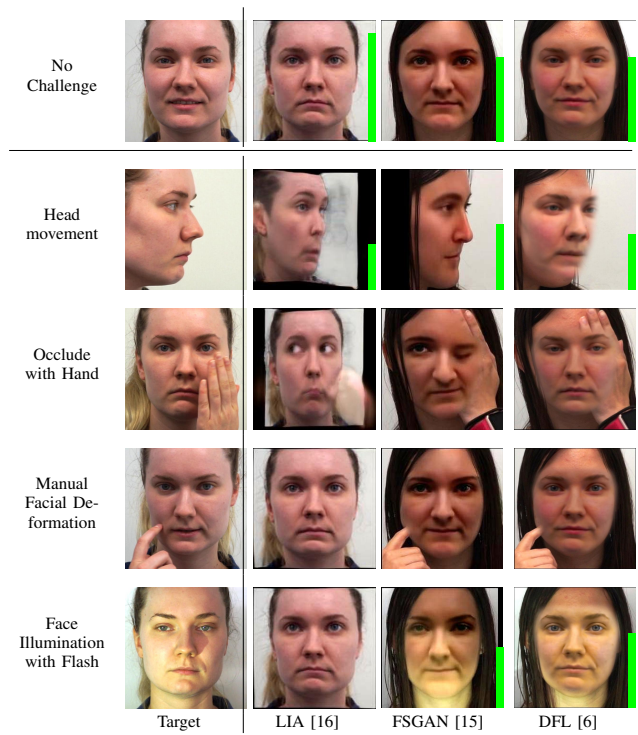

    \subsection{Dataset Collection and Curation}
    \label{subsec:chal_dataset}
    From the challenge categories described above, we selected eight instances for data collection and further evaluations. Table~\ref{tab:video_counts} lists the selected challenges, which span the four identified categories. 
    \vspace{0.5em}
    
    \noindent \textbf{Original Videos: Building a Unique Dataset.} In the absence of publicly available deepfake datasets incorporating challenge-response mechanisms, we created a novel dataset tailored for this study. The dataset features a group of 47 participants, encompassing various demographics (refer to Fig.~\ref{fig:demographics}). Each participant was recorded performing specified challenges in a talking-head video-style, with only one individual appearing per video to simplify the evaluation process. We outline the step-by-step instructions provided to the participants while recording in \S\ref{subsec:data_instructions}.
    
    The dataset comprises of manually recorded Full HD videos. This results in an expansive archive of almost 50 GB of 409 original video data. Table~\ref{tab:video_counts} provides a breakdown of individual original challenges. Numbers $<47$ indicate error in recording.
    
    Each participant contributed around 5 to 6 minutes of total video footage. Since this work focuses primarily on headshots, we isolated this area from the full-view footage for each participant. The average dimensions of the headshots—which include the head, neck, upper shoulders, and some margin—were approximately $500 \times 570$ pixels, with minor variations across participants. For uniformity across evaluation, we extracted these headshots and resampled them to a standard resolution of $512 \times 512$.

   \vspace{0.5em}
    \noindent \textbf{Deepfake Videos.} With the original video dataset in place, we proceeded to generate deepfake counterparts using three RTDF pipelines as follows:

     \begin{table}[t!]
        \centering
        \caption{Counts of challenge videos.}
   
        \begin{tabular}{lcccc}
        \toprule
           \textbf{\diagbox[innerwidth=2.5cm]{Challenge}{Generator}} & \textbf{Original} & \textbf{LIA} & \textbf{FSGAN} & \textbf{DFL}  \\
        \midrule
            No Challenge & 47 & 2,162 & 2,103 & 2,162 \\
            Head Movement & 46 & 2,068 & 1,953 & 2,115 \\
            Occlude w/hand & 45 & 2,068 & 1,954 & 2,115 \\
            Occlude w/Sunglass & 45 & 2,068 & 1,952 & 2,115 \\
            Occlude w/Facemask & 44 & 2,021 & 1,916 & 2,068 \\
            Manual Deform & 44 & 2,021 & 1,907 & 2,068 \\
            Protrude Tongue & 45 & 2,068 & 1,944 & 2,115 \\
            Alter Expression & 46 & 2,115 & 1,998 & 2,162 \\
            Flash & 47 & 2,162 & 2,088 & 2,162 \\
            \hline
            \textbf{Total} & \textbf{409} & \textbf{18,847} & \textbf{17,815} & \textbf{19,176} \\
                \hline
            \end{tabular}
        \label{tab:video_counts}
    \end{table}

    \begin{itemize}[nosep,leftmargin=*]
        \item \emph{LIA (Latent Image Animator)}: This pipeline is a facial reenactment method outlined in~\cite{wang2022latent}. Given its target-agnostic nature—meaning it does not require specific target data during inference—we employed a pre-trained model for our experiments.
        
        \item \emph{FSGAN (Face Swapping Generative Adversarial Network)}: This corresponds to the second version of FSGAN~\cite{nirkin_fsganv2_2022}. Similar to LIA, this model is also target-agnostic, hence we utilized a pre-trained model made available by the authors for our study.
        
        \item \emph{DFL (DeepFaceLab)}: Notorious for generating hyper-realistic deepfakes~\cite{iperov_deepfacelab_2021}, this pipeline serves as a baseline for in-the-wild deepfake videos. For our study, we trained \textit{individual DFL deepfake generators for each participant} using their `no challenge' videos. These videos records the participants in a range of frontal angles while they sit naturally, aiming to mimic the kind of data readily accessible online for non-celebrity individuals. Training continued until convergence for approximately around 300,000 iterations.
    \end{itemize}
    
    \vspace{0.5em}
    \noindent\emph{Inference Procedure for Deepfake Video Evaluation.} To build a comprehensive set of deepfake videos containing challenges, each of the 47 participants was designated as the target. In contrast, the remaining participants acted as imposters. This configuration yielded a total of $47 \times 46 = 2,162$ unique imposter-target pairs. We repeat this procedure across all three RTDF pipelines and create a rich evaluation set for every challenge (for computing details, refer to Section \S\ref{sec:compute}). Table~\ref{tab:video_counts} specifies the counts of videos generated per challenge and pipeline, totalling 500 GB of deepfake videos. Some counts are lower than 2,162 since some imposter-target pairs were missing or resulted in erroneous deepfake videos.

\vspace{0.5em}
\noindent\textbf{Ethics.} All participants signed consent forms that permit us to release their data to accredited academic institutions for non-commercial research purposes. Participants retain the option to withdraw their consent at any time. To mitigate privacy risks, we prohibit the unauthorized distribution and public display of participants' faces. Our institutional review board oversaw these provisions under IRB-FY2022-6482 and formalized them in a data release form, enforcing compliance in future research.

   \section{Evaluation}
   To comprehensively gauge the effectiveness of \textsc{Gotcha} for RTDF detection, we used both human assessments and automated scoring using ML algorithms. Below, we provide details of each evaluation strategy. 

    \subsection{Human Evaluation} \label{subsec:human_eval} 
    
    A key characteristic of the proposed challenge-response approach is that the degradations caused by the challenges are easily perceptible to the human eye, and hence, they enable explainable evaluations. To corroborate this statement, we conducted a human study. We investigated the question: \emph{\textbf{Q1.} Can humans detect deepfakes and can challenges enhance this ability?} and to ensure that the responses were not spurious, we check whether \emph{\textbf{Q2.}  Can humans identify deepfake artifacts successfully?} A preview of the experiment is available at \url{https://app.gorilla.sc/openmaterials/693684}.

    \vspace{0.5em}
    \noindent \textbf{Dataset subselection:} As evaluating the full dataset is unnecessary, we select a representative subset of 150 videos, each of $10 \pm 4$ seconds duration, as follows:
    \begin{itemize}[nosep,parsep=2pt,leftmargin=*]
        \item \emph{Challenge Category:}  We selected a single instance for each challenge category, i.e., face occlusion, head movement, facial deformation, and face illumination, along with `no challenge' samples.

        \item \emph{Face Identity:} We manually selected ten imposter-target pairs per challenge. In order to make human evaluation more challenging, we limit to imposter-target pairs with matching sex, complexion, and facial structure, thereby eliminating trivial inconsistencies.

        \item \emph{RTDF:} We consider Original, FSGAN~\cite{nirkin_fsganv2_2022} and DFL~\cite{iperov_deepfacelab_2021} samples for evaluation due to their lower artifact rates. LIA~\cite{wang2022latent} is excluded as it fails to perform most challenges (except head movement), making it trivially detectable (see Fig.~\ref{fig:challenges}). 
    \end{itemize}

\begin{figure}[ht]
        \centering
        \includegraphics[width=\columnwidth]{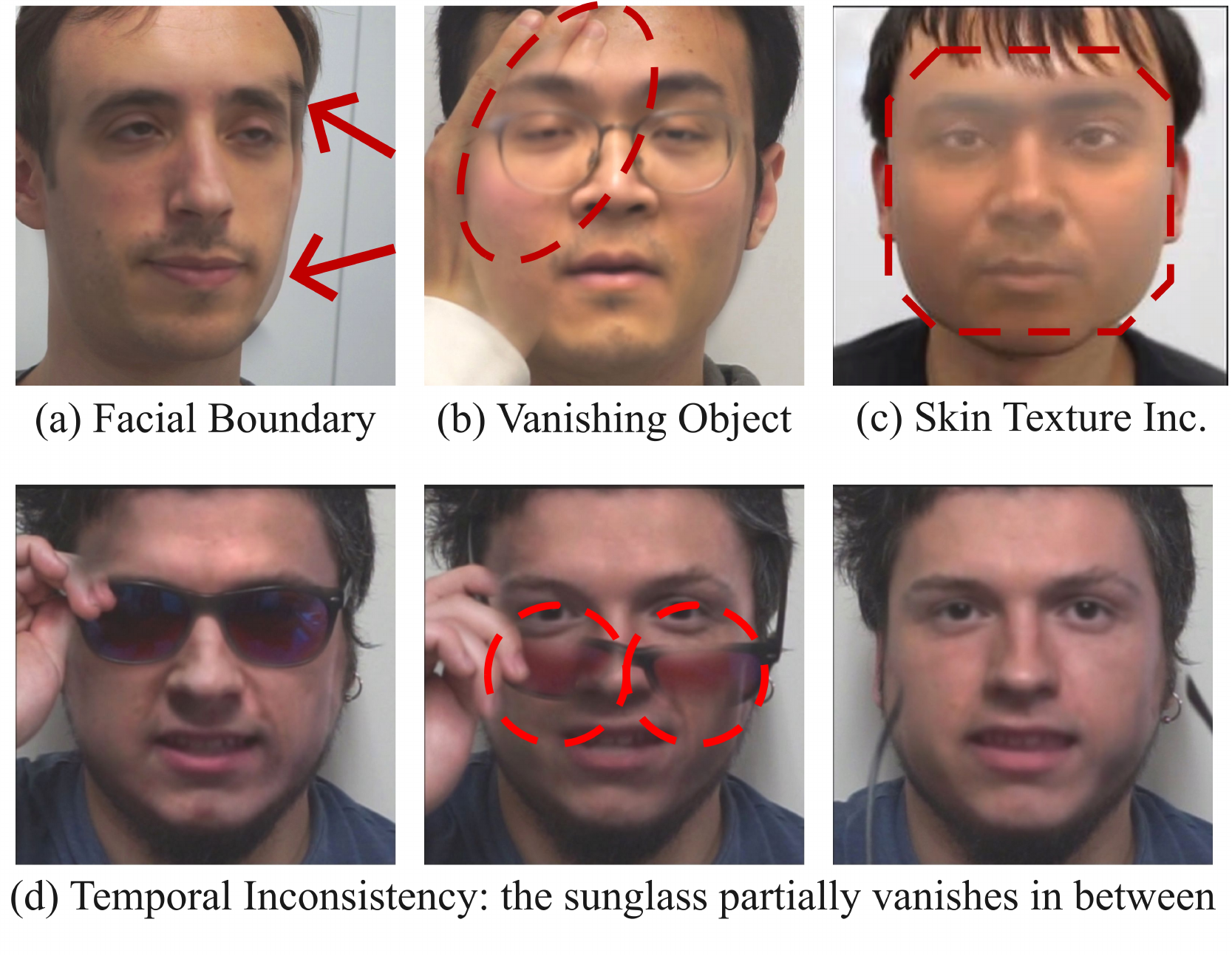}
        \vspace{-1em}
        \caption{Artifacts defined for human evaluation. (a) has a boundary artifact near the right brow, in (b) the hand vanishes behind the face, (c) has hazy face detail, and (d) the sunglass starts vanishing briefly. Red objects indicate artifact locations. Video version at \url{http://govindm.me/gotcha-figures/}.}
        \label{fig:visible_artifacts}
    \end{figure}

    \vspace{0.5em}
    \noindent \textbf{Screening:}  We began by familiarizing each participant with different types of deepfake artifacts they might encounter. Instructions included definitions and example videos of `no artifact' and four suggestive artifacts -- Facial Boundary Artifacts, Vanishing Object, Skin Texture Inconsistency, and Temporal Inconsistency. For examples see Fig.~\ref{fig:visible_artifacts} and for definitions see \S\ref{sec:human_eval_details}, or 
    
    After instructions, we tested each participant by asking them to select the most apparent artifact present in 10 previously unseen videos. Out of a pool of 60 recruited participants (males = females = 30), 43 participants qualified passing the test and proceeded to the main task. 
      \begin{figure}[t!]
        \centering
        \includegraphics[width=0.9\columnwidth]{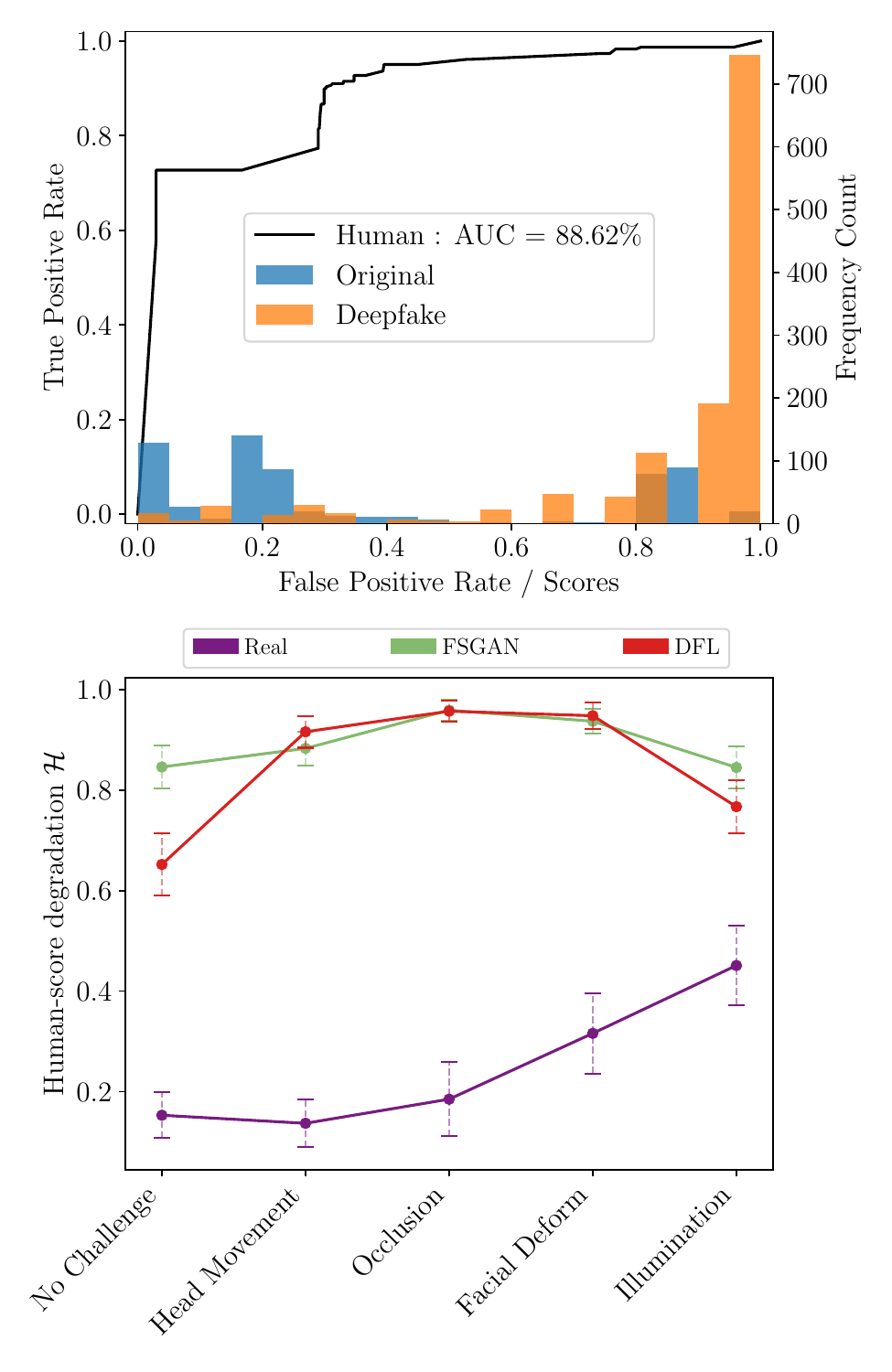}
        \caption{\textbf{(Top)} ROC curve of human-score degradation for original and deepfake videos along with histogram of both categories spread across the scores. (\textbf{Bottom}) Boxplot of mean degradation scores with 95\% confidence interval against the considered challenge categories.} 
        \label{fig:human_scores}
    \end{figure}
    
    \vspace{0.5em}
     \noindent \textbf{Main Task.} We tasked each qualifying participant to evaluate 45 videos randomly sampled from the subset of 150 videos. The experiment resulted in an average of 12.9 evaluations per video sample and accumulated 129 evaluations per challenge per RTDF. Each evaluation comprised of four responses (see their screenshots in \S\ref{sec:human_eval_details}):
    \begin{itemize}[nosep,parsep=2pt,leftmargin=*]
        \item \emph{Compliance to challenge} ($\mathcal{C}$): A binary choice determining \emph{whether the subject in the video executed the action specified by a text description of the challenge}.
        \item \emph{Realism of the video} ($\mathcal{R}$): A slider ranging from 0~to~100\% in steps of 5\%, captured \emph{how fake the video appears}. Higher scores indicate a less convincing video.
        \item \emph{Identification of artifacts}: A multi-selection question asked participants to identify \emph{which of the artifacts listed above were present in the video}. 
        \item \emph{Localization of artifacts}: A grid with 18 cells was overlayed over the face and neck regions, and participants needed to select \emph{the locations of visible artifacts}.
    \end{itemize}

\vspace{0.5em}
\noindent \textbf{Metric and Visualizations.} 
    Using the responses to Compliance~$\mathcal{C}$ and Realism~$\mathcal{R}$, we define the \emph{mean human-score degradation} $\mathcal{H}$ as:
    {\small
    \begin{equation*}
        \mathcal{H}_\mathcal{P}(c) = \frac{1}{n} \sum_{i=1}^n \max\left(1 - \mathcal{C}(V^i_\mathcal{P}(c)), 1-\mathcal{R}(V^i_\mathcal{P}(c))\right),
    \end{equation*}}
     where $V (c)$ is a video of challenge $c$ either original or generating using RTDF pipeline $\mathcal{P}$, and $n$ is the response count. This metric assigns a 100\% degradation value to a video that fails to comply with the given challenge or a lower score based on its realism value.
    
    Fig.~\ref{fig:human_scores} presents the human performance in detecting deepfakes using the above metric. The top part shows an ROC curve with the frequency count of responses attributed to original and deepfake videos. The bottom part contains mean human scores $\mathcal{H}$ for original videos, DFL, and FSGAN across challenge classifications, averaged with $n = 129$ responses.
    
\begin{figure}[t!]
        \centering
        \includegraphics[width=\columnwidth]{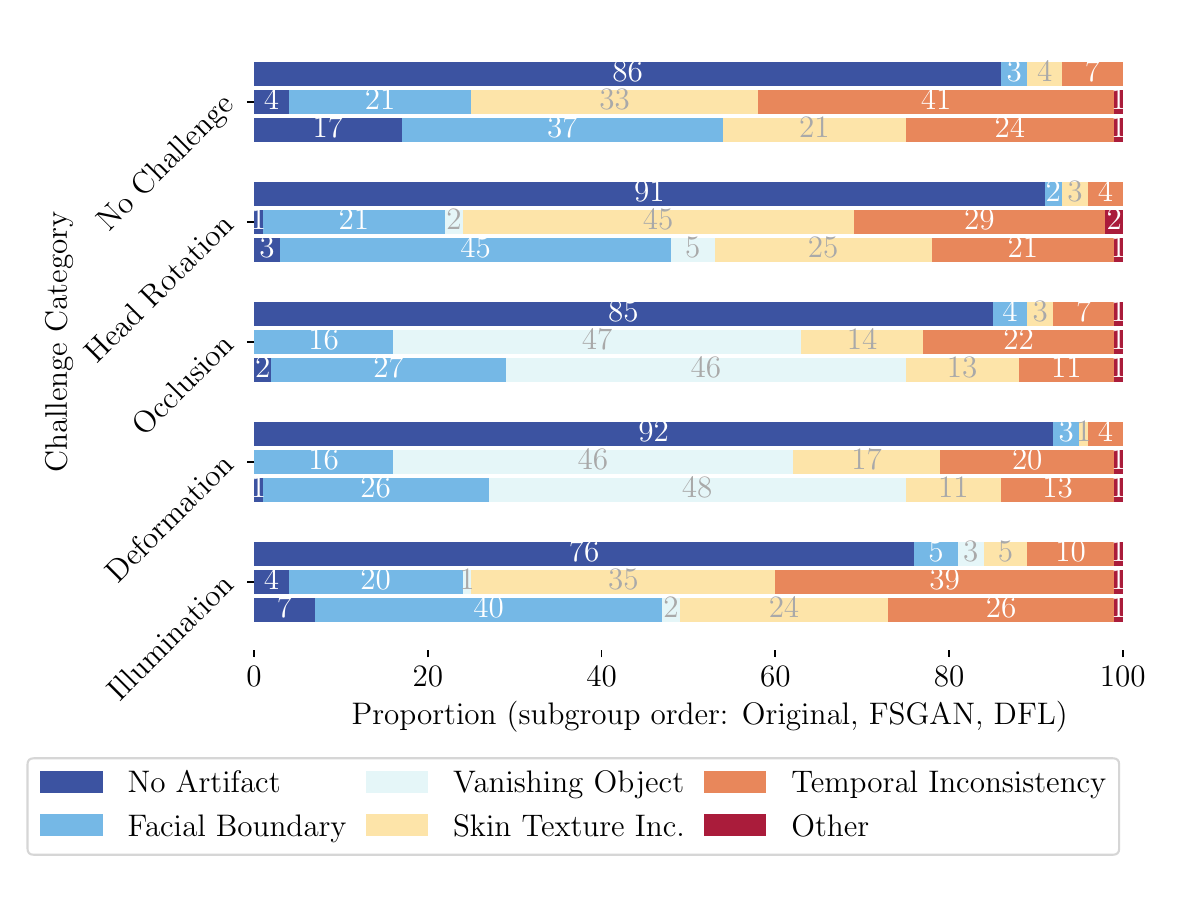}
        \caption{Proportion of artifacts present in each challenge per generator, tagged by human-evaluators. Each groups of three is in the order - Original, FSGAN, DFL.} 
        \label{fig:justification}
    \end{figure}

\begin{figure}[t!]
        \centering
        \includegraphics[width=1.05\columnwidth]{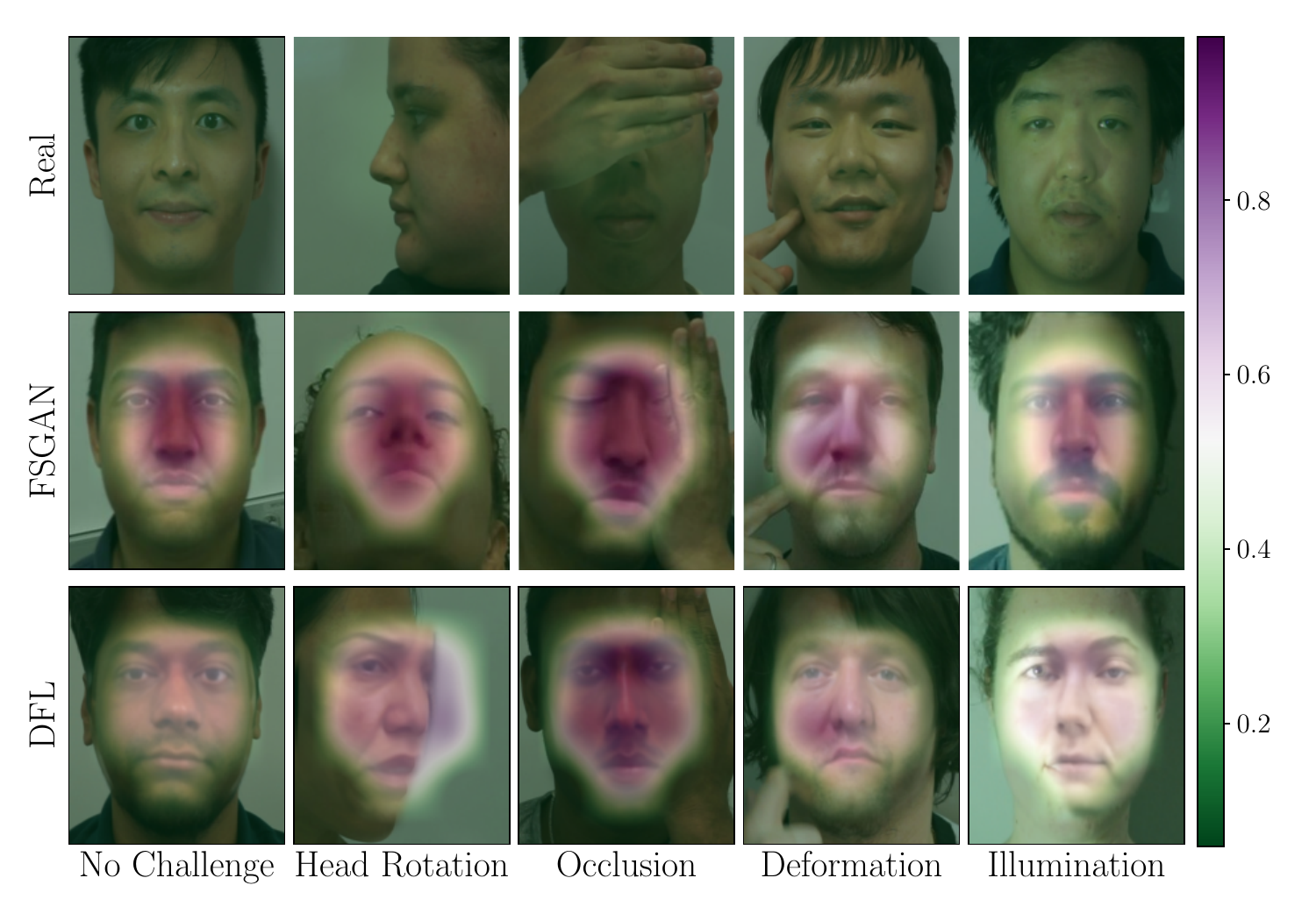}
        \caption{Distribution of artifact locations selected by human-evaluators. The background images are representative examples of each challenge.} 
        \label{fig:localization}
    \end{figure}

    Fig.~\ref{fig:justification} illustrates the distribution of artifacts (including `no artifact') tagged to a challenge category. Fig.~\ref{fig:localization} overlays a saliency map-like distribution on a representative challenge image to indicate the locations of detected artifacts. We reiterate that we provided each evaluator with bounding boxes spread across the neck and face region (see Fig.~\ref{fig:sdfl_human_eval}~d). We added, normalized, and smoothened all responses for videos of a challenge and RTDF to produce the illustrated distribution.

\vspace{0.5em}
\noindent \textbf{Inferences.} 
    \noindent \emph{Challenges help humans to detect RTDFs} 
    Humans consistently scored RTDFs to have higher degradation compared to original videos with AUC of 88.6\% resulting in an accuracy of 81.2\% (threshold~=~37\%). The introduction of a challenge caused a 25\% rise in DFL's degradation score and a 12\% increase in original videos. In contrast, FSGAN's score rose by only 6\%, likely due to its already high baseline scores of `no challenge' videos. This low gain is attributable to textural artifacts in the target-agnostic FSGAN videos, even without challenges. In contrast, DFL videos appeared more natural as we trained them with target data. 
    
    Interestingly, challenges involving illumination introduced human error, causing evaluators to misidentify some original videos to be RTDFs and vice versa. This error in judgment was likely caused by confusion in discerning lighting changes when a flash was turned on-and-off.
    
    \vspace{0.5em}
    \noindent\emph{Humans were effective in identifying RTDF artifacts.}
    Observing Fig.~\ref{fig:justification}, human evaluators consistently labeled original videos as having `no artifact' while correctly identifying at least one type of artifact in RTDF videos with an accuracy of 92.7\%. Specifically, for challenges involving occlusion and facial deformation, the evaluators accurately tagged the RTDF videos as containing a `vanishing object' 46.8\% of the time. This accuracy rate increased to 67.8\% when we also considered the signature artifacts specific to each RTDF generator, namely, `skin texture inconsistency' for FSGAN and `facial boundary artifacts' for DFL. When combining the above observations, the overall human accuracy for artifact identification reached 80.2\%.

    Observing Fig.~\ref{fig:localization}, original videos get attributed with fainter distributions, suggesting they have fewer artifacts, while RTDF videos indicate a concentration of more pronounced artifacts. This pattern is consistent across both FSGAN and DFL. As more noticeable artifacts are spread specifically in the inner face region, they suggest that human evaluators are also localizing these imperfections. 
    
    \vspace{0.5em}
    \noindent \textbf{Takeaway.} Our human-study yields affirmative answers to the questions posed. While the efficacy of challenges in improving detection could be contingent on the quality of the generator output, they undeniably make artifacts more visible, thereby aiding human evaluators in assessing the authenticity of video calls.
    
   \subsection{Automated Evaluation}
    \label{subsec:chal_eval}
   A natural question arises whether RTDFs can be detected in a scalable way. To this end, we evaluated traditional deepfake detection techniques and go on to develop our own in-house detector. 
    
    In order to automatically differentiate deepfakes from original videos, we need a fidelity score model along detector to infer compliance. The primary task of a fidelity score model, denoted as $f$, would be to measure the difference—or degradation—in the quality of an impersonation generated by an imposter with respect to the genuine target, particularly during a challenge scenario. Complementary to fidelity, compliance detection, denoted as $\mathcal{C}$, would inform whether a video contains the corresponding challenge. 
    
    Let $V_\mathcal{P}(imp, tg, c)$ represent a deepfake video wherein an imposter $imp$ employs an RTDF pipeline $\mathcal{P}$ to impersonate a target $tg$ while undergoing challenge $c$. Using the same notation, genuine original video are represented by $V_{\text{orig}}(\phi, tg, c)$. We compute compliance~$\mathcal{C}$ and fidelity loss~$\Delta\mathcal{F}$ and combine them into the \textbf{normalized machine-scored degradation} $\mathcal{M}$, an aggregate measure of this quality difference:

    \vspace{-1em}
    {\small
    \begin{gather*}
    \text{fake}_i(\mathcal{P}, c, imp, tg) = V^i_{\mathcal{P}}(imp, tg, c) \\
    \text{orig}_i(c, tg) = V^i_{\text{orig}}(\phi, tg, c)\\
    \Delta\mathcal{F}(\text{fake}_i, \text{orig}_i) =  \frac{w}{n}\sum_{i=1}^{n/w} \frac{f(\text{fake}_i) - f(\text{orig}_i)}{f(\text{orig}_i)} \\ 
    \mathcal{C}(\text{fake}_i, \text{orig}_i) = 
    \left\{
    \begin{array}{ll}
    1 & \text{if } \mathcal{C}_c(\text{fake}_i, \text{orig}_i)\text{ is True} \\
    0 & \text{otherwise}
    \end{array}
    \right.\\
    \mathcal{M}_{\mathcal{P}}(c) =  \sum_{\forall imp,tg}\text{max}(\Delta\mathcal{F}(\text{fake}_i, \text{orig}_i),  1 - \mathcal{C}(\text{fake}_i, \text{orig}_i)).
    \end{gather*}
    }
    
    \noindent Here, $n$ stands for the total number of frames in the video,  $V^i$ is the $i^{th}$ fragment of the video with $w$ frames, and $\mathcal{C}_c$ is a challenge-specific compliance detector.
    \begin{figure}[t!]
        \centering
        \includegraphics[width=0.8\columnwidth]{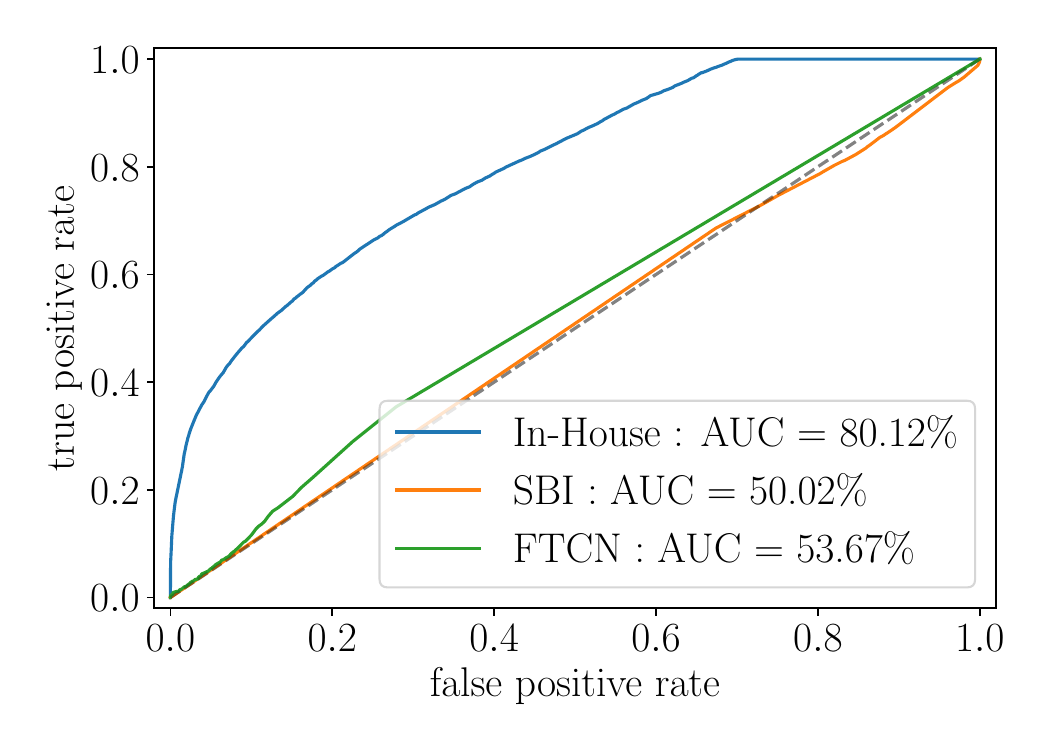}
        \caption{ROC curves for deepfake detectors, SBI~\cite{shiohara2022detecting} and FTCN~\cite{zheng2021exploring}, and our in-house fidelity model. The Area-Under-Curve (AUC) measures the degree of separation between original and deepfake video scores; a higher AUC implies higher discrimination, thus better performance.}
        \label{fig:detector_roc}
    \end{figure}

     \begin{table*}[ht!]
        \caption{Compliance $\mathcal{C}$ and Fidelity Loss $\Delta\mathcal{F}$ along with Machine-scored Degradation $\mathcal{M}_\mathcal{P}(c) = \text{max}(\Delta\mathcal{F}, 1 - \mathcal{C})$ in \%. Standard deviations are indicated as subscripts of Means. Arrows signify direction of increase of failure rate in deepfakes. The challenges are sorted in ascending order of their efficacy. }
       \centering
       \resizebox{2.1\columnwidth}{!}{\begin{tabular}{L{2.8cm}L{2.95cm}|C{0.65cm}C{0.75cm}C{0.62cm}|C{0.65cm}C{0.75cm}C{0.62cm}|C{0.65cm}C{0.75cm}C{0.62cm}|C{0.70cm}}
    \toprule
       \multirow{2}{2.7cm}{\textbf{\diagbox[innerwidth=2.7cm]{Challenge \(c\)}{RTDF \(\mathcal{P}\)}}} & \multirow{2}{2.90cm}{\textbf{Compliance Detection Strategy}} & \multicolumn{3}{c}{\textbf{DFL}} & \multicolumn{3}{c}{\textbf{FSGAN}} & \multicolumn{3}{c}{\textbf{LIA}} & \textbf{Chal.}\\
        & & $\mathcal{C}\downarrow$ & $\Delta\mathcal{F}\uparrow$ & $\mathcal{M}\uparrow$ & $\mathcal{C}\downarrow$ & $\Delta\mathcal{F}\uparrow$ & $\mathcal{M}\uparrow$ & $\mathcal{C}\downarrow$ & $\Delta\mathcal{F}\uparrow$ & $\mathcal{M}\uparrow$ & \textbf{Avg.}\\
    \midrule
	No Challenge & - & 100.0 &10.3 $ ^{\pm 7.3}$ &10.3 $ ^{\pm 7.3}$ &100.0 &12.1 $ ^{\pm 6.9}$ &12.1 $ ^{\pm 6.9}$ &100.0 &16.6 $ ^{\pm 5.6}$ &16.6 $ ^{\pm 5.6}$ & 13.0 \\ 
	Head Movement & Change in Yaw and Pitch & 97.6 &17.0 $ ^{\pm 3.5}$ &18.4 $ ^{\pm 11.1}$ &98.1 &16.8 $ ^{\pm 4.0}$ &18.2 $ ^{\pm 11.4}$ &72.8 &23.6 $ ^{\pm 5.9}$ &43.0 $ ^{\pm 33.8}$ & 26.5 \\ 
	Illuminate w/ Flash & Peaks in Face Intensity & 51.6 &19.7 $ ^{\pm 10.5}$ &58.8 $ ^{\pm 40.8}$ &83.3 &17.7 $ ^{\pm 11.0}$ &31.2 $ ^{\pm 32.5}$ &51.6 &17.1 $ ^{\pm 9.7}$ &57.2 $ ^{\pm 42.1}$ & 49.1 \\ 
	Alter Expression* & Expression Recognition & 30.1 &14.7 $ ^{\pm 9.4}$ &65.9 $ ^{\pm 42.2}$ &35.8 &13.9 $ ^{\pm 7.3}$ &57.8 $ ^{\pm 43.2}$ &34.9 &16.4 $ ^{\pm 9.2}$ &60.2 $ ^{\pm 42.0}$ & 61.3 \\ 
	Occlude w/ Sunglasses* & Face Segmentation (change in face coverage) & 51.8 &11.6 $ ^{\pm 5.0}$ &53.3 $ ^{\pm 44.2}$ &7.2 &10.3 $ ^{\pm 4.4}$ &93.6 $ ^{\pm 23.1}$ &51.8 &14.3 $ ^{\pm 7.3}$ &56.2 $ ^{\pm 42.7}$ & 67.7 \\ 
	Protrude Tongue* & Object Detector & 20.3 &15.5 $ ^{\pm 15.3}$ &81.5 $ ^{\pm 34.4}$ &19.9 &16.1 $ ^{\pm 13.7}$ &81.7 $ ^{\pm 35.4}$ &20.3 &17.9 $ ^{\pm 12.6}$ &82.7 $ ^{\pm 33.2}$ & 82.0 \\ 
	Occlude w/ Facemask* & Face Segmentation (change in face coverage) & 22.0 &33.0 $ ^{\pm 15.2}$ &81.1 $ ^{\pm 34.1}$ &21.7 &18.4 $ ^{\pm 8.5}$ &83.2 $ ^{\pm 32.0}$ &22.0 &39.4 $ ^{\pm 18.7}$ &83.5 $ ^{\pm 31.5}$ & 82.6 \\ 
	Manual Deform & Finger-on-Face Detector & 20.3 &24.0 $ ^{\pm 21.2}$ &82.7 $ ^{\pm 32.1}$ &19.8 &23.9 $ ^{\pm 20.8}$ &82.9 $ ^{\pm 33.3}$ &20.3 &21.3 $ ^{\pm 16.9}$ &82.5 $ ^{\pm 33.9}$ & 82.7 \\ 
	Occlude w/Hand & Face Segmentation (change in face coverage) & 11.2 &12.4 $ ^{\pm 5.5}$ &88.6 $ ^{\pm 29.4}$ &6.2 &12.6 $ ^{\pm 4.7}$ &94.6 $ ^{\pm 21.0}$ &11.2 &19.9 $ ^{\pm 9.3}$ &91.0 $ ^{\pm 25.4}$ & 91.4 \\ 
    \midrule
    RTDF avg. (chal only)           & -  & 38.1 & 18.5 & 66.3 & 36.5 & 16.2 & 67.9 & 35.6 & 21.2 & 69.5 & - \\
    \bottomrule
    \multicolumn{3}{l}{\smaller *unseen challenge during training}
\end{tabular}

}
        \label{tab:auto_score}
    \end{table*}

    \vspace{0.5em}
    \noindent \textbf{Realizing a Fidelity Score Model.} Initially, our evaluation incorporated well-established offline deepfake detection algorithms, aiming to adapt them as the fidelity score \( f \). Specifically, we considered Self-Blending Images (SBI)~\cite{shiohara2022detecting} and Fully Temporal Convolution Network (FTCN)~\cite{zheng2021exploring} as candidates, due to their promising achievements on standard deepfake datasets such as Celeb-DF~\cite{celebdf},  DFDC~\cite{dfdc}, and FFIW~\cite{zhou2021face}. However, they under-performed on a set of 620 original and 25,400 deepfake videos derived from our dataset (see Fig.~\ref{fig:detector_roc}). We attribute the low performance of previous methods to the introduction of novel artifacts by the challenges, which are largely absent in their respective training corpora. Hence, we created a customized model to score the fidelity of the videos more reliably.

    Like prior works, we use an ML-based approach utilizing a 3D convolutional neural network as backbone, specifically, a 3D-ResNet18~\cite{hara2017learning}. 3D-ResNets have been used for predominantly for action recognition in videos and can be used to model temporal information, making them suitable for tasks that require understanding the temporal dynamics in videos. This model quantifies the loss in fidelity induced by each challenge while being trained only on a subset of our dataset and outperforms others (see Fig.~\ref{fig:detector_roc}). The training subset contains the following subselections:
    \begin{itemize}[nosep,parsep=2pt,leftmargin=*]
        \item Original and deepfake videos with four challenges, one from each category and `no challenge' (the first five challenges listed in Table~\ref{tab:auto_score}),
        \item 32 frames per sample,
        \item 35 out of 47 target identities, and
        \item Deepfake videos created using only DFL.
    \end{itemize}

    The training objective was to optimize a contrastive loss that draws the embeddings of original samples closer together during training while pushing deepfake video embeddings further apart~\cite{kopuklu2021driver}. Using $f$, we compute the deviation in fidelity $\Delta\mathcal{F}$ of a deepfake video from its ground truth. 

    \vspace{0.5em}
    \noindent \textbf{Compliance Detectors.} While the fidelity score $f$ can measure video artifacts, it does not account for cases where an imposter neglects the request to execute the challenge or the deepfake generators silently fail to replicate the intended action. There are minimal artifacts in such cases despite being a fake video. Hence, we verify compliance to each challenge $\mathcal{C}$ to account for them.

    As each challenge $c$ relies on a different signal for verifying compliance, we deploy a set of challenge-specific compliance detectors $\mathcal{C}_c$, based on the following strategies:
    
    \vspace{0.25em}
    \begin{itemize}[nosep,parsep=2pt,leftmargin=*]
    \item \textit{Occlusion:} We use MediaPipe~\cite{lugaresi2019mediapipe} to segment any visible part of the face and compute face coverage (ratio of visible face to the whole frame). For hand occlusion scenarios, where participants place their hands on their faces four times while facing the camera, we fit a sinusoid to the face coverage data and consider the video complied if the sinusoid has significant amplitude.

    \vspace{-0.25em}
    \hspace{0.5em} Also, we consider the video compliant for facemasks and sunglass occlusions if the mean face coverage difference from the exact original video deviates by less than 1\%.

    \item \textit{Head Movements:} We predict yaw and pitch for each frame in the video~\cite{yawpitch_compliance}. The video is compliant if the pitch and yaw values indicate significant head movement ($>10$ or $<-10$) in all four directions.

    \item \textit{Facial Deformations:} We address challenges involving manual deformation, such as poking the cheek with a finger or protruding the tongue using objects and face detectors. We use the detectors to identify and locate the position of respective objects and grant compliance if these objects are detected and their position intersects with the correct facial region (right/left face for the finger, the lower face for the tongue).

    \item \textit{Expressions:} Facial expressions are predicted using an EfficientNet model trained on the AffectNet dataset~\cite{fer_compliance}, followed by computing Shannon's entropy. A video complies with altering facial expressions if the difference in facial expression entropy between the ground truth and the deepfake video is less than 0.20.

    \item \textit{Illumination:} Face intensity changes (in grayscale) are computed across the video, focusing on the presence of sharp peaks. The video is compliant if the difference in the number of such peaks between deepfake and original videos is less than 10\%.
    \end{itemize}

    The final machine-scored degradation derived from fidelity and compliance achieves an AUC of 80.1\%. Refer to \S\ref{subsec:detector} for training and architecture details. Note that we do not assert that this model is the new SoTA. We rather imply that deepfake detection is a simple problem to pose but a complex problem to solve in practice.
    
    We used full videos during scoring, each comprising hundreds of frames per sample. Table~\ref{tab:auto_score} describes the normalized machine-scored degradation score of each evaluated RTDF generator \emph{relative to their original counterparts}. Although we split the seen identities during the training of fidelity score to follow best practices for evaluating ML models, we show results on all identities.

    \vspace{0.5em}
    \noindent\textbf{Observations.} The `no challenge' condition typically resulted in lower scores than most challenge scenarios, with them scoring 100\% compliance due to cooperating participants. Among the evaluated RTDF pipelines, the facial-reenactment-based LIA fared the worst across all challenges. DFL and FSGAN came close while struggling against occlusion challenges, likely due to their models' lack of segmentation capabilities. 

    The observation that the fidelity score function captured the fidelity loss across all challenges, despite some not seen during training (asterisk marked in Table~\ref{tab:auto_score}), suggests that it has inherent generalizable capabilities.
    
    Compliance and realism provide complementary signals, with compliance playing a significant role especially in cases of occlusions and manual deformations. On average, failure to comply to a challenge contributes three times the degradation compared to loss in fidelity.
    
    \vspace{0.5em}
    \noindent\textbf{Takeaway.} We infer that a machine-assisted scoring method can aid in detecting RTDFs by measuring their deviation from pre-determined original videos, with the effect of challenges disproportionately impacting deepfake videos more.
    

    \vspace{0.5em}
    \noindent\textbf{Summary.} Human and Machine evaluations showcase the effectiveness of \textsc{Gotcha} in drawing out the inherent weaknesses in real-time deepfake generation in an interpretable and scalable way. Interestingly, both evaluations have a perfect alignment in their ordering of challenge categories efficacy:
    
    \vspace{0.25em}
    \noindent No challenge $<$ Face Illumination $<$ Head Movement $<$ Facial Deformation $<$ Occlusions.
    \vspace{0.25em}
    
    The prescribed challenges offer reliable protection against deepfakes with a mere 10 seconds of engagement. Moreover, they support both human-interpretative explanations and scalable algorithmic evaluations.

\section{Defenses against Real-Time Deepfakes}
    Our results indicate that  \textsc{Gotcha}  can be an efficient tool to counter the  threat of RTDFs. When effectively implemented, a single challenge can unmask a deepfake within a brief time-frame of 15 seconds. 
    
    However, this effectiveness presupposes a naïve threat. In a more realistic scenario, a savvy imposter could be aware of the nature of the proposed challenges and  adapt accordingly. For example, they could curate pre-fabricated responses or advance the technological capabilities of deepfake algorithms to evade detection.
    
    On the other hand, defenders have limited situational awareness, i.e., they are versed with the general attributes of deepfake generation but are blind to specifics about the person they are striving to authenticate. Thus, they can rely only on the inherent limitations of the generation process and make educated assumptions about the data an imposter could potentially access. 

    In the rest of this section we discuss the inherent limitations of the imposter, and the countermeasures they can employ to evade detection.   Armed with the limited resources at our disposal, we outline simple but effective mitigation strategies a defender can employ to confound adaptive imposters. These include randomising each challenge instance and deploying a sequence of challenges. However, it is essential to acknowledge that such mitigation adds complexity to the authentication process, which may affect user experience. We explore these tradeoffs in subsequent discussions.

    \subsection{Limitation of Imposters}
     \label{sec:insights}
      In this subsection, we discuss inherent limitations of the deepfake generators. First we look at the typical architecture of an RTDF and examine the limitation of the essential components of a standard design. Second, we look at the effect of the availability of data on the intended target  on the quality of the impersonation. Both these factors  become important tools for the design of challenges to help expose RTDFs even against a resourceful and adaptive adversary.  

\begin{figure}[t!]
\centering
    \subfloat[Facemask]{%
        \begin{minipage}{0.25\columnwidth}
            \includegraphics[width=\columnwidth]{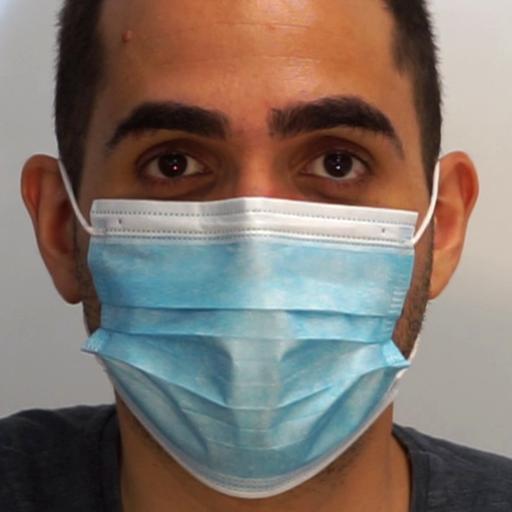}\\
            \includegraphics[width=\columnwidth]{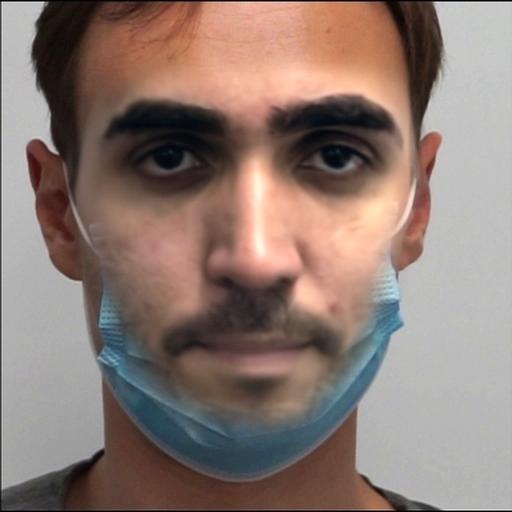}
        \end{minipage}
    }%
    \subfloat[Occlusion]{%
        \begin{minipage}{0.25\columnwidth}
            \includegraphics[width=\columnwidth]{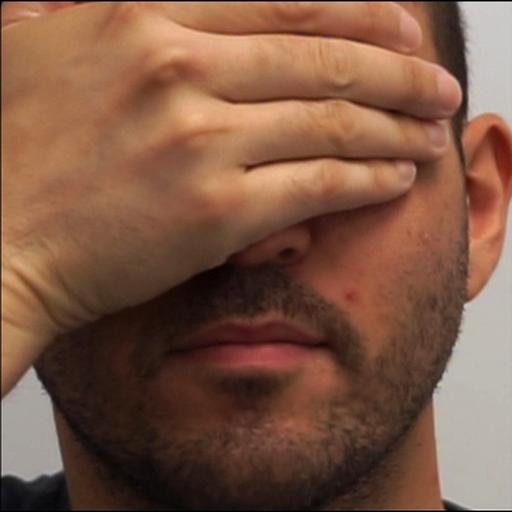}\\
            \includegraphics[width=\columnwidth]{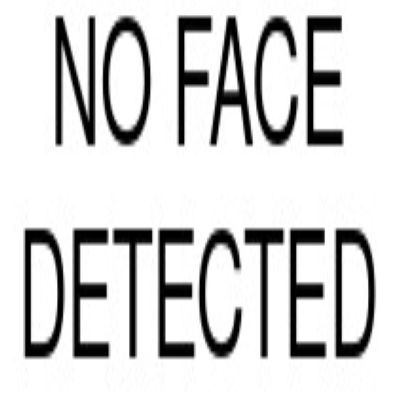}
        \end{minipage}
    }%
    \subfloat[Look up]{%
        \begin{minipage}{0.25\columnwidth}
            \includegraphics[width=\columnwidth]{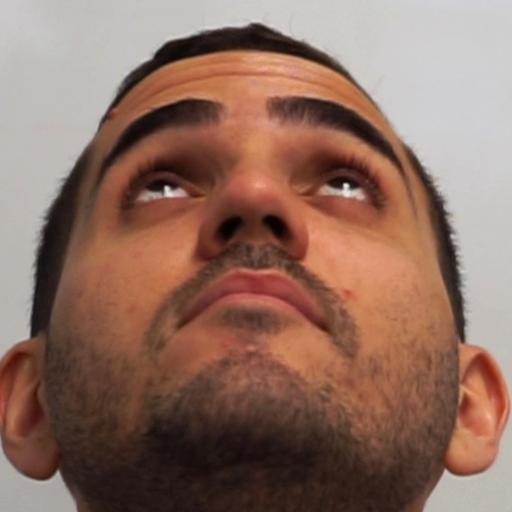}\\
            \includegraphics[width=\columnwidth]{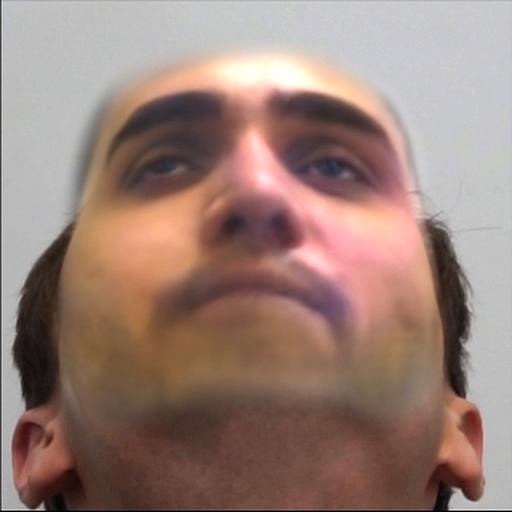}
        \end{minipage}
    }
    \\
    \subfloat[Deformation]{%
        \begin{minipage}{0.25\columnwidth}
            \includegraphics[width=\columnwidth]{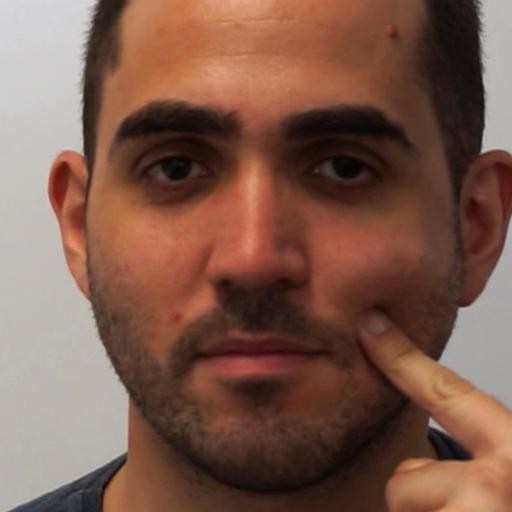}\\
            \includegraphics[width=\columnwidth]{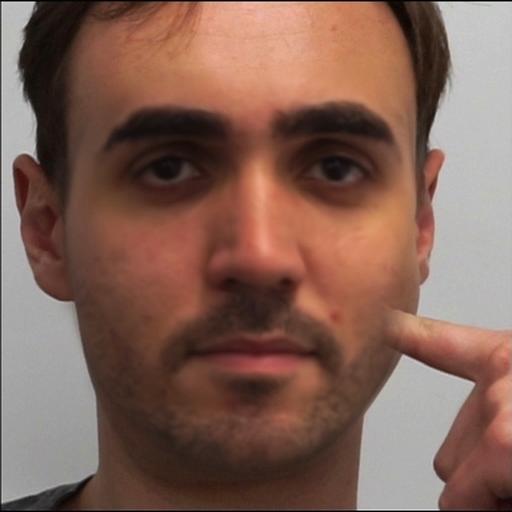}
        \end{minipage}
        }%
    \subfloat[Expressions]{%
        \begin{minipage}{0.25\columnwidth}
            \includegraphics[width=\columnwidth]{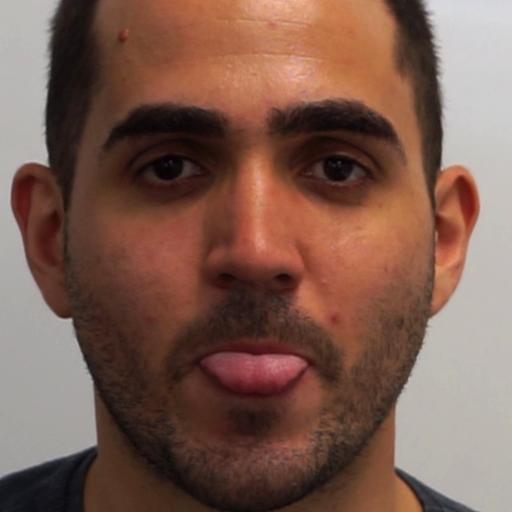}\\
            \includegraphics[width=\columnwidth]{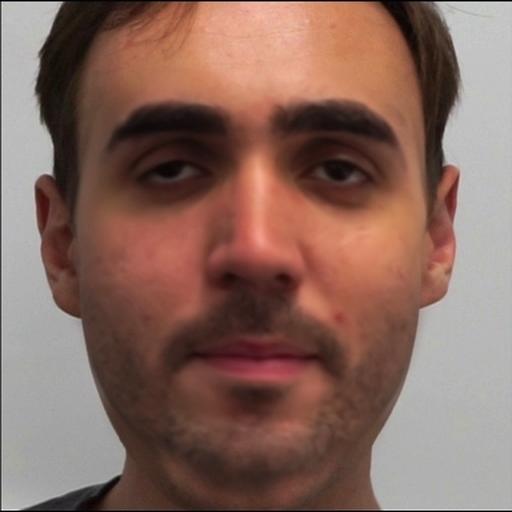}
        \end{minipage}
    }%
    \subfloat[Illumination]{%
        \begin{minipage}{0.25\columnwidth}
            \includegraphics[width=\columnwidth]{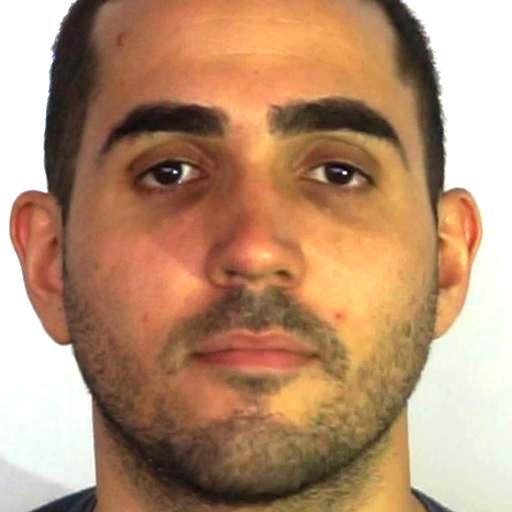}\\
            \includegraphics[width=\columnwidth]{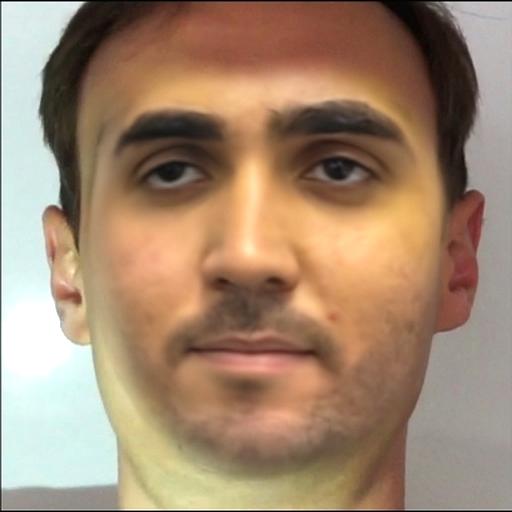}
        \end{minipage}
        }
\caption{This figure showcases a range of tasks more expansive than those displayed in Fig.~\ref{fig:challenges}. The top row presents the ground truth target images, whereas the bottom row contains the corresponding deepfake predictions of target images.} 
\label{fig:occs}
\end{figure}

    \vspace{0.5em}
    \noindent \textbf{Inherent limitations of RTDF generation pipeline} Traditional literature considers RTDF pipelines as monolithic entities; however, we examined how to degrade each component individually. The following description provides insights into the vulnerabilities of typical components found in a generic deepfake generator (refer to \S\ref{sec:rtdf_pipeline}). 
    The examples provided in this  subsection are based on experimentation done on DeepFaceLab~\cite{iperov_deepfacelab_2021}.

    \vspace{0.5em}
    \noindent \emph{Face Detection, Landmark Detection, Face Alignment.} 
        These components are quite robust due to their ability to perform effectively on diverse inputs (including faces in profile, distorted faces, and facial expressions). This robustness is primarily due to the abundance of high-quality face datasets available for training, and the maturity of computer vision techniques.
        
        We find that occluding eyes significantly degrades the performance of these modules, as this region is a critical feature for these components (see Fig.~\ref{fig:occs}~b). Additionally, since face landmark detection only provides limited facial structural information, any significant facial deformation or depression can cause noticeable anomalies. E.g., even with MediaPipe~\cite{lugaresi2019mediapipe}, which provides 468 landmarks for a full frontal face, RTDFs are unable to accurately capture significant facial depressions (Fig.~\ref{fig:occs}~d).

        Despite their robustness, experiments reveal that under certain conditions, these modules can still produce unexpected results, which can then be used to the defender's advantage. These modules can mistakenly overfit to ``face-like'' features in a given scene (such as a cartoon of a face). In these cases, the rest of the deepfake generation process gives garbled outputs. 
        
        In summary, while the above modules themselves serve as essential pre-processing steps for the next components, they are still vulnerable to challenges that induce out-of-distribution data.
  
    \vspace{0.5em}
    \noindent \emph{Face-swapper} module contains most of the target-specific facial information. Hence, this module is freshly trained on target data, making it \emph{the most vulnerable component} to out-of-distribution data. Challenges designed to degrade earlier components eventually affect the face-swapper, which is inadequate for handling degraded inputs. 
 
    Experiments indicate that numerous types of challenges prove effective in disrupting the face-swapper, including (a) structured light illumination, (b) object occlusion, and (c) facial deformations, to name a few. In Fig.~\ref{fig:occs}~d, the generated expressions do not match the original expressions (also corroborated by~\cite{mazaheri2022detection}); the deepfake generator fails to copy the distortions faithfully. 

   \vspace{0.5em}
    \noindent \emph{Segmentation.}
    An RTDF pipeline optionally includes a face-segmentation module that improves swap quality. Experiments reveal that if this module is absent, the resulting RTDF is brittle to occlusions, producing unnatural artifacts. E.g., Fig.~\ref{fig:occs}~(a) shows that an occluding object (here, face-mask) blends with the face itself. 

  \vspace{0.5em}
    \noindent \emph{Blending.}
    Blending is a multi-stage post-processing task that overlays the swapped prediction of the target face on the imposter's face. These operations range from smoothing, color correction, and positioning to scaling. Color correction (CC) is a vital blending step that samples a color from the imposter's outer face and distributes it over the predicted target face, or vice-versa.

    Several hyper-parameters control all the above operations. Any imposter would begin an interaction with a well-calibrated set of hyper-parameters that suits the target-imposter pair and environmental conditions (e.g., lighting). Any inconsistency or abrupt change in these settings can lead to blending artifacts.

    Since ambient challenges can introduce abrupt changes in the environment, these hyper-parameters become uncalibrated in the changed setting. For instance, in Fig.~\ref{fig:occs}~f, the flash effect illuminates a smaller portion causing a boundary effect on the fake face. Similarly, in Fig.~\ref{fig:occs}~a, the segmentation module inaccurately includes a surgical mask as part of the face, and the CC module imputes incorrect colors, resulting in clear artifacts.
    
    Temporal incoherence is another factor affecting blending. Most RTDFs generate predictions frame-by-frame independently, or with dependencies on a limited number of past frames (usually $\leq$ 10), as it throttles throughput. In scenarios where an object moves rapidly in front of the face, the face-swap can fluctuate across the imposter's face, when observed frame-wise (see Fig.~\ref{fig:visible_artifacts}~d). This act results in visible temporal inconsistencies in deepfakes.

        \begin{figure}[t]
      \centering
      \includegraphics[width=\columnwidth]{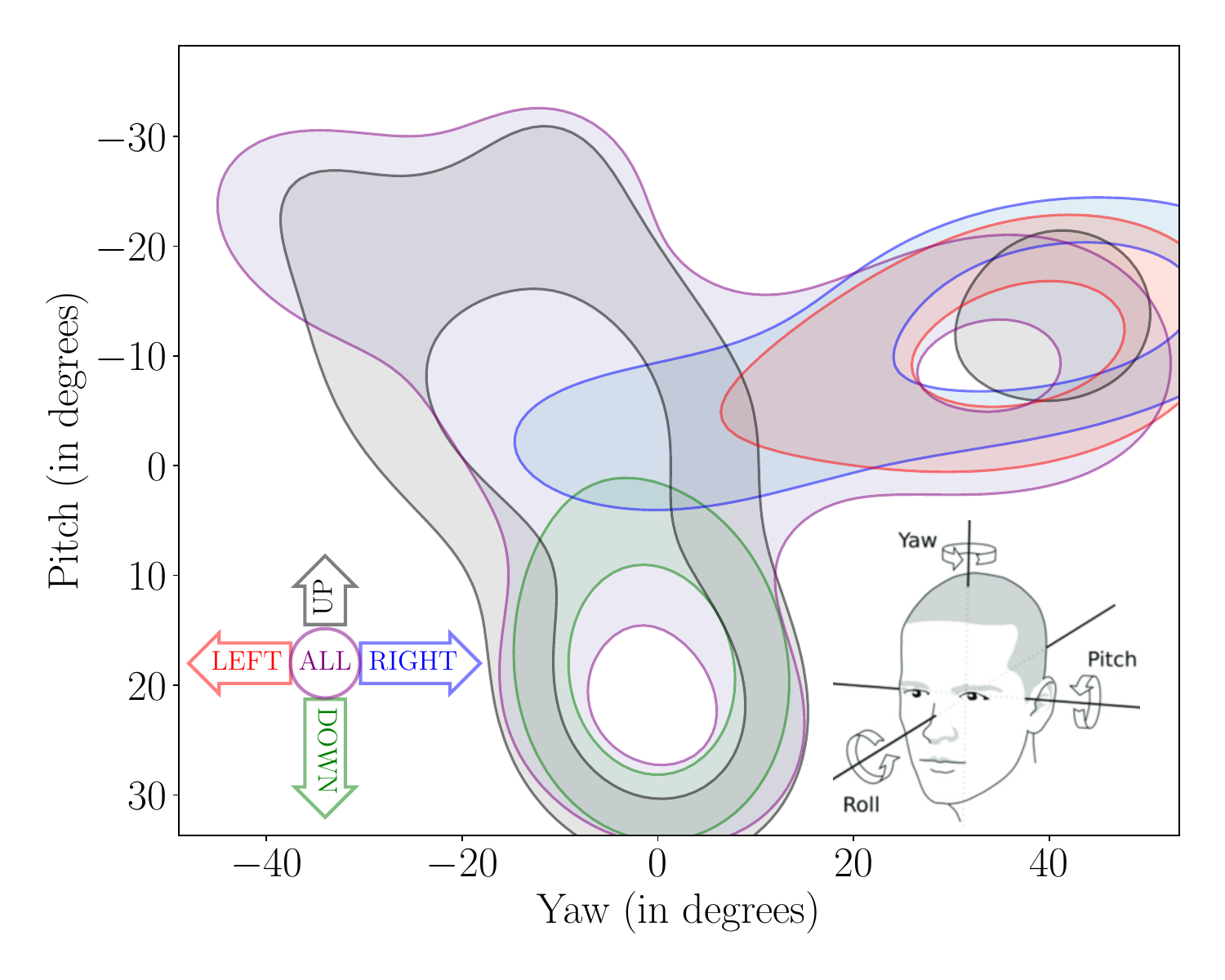}
      \caption{Comparative Analysis of Model Performance Based on Data Diversity: face-swap models trained on direction-specific data exhibit limited adaptability across multiple directions. Conversely, a model trained on a diverse dataset, incorporating movements from all directions, demonstrates robust performance across all test conditions. Each directional model is represented by two successive contours, which signify equally `good' quality of deepfake generation in its support region. The contours only partially cover the space, highlighting the importance of diverse training data for unbiased generation. The holes would be successive (third, onwards) contours which are not shown to keep the figure discernible.}
      \label{fig:data_diverse}
    \end{figure}
    
  \vspace{0.5em}
    \noindent \textbf{Dependency on Data Diversity.} We hinted to this as a hurdle to generation in \S\ref{sec:constraints}. To further establish its importance, we conduct an experiment. In the experiment, we recorded five videos of an individual with specific head movements -- one covering each of the four directions, namely, left, right, up, and down, and a fifth covering all of them. A pre-trained DFL model was further trained individually per video, resulting in four `directional' models and one `all' direction model.

    Fig.~\ref{fig:data_diverse} illustrates contours of equal fidelity for the five models when used to generate the individual facing all four directions. Ideally, if the result was unbiased, every contour should have covered the figure maximally, like the `all' contour. However, the directional models, constrained by their training data, exhibited limited support. These results show that a lack of diversity in data can result in biased outputs. In contrast, a diverse dataset enables robust model performance.

\subsection{Countermeasures}
    \label{sec:adaptive}
     We consider potential countermeasures an imposter could employ to reduce the effectiveness of \textsc{Gotcha}. In the threat model, we assumed that the nature of the challenges is \emph{not a secret}. Hence, imposters can anticipate a challenge, and could replay pre-fabricated responses or use a more sophisticated deepfake generator. This subsection discusses why such an adaptation would be complicated and how a defender can fortify against them:
  
    \vspace{0.5em}
    \noindent \textbf{Imposter pre-fabricates responses.} An adaptive imposter can fabricate a  response to an anticipated challenge, and weed out any artifacts using operations akin to professional video editing, which are unrealistic to perform live. 

    In order to thwart such efforts, the defender can \textit{include randomness when defining challenges}, thus eliminating non-randomizable tasks (e.g., changing iris color). Supposing the defender randomises each instance of a challenge, an imposter's ability to anticipate or pre-calculate responses gets significantly hampered~\cite{liSeeingLivingRethinking2022a}. 
    
    Consider, for instance, that the defender picks a challenge category involving hand occlusions. This challenge category includes positioning a finger before the face or resting a hand on the face. They could individually randomise each challenge by altering its parameters, such as the angle of the hand relative to the face, pose, or distance from the face. 
        
    A practical method to introduce these challenges could involve presenting an animated puppet head during the interaction, which carries out the challenge. The user's task would then be to mimic the actions of this puppet. Fig.~\ref{fig:laptop_pattern} illustrates one such scenario. The degree of difficulty introduced by randomness may vary across different challenges. However, it universally makes real-time response pre-fabrication more challenging. 
    
    Incorporating randomness allows assembling a diverse suite of unpredictable instances even of the same challenge, bolstering their resilience against adversarial attempts to predict or preempt.

    \vspace{0.5em}
    \noindent \textbf{Imposter uses a more sophisticated RTDF.} An imposter could refine their RTDF generator to bypass specific challenges, either by using more training data or optimizing their network architecture. 

    In our extended threat model, we consider such advanced RTDFs, mainly when abundant target data is available. Such a case often occurs for public figures or when the targets themselves act as accomplices, thereby defeating any ``what-you-know'' authentication schemes by sharing personal information with the imposter.
    
    To offset this vulnerability, defenders can employ \emph{a sequence of challenges}. The underlying premise is that while an imposter might successfully bypass a single challenge, evading an entire sequence is considerably more difficult.
   
\begin{figure}[t!]
    \centering

\captionsetup[subfloat]{labelformat=empty}
 \subfloat[No Challenge ]{%
          \begin{tikzpicture}
            \node[anchor=south west,inner sep=0] (image) at (0,0) { \includegraphics[width=0.23\columnwidth]{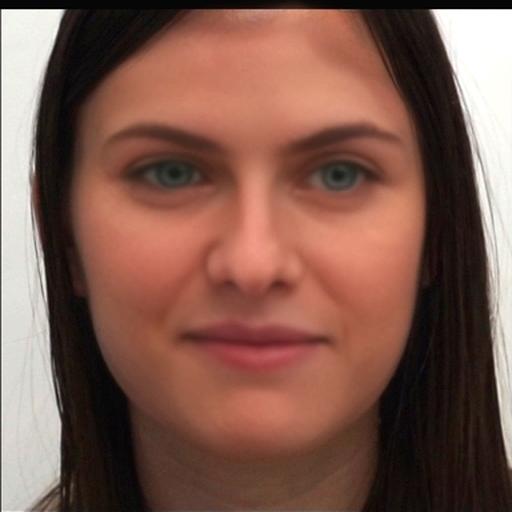}};
            \begin{scope}[x={(image.south east)},y={(image.north west)}]
                \fill[green] ([xshift=-1.5mm]image.south east) rectangle ([yshift=9.9mm]image.east);
            \end{scope}  \end{tikzpicture}%
            }\hspace{0.012em}
            \subfloat[Head movement]{%
            \begin{tikzpicture}
            \node[anchor=south west,inner sep=0] (image) at (0,0) { \includegraphics[width=0.23\columnwidth]{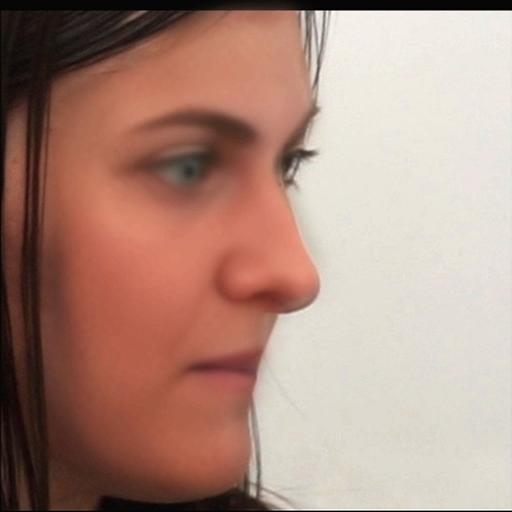}};
            \begin{scope}[x={(image.south east)},y={(image.north west)}]
                \fill[green] ([xshift=-1.5mm]image.south east) rectangle ([yshift=9.9mm]image.east);
            \end{scope}\end{tikzpicture}%
            }\hspace{0.012em}
            \subfloat[Occlude w Hand]{%
            \begin{tikzpicture}
            \node[anchor=south west,inner sep=0] (image) at (0,0) { \includegraphics[width=0.23\columnwidth]{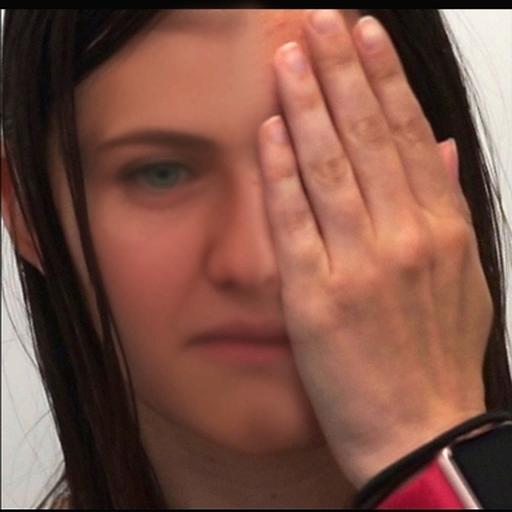}};
            \begin{scope}[x={(image.south east)},y={(image.north west)}]
                
                \fill[green] ([xshift=-1.5mm]image.south east) rectangle ([yshift=4.95mm]image.east);
            \end{scope} \end{tikzpicture}%
            }
            
          \subfloat[Occlude with Object]{%
          \begin{tikzpicture}
            \node[anchor=south west,inner sep=0] (image) at (0,0) { \includegraphics[width=0.23\columnwidth]{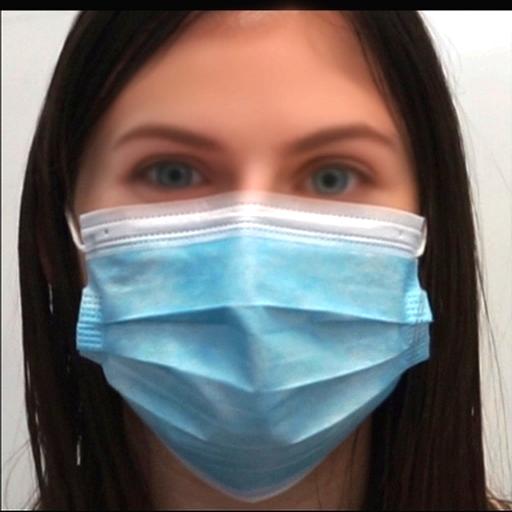}};
            \begin{scope}[x={(image.south east)},y={(image.north west)}]
                \fill[green] ([xshift=-1.5mm]image.south east) rectangle ([yshift=9.9mm]image.east);
            \end{scope}  \end{tikzpicture}%
            }\hspace{0.012em}
          \subfloat[Manual Deformation]{%
          \begin{tikzpicture}
            \node[anchor=south west,inner sep=0] (image) at (0,0) { \includegraphics[width=0.23\columnwidth]{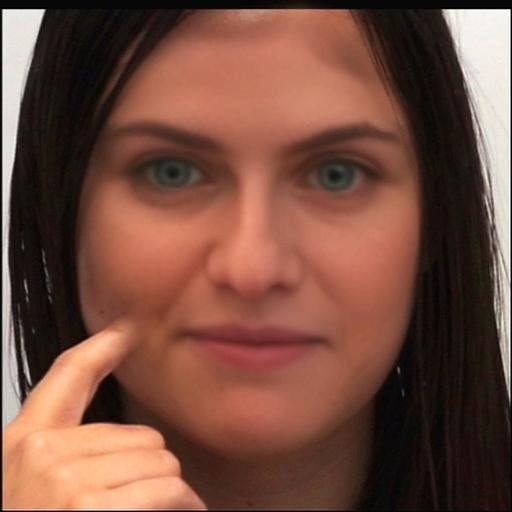}};
            \begin{scope}[x={(image.south east)},y={(image.north west)}]
                
            \end{scope}
        \end{tikzpicture}%
        }\hspace{0.012em}
        \subfloat[Face Illumination ]{%
        \begin{tikzpicture}
            \node[anchor=south west,inner sep=0] (image) at (0,0) { \includegraphics[width=0.23\columnwidth]{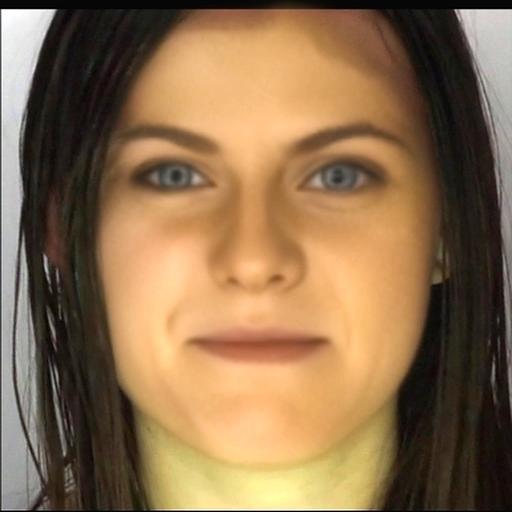}};
            \begin{scope}[x={(image.south east)},y={(image.north west)}]
                
                \fill[green] ([xshift=-1.5mm]image.south east) rectangle ([yshift=4.95mm]image.east);
            \end{scope} \end{tikzpicture}%
            }
\caption{Sample challenge images obtained using a adaptive variant of DFL~\cite{iperov_deepfacelab_2021} trained with diverse data for long training time along with a segmentation module. As observed this version is able to do several challenges with lesser artifacts. Green bars are a metaphor to fidelity score, in comparison to Fig.~\ref{fig:challenges}. Taller bars imply higher fidelity, and the missing bar denotes failure. Video version at \url{http://govindm.me/gotcha-figures/}.}

\label{fig:hdfl_challenges}
\end{figure}
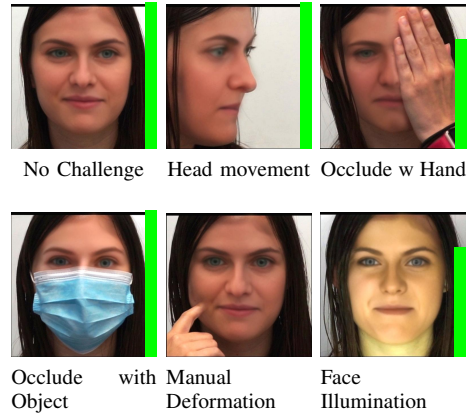
    To empirically substantiate this approach, we conducted experiments with an advanced variant of DFL~\cite{iperov_deepfacelab_2021}, which we treated as an adaptive adversary. In this configuration, we enhanced the face-swapping capabilities by training on a more diversified dataset and incorporated a separate, target-specific segmentation module. The modified RTDF underwent over 2M training iterations on a comprehensive dataset, and an additional 1M pretraining iterations on FFHQ~\cite{karras2019style}, cumulatively requiring about three weeks of training.
    
    This advanced RTDF variant proved substantially resilient to the challenges, especially those based on occlusions. However, such results are constrained to individuals with abundant data availability, typically celebrities (see Fig.~\ref{fig:hdfl_challenges}).
    
    We gauged the efficacy of deploying a sequence of four challenges using human evaluations and automated detection algorithms. These assessments replicated the methodologies outlined in \S\ref{subsec:human_eval} and \S\ref{subsec:chal_eval}, but included a condition where \emph{evaluators encountered multiple challenge videos for the same identity in conjunction with any prior challenges}. Evaluations were conducted across ten identities, incorporating four challenges plus a 'no-challenge' control. The human evaluation involved an average of 13 responses per video sample.
    
\begin{figure}[t!]
        \centering
        \includegraphics[width=0.95\columnwidth]{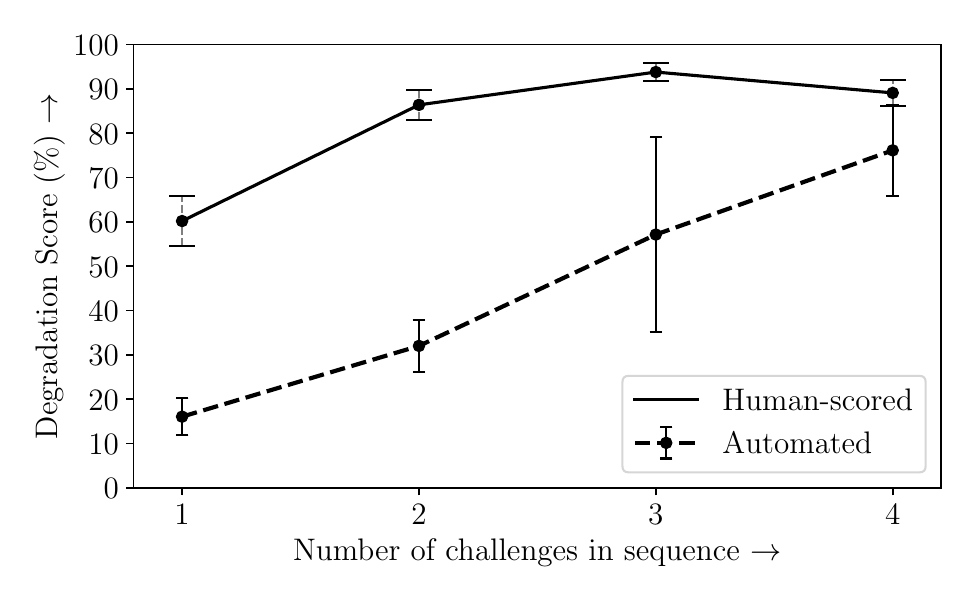}
        \caption{Degradation scored by humans and the in-house model of an adaptive deepfake generator as the challenge sequence progresses. Human scores are given by $\mathcal{H}$ and its 95\% confidence interval, while model-score are given by $\mathcal{A}$ and its standard deviation.}
        \label{fig:cas_conf}
    \end{figure}

    Fig.~\ref{fig:cas_conf} illustrates the mean degradation scores — automated and human — when provided with a sequence of challenges. As the imposter navigates through the sequence, both scores escalated close to 90\% (for reference, in \S\ref{subsec:human_eval} human threshold was 37\%). This monotonic boost validates the usefulness of this strategy in authenticating even an adaptive imposter in under a minute.

    As previously articulated, challenges exploit vulnerabilities in specific components of RTDF generators. Consequently, an advanced RTDF would need to universally fortify all its components to sidestep the gamut of possible challenges entirely. 
        
    Of course, we do not rule out the existence of such an RTDF. Hence, we acknowledge that challenge-response systems have the inherent uncertainty of remaining effective in the face of rapid technological advancements. However, the resilience of existing \textsc{Captcha} solutions offers an optimistic precedent. As deepfakes mature, \textsc{Gotcha} emerges as a promising, proactive defense mechanism in the ever-evolving battle against RTDFs.
  
    \subsection{Usability}
    \label{subsec:chal_design}

\begin{figure}[t!]
        \centering
        \includegraphics[width=0.85\columnwidth]{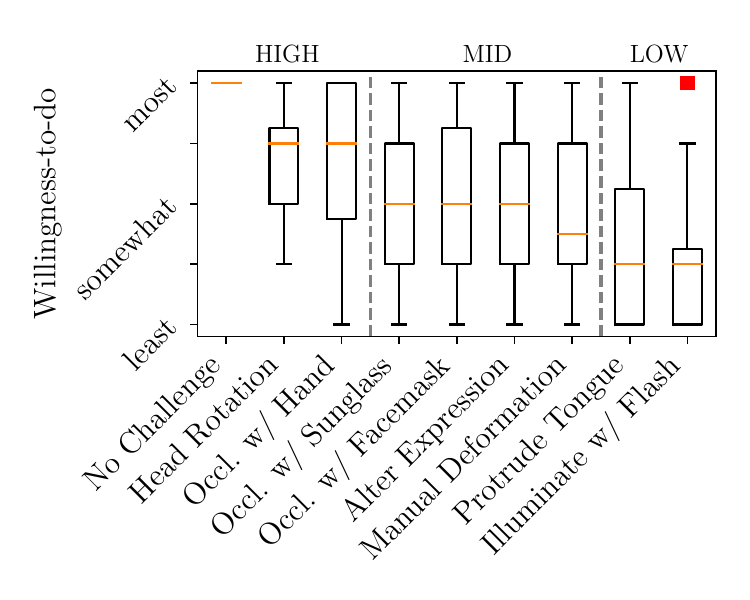}
        \caption{Boxplot of Usability ratings on a five-point Likert scale. Users responded to `How willingly would you do this task in an online meeting, for verifying yourself?'. The challenges are ordered by decreasing median rating (n = 25). The dotted lines show the separation of high, moderate, and low usable challenges.} 
        \label{fig:usability_degradation}
    \end{figure}

    In the preceding discussion, we advocated for the randomization of challenges and the implementation of challenge sequences to strengthen security. While these measures enhance robustness against imposters, they inherently increase the complexity of the authentication process, potentially at the expense of user experience. Thus, designing practical challenges necessitates a nuanced balance between security and usability, considering user-specific contexts, and the optional but beneficial presence of visible artifacts. We discuss them as follows:
     
    \vspace{0.5em}
    \noindent\textbf{User-cooperation.} In practical scenarios, a defender needs user cooperation. As the correct response to a challenge is necessary for verification, users must willingly perform the requested actions. Thus, a challenge must be minimally disruptive to the user experience and maximally reveal anomalies in an imposter's video. To gauge the participants' willingness to perform such challenges in a practical scenario like online meetings, we administered a post-participation survey during data collection (Fig.~\ref{fig:usability_degradation}).

    Some challenges inherently possess superior usability; for example, simple actions like removing glasses or gesturing could be adopted more readily than disruptive actions like spilling water or poking cheeks. We compared Fig.~\ref{fig:usability_degradation} with Fig.~\ref{fig:human_scores} and observed that moderately usable challenges were the most effective in inducing degradation. Also, Table~\ref{tab:taxonomy} symbolizes subfactors such as comprehensibility, appropriateness, physical effort, equipment necessity, and human detectability for each category to help make more informed deployment decisions, described further in \S\ref{subsec:usability}. 
    
    The sensitivity of the call also influences user cooperation. Defenders have more leeway to enforce compliance in critical identity verification scenarios, such as job interviews or online exams. Conversely, automated detection can address concerns about challenge appropriateness and potential participant reluctance in less sensitive settings like business meetings. \textsc{Gotcha} offers real-time automated evaluation without the need to keep recordings. It reassures users that a machine will independently assess their responses, which are then promptly deleted post-verification. To alleviate selfie anxiety in participants, \textsc{Gotcha} may only display the outlines of participants' faces and remove extensive detail~\cite{selfie_conscious}. 

    \vspace{0.5em}
    \noindent\textbf{User-context.} A well-designed challenge should account for the situational context in which it will be deployed. The efficacy and appropriateness of a challenge may depend on many factors, such as the user's specific actions, environmental lighting conditions, network bandwidth, and pertinent security requirements. 

    \vspace{0.5em}
    \noindent\textbf{Human Evaluation Constraints.} When designing challenges that permit human evaluation, it is crucial to consider the limitations of human response time and perceptual acuity. For instance, rapid challenges, such as those based on fast illumination changes, could confound human evaluators, making real-time assessment impractical. Thus, challenges should be tailored to be discernible and interpretable within human reaction times and, preferably, induce visible artifacts that facilitate easier identification of deepfakes.

\section{Related Work}
\label{sec:related_work}
Our work is related to several topics. We describe where we intersect and where we differ with each of them:

\vspace{0.5em}
\noindent \textbf{Deepfake Detection.} In recent years, automatic deepfake detection has burgeoned due to an arms race with deepfake generators. They target tell-tale artifacts such as distinct patterns in the frequency domain \cite{marraGANsLeaveArtificial2019}, eye features \cite{guo2022eyes}, inner \& outer facial features \cite{dong2022protecting}, expressions \cite{mazaheri2022detection}, and biological signals. Notably, Vahdati et al.~\cite{drexelcvprw} detect real-time facial reenactment deepfakes by calculating errors caused by inconsistent facial landmarks between the driving video and a frame reconstruction of the speaker. Despite competition against generators and the constant threat of becoming obsolete, several methods are technically impressive. Hence, extending them to real-time detection of an incoming feed is possible. 

We found that in practice, however, existing automatic detectors are limited; they too are prone to the same impediments that the generators suffer from and more, namely, poor dataset quality, inconsistent pre-processing, computational resource constraints, and susceptibility to adversarial perturbations (\S\ref{sec:background}). These observations were corroborated by prior work suggesting that these models learn non-interpretable, spurious features~\cite{zhou2021face,detectorsfail,chandrasegaran2021closer}. We also empirically confirm that well-known detectors~\cite{shiohara2022detecting,zheng2021exploring} performed unreliably on our dataset (\S\ref{subsec:chal_eval}).

\textsc{Gotcha} is not a deepfake detector; rather, it is a way to strengthen deepfake detection using challenges and making more confident detection decisions.

\vspace{0.5em}
\noindent \textbf{Deepfake Datasets.} Parallel to the evolution of deepfake detectors, researchers curated several deepfake datasets, such as DFDC~\cite{dfdc}, FFIW~\cite{zhou2021face} and FaceForensics++~\cite{roessler2019faceforensicspp}. While such legacy datasets contain diverse deepfake videos regarding environment settings, synthetic methods, and face multiplicity, they do not feature the diverse set of actions assessed in this work in a principled manner. Thus, we could not use traditional deepfake datasets and had to create our own. This decision underscores our commitment to accurately assessing our work to detect deepfakes in novel scenarios.

\vspace{0.5em}
\noindent \textbf{Human Deepfake Evaluators.} To provide a baseline for comparison, some studies evaluate  performance of humans in detecting deepfakes. Groh et al.~\cite{grohDeepfakeDetectionHuman2022} conducted an online study with 15,016 participants and concluded that detectors and humans perform overall comparably, with 82\% of humans outperforming the leading detector. Also, when they work together, automated detectors boost human performance. In follow-up work, when Josephs~et~al.~\cite{josephs2023artifact} amplified deepfake artifacts, human accuracy and confidence improved, indicating that humans are good at detecting artifacts if they are discernible and are within their reaction speed. \textsc{Gotcha}'s challenges take priority to make the deepfake artifacts human-visible. 

\vspace{0.5em}
\noindent \textbf{Challenge-Response-based Authentication Systems.}
\noindent \emph{Liveness Detection.} 
Uzun et al.~\cite{uzunRtCaptchareal-timeCAPTCHA2018} used challenge-response to detect whether a user is live or not by asking a caller to read out a random text, then verify that (1) they read the correct text and (2) matches their voice and face against a pre-enrolled user. Li et al.~\cite{liSeeingLivingRethinking2022a} assessed that practical facial liveness detection systems do not protect against deepfakes, and as deepfakes have evolved into being real-time, an imposter can trivially bypass the first verification step.

\vspace{0.5em}
\noindent \emph{Biometric Authentication.} Sluganovic et al.~\cite{sluganovic2016using} track saccade movements after a marker pops on the screen and compares the movement with those of the enrolled user. However, \textsc{Gotcha} does not require extensive biometric enrollment and works under a stronger threat model.

\vspace{0.5em}
\noindent \emph{Real-time Deepfakes.} Yasur et al.~\cite{yasur2023deepfake} addressed deepfake detection on audio calls using a similar approach and created their challenges, such as clapping, coughing, and playback, based on four constraints -- realism, identity, task, and time. They evaluated their challenges on five audio deepfake generators using a private dataset, ASVspoof21~\cite{yamagishi2021asvspoof} and realistic-in-the-wild dataset~\cite{muller2022does}. They concluded that such challenges degrade audio quality and force the generator to expose itself. \textsc{Gotcha} corroborated their claims but for video calls. 

Active Illumination~\cite{gerstnerDetectingreal-timeDeepFake} works by projecting hues from the user's screen and then correlating these with the hue reflected off the user's face. This approach assumes deepfake systems cannot accurately generate complex hue illuminations in real time. It analyzes the correlation frame-by-frame, with low correlations indicating it is a deepfake and vice-versa. Corneal Reflection~\cite{guo2023detection}, similarly, relies on projecting a specific shape and capturing its reflection in the user's cornea for verification. The technique assumes that webcams can consistently capture eye details and that deepfake algorithms cannot mimic such reflections convincingly. The above works also address video RTDFs and are example challenges of \textsc{Gotcha}.

\section{Limitations and Conclusion}
\noindent\textbf{Limitations and Future Work.} While challenges effectively aid in detecting real-time deepfakes, they have limitations, which we will explore in future work.

    \vspace{0.25em} 
    \noindent \emph{Demographic Variations:} Our dataset is limited to 47 identities, covering a variety of races, genders, and ages (as shown in Fig.~\ref{fig:demographics}). While diverse, this sample size cannot account for the full spectrum of human facial diversity. We acknowledge that this limitation could introduce bias into our qualitative and quantitative evaluations. 
    
    \vspace{0.25em} 
    \noindent \emph{Contextual Variability:} Our study relies on data collected in a controlled setting with cooperative participants. Actual conditions could present a host of variables we have not accounted for—such as poor network connections, indoor vs. outdoor settings, or adversarial tactics deployed by imposters to confound detection attempts.

     \vspace{0.25em}
    \noindent \emph{Machine-assisted Scoring Improvements:} The automated scoring function comprises of a 3D-CNN for fidelity scoring and a challenge-specific compliance detector. Potential avenues for enhancement include augmenting it with a single compliance detector, which can work for all challenges, and using a hyper-parameter optimized architecture along with uncertainty quantifiers to elevate its accuracy and reliability. 
    
\vspace{0.5em}    
\noindent\textbf{Conclusion.} We presented \textsc{Gotcha}, a challenge-response approach designed to authenticate live video interactions in an environment increasingly susceptible to real-time deepfakes. By exploiting inherent weaknesses in contemporary deepfake generation algorithms, \textsc{Gotcha} employs a carefully curated set of challenges that produce easily identifiable and human-visible artifacts in videos generated by state-of-the-art deepfake pipelines.

Through validation on a novel and unique challenge dataset, with the help of human evaluators and using a fidelity score model, we empirically establish the efficacy of \textsc{Gotcha}. The observations made during these assessments, and backed by insights, we offer a nuanced understanding of \textsc{Gotcha}'s robustness against adaptive adversaries, and delineate a security-usability tradeoff.

\subsection*{Resource Availability}
\noindent\textbf{Dataset.} The full dataset of original recorded challenges, corresponding deepfakes, and derivative deepfake generators are released under a CC-BY-NC-SA 4.0 license. This data will be available exclusively to accredited institutions/organizations for non-commercial research.

\vspace{0.5em}
\noindent\textbf{Code.} We release the code to train the fidelity score function (and dataset) at \url{https://github.com/mittalgovind/GOTCHA-Deepfakes}.

\vspace{0.5em}
\noindent\textbf{Human Evaluation Instruments.} We also release the instruments utilized during human evaluations for easy cloning and preview at \url{https://app.gorilla.sc/openmaterials/693684}.

\subsection*{Acknowledgement}
\noindent This material is partially supported by the National Science Foundation under Grant No. 1956200. The authors were partially supported by the AI Research Institutes Program supported by NSF and USDA-NIFA under grant no. 2021-67021-35329, NSF SaTC grant 2154119, and a Cyber NYC gift from Google Research.

{\bibliographystyle{ieeetr}
\bibliography{egbib}}

\begin{thebibliography}{10}

\bibitem{videoconf}
GETVOIP, ``{The State of Video Conferencing in 2023}.'' \url{https://getvoip.com/blog/state-of-conferencing/}, 2023.
\newblock [Online; accessed 22-May-2023].

\bibitem{sensitykyc}
Sensity, ``{Deepfakes vs biometric KYC verification}.'' \url{https://sensity.ai/blog/deepfake-detection/deepfakes-vs-kyc-biometric-verification/ }, 2022.
\newblock [Accessed: 22-May-2023].

\bibitem{deepfake_scam0}
NPR, ``{Deepfake video of Zelenskyy could be 'tip of the iceberg' in info war, experts warn}.'' \url{https://www.npr.org/2022/03/16/1087062648/deepfake-video-zelenskyy-experts-war-manipulation-ukraine-russia}.
\newblock [Accessed: 22-May-2023].

\bibitem{deepfake_scam1}
Reuters, ``{Deepfake scam in China fans worries over AI-driven fraud}.'' \url{https://www.reuters.com/technology/deepfake-scam-china-fans-worries-over-ai-driven-fraud-2023-05-22/}.
\newblock [Accessed: 22-May-2023].

\bibitem{mollickquickfake}
E.~Mollick, ``{A quick and sobering guide to cloning yourself}.'' \url{https://www.oneusefulthing.org/p/a-quick-and-sobering-guide-to-cloning}.
\newblock [Accessed: 30-Apr-2023].

\bibitem{iperov_deepfacelab_2021}
I.~Perov, D.~Gao, N.~Chervoniy, K.~Liu, S.~Marangonda, C.~Umé, M.~Dpfks, C.~S. Facenheim, L.~RP, J.~Jiang, S.~Zhang, P.~Wu, B.~Zhou, and W.~Zhang, ``{DeepFaceLab}: {Integrated}, flexible and extensible face-swapping framework,'' {\em arXiv:2005.05535 [cs, eess]}, June 2021.
\newblock arXiv: 2005.05535.

\bibitem{bloomberg1}
Bloomberg, ``{Deepfake Imposter Scams Are Driving a New Wave of Fraud}.'' \url{https://www.bloomberg.com/news/articles/2023-08-21/money-scams-deepfakes-ai-will-drive-10-trillion-in-financial-fraud-and-crime}.
\newblock [Accessed: 26-Aug-2023].

\bibitem{fbiworries}
CyberScoop, ``{The growth in targeted, sophisticated cyberattacks troubles top FBI cyber official }.'' \url{https://cyberscoop.com/fbi-worries-about-future-cyber-threats}.
\newblock [Accessed: 25-May-2023].

\bibitem{deepfake_scam2}
Guardian, ``{Kremlin critic Bill Browder says he was targeted by deepfake hoax video call}.'' \url{https://www.theguardian.com/world/2023/may/25/kremlin-critic-bill-browder-says-he-was-targeted-by-deepfake-hoax-video-call}.
\newblock [Accessed: 25-May-2023].

\bibitem{marraGANsLeaveArtificial2019}
F.~Marra, D.~Gragnaniello, L.~Verdoliva, and G.~Poggi, ``Do {GANs} {Leave} {Artificial} {Fingerprints}?,'' in {\em 2019 {IEEE} {Conference} on {Multimedia} {Information} {Processing} and {Retrieval} ({MIPR})}, pp.~506--511, Mar. 2019.

\bibitem{guo2022eyes}
H.~Guo, S.~Hu, X.~Wang, M.-C. Chang, and S.~Lyu, ``Eyes tell all: Irregular pupil shapes reveal gan-generated faces,'' in {\em ICASSP 2022-2022 IEEE International Conference on Acoustics, Speech and Signal Processing (ICASSP)}, pp.~2904--2908, IEEE, 2022.

\bibitem{dong2022protecting}
X.~Dong, J.~Bao, D.~Chen, T.~Zhang, W.~Zhang, N.~Yu, D.~Chen, F.~Wen, and B.~Guo, ``Protecting celebrities from deepfake with identity consistency transformer,'' in {\em Proceedings of the IEEE/CVF Conference on Computer Vision and Pattern Recognition}, pp.~9468--9478, 2022.

\bibitem{mazaheri2022detection}
G.~Mazaheri and A.~K. Roy-Chowdhury, ``Detection and localization of facial expression manipulations,'' in {\em Proceedings of the IEEE/CVF Winter Conference on Applications of Computer Vision}, pp.~1035--1045, 2022.

\bibitem{boccignone2022deepfakes}
G.~Boccignone, S.~Bursic, V.~Cuculo, A.~D’Amelio, G.~Grossi, R.~Lanzarotti, and S.~Patania, ``Deepfakes have no heart: A simple rppg-based method to reveal fake videos,'' in {\em International Conference on Image Analysis and Processing}, pp.~186--195, Springer, 2022.

\bibitem{nirkin_fsganv2_2022}
Y.~Nirkin, Y.~Keller, and T.~Hassner, ``Fsganv2: Improved subject agnostic face swapping and reenactment,'' {\em IEEE Transactions on Pattern Analysis and Machine Intelligence}, vol.~45, no.~1, pp.~560--575, 2022.

\bibitem{wang2022latent}
Y.~Wang, D.~Yang, F.~Bremond, and A.~Dantcheva, ``Latent image animator: Learning to animate images via latent space navigation,'' in {\em International Conference on Learning Representations}, 2022.

\bibitem{reinhard2001color}
E.~Reinhard, M.~Adhikhmin, B.~Gooch, and P.~Shirley, ``Color transfer between images,'' {\em IEEE Computer graphics and applications}, vol.~21, no.~5, pp.~34--41, 2001.

\bibitem{athar2022rignerf}
S.~Athar, Z.~Xu, K.~Sunkavalli, E.~Shechtman, and Z.~Shu, ``Rignerf: Fully controllable neural 3d portraits,'' in {\em Proceedings of the IEEE/CVF Conference on Computer Vision and Pattern Recognition}, pp.~20364--20373, 2022.

\bibitem{dhariwal2021diffusion}
P.~Dhariwal and A.~Nichol, ``Diffusion models beat gans on image synthesis,'' {\em Advances in Neural Information Processing Systems}, vol.~34, pp.~8780--8794, 2021.

\bibitem{ramesh2022hierarchical}
A.~Ramesh, P.~Dhariwal, A.~Nichol, C.~Chu, and M.~Chen, ``Hierarchical text-conditional image generation with clip latents,'' {\em arXiv preprint arXiv:2204.06125}, 2022.

\bibitem{synthesia}
Synthesia, ``{Text-to-synthetic videos}.'' \url{https://www.synthesia.io/}.
\newblock [Accessed: 25-May-2023].

\bibitem{thies2020neural}
J.~Thies, M.~Elgharib, A.~Tewari, C.~Theobalt, and M.~Nie{\ss}ner, ``Neural voice puppetry: Audio-driven facial reenactment,'' in {\em Computer Vision--ECCV 2020: 16th European Conference, Glasgow, UK, August 23--28, 2020, Proceedings, Part XVI 16}, pp.~716--731, Springer, 2020.

\bibitem{siarohin2019first}
A.~Siarohin, S.~Lathuili{\`e}re, S.~Tulyakov, E.~Ricci, and N.~Sebe, ``First order motion model for image animation,'' {\em Advances in Neural Information Processing Systems}, vol.~32, 2019.

\bibitem{colortransferrct}
E.~Reinhard, M.~Adhikhmin, B.~Gooch, and P.~Shirley, ``Color transfer between images,'' {\em IEEE Computer Graphics and Applications}, vol.~21, no.~5, pp.~34--41, 2001.

\bibitem{googlecolab}
Unite.AI, ``{Google Has Banned the Training of Deepfakes in Colab}.'' \url{https://www.unite.ai/google-has-banned-the-training-of-deepfakes-in-colab/ }, 2022.
\newblock [Accessed: 22-May-2023].

\bibitem{stypulkowski2023diffused}
M.~Stypu{\l}kowski, K.~Vougioukas, S.~He, M.~Zi{\k{e}}ba, S.~Petridis, and M.~Pantic, ``Diffused heads: Diffusion models beat gans on talking-face generation,'' {\em arXiv preprint arXiv:2301.03396}, 2023.

\bibitem{GAvatar2023}
X.~Li, S.~De~Mello, S.~Liu, K.~Nagano, U.~Iqbal, and J.~Kautz, ``Generalizable one-shot neural head avatar,'' {\em Arxiv}, 2023.

\bibitem{kim2023dcface}
M.~Kim, F.~Liu, A.~Jain, and X.~Liu, ``Dcface: Synthetic face generation with dual condition diffusion model,'' in {\em Proceedings of the IEEE/CVF Conference on Computer Vision and Pattern Recognition}, pp.~12715--12725, 2023.

\bibitem{airbnb}
``{How Airbnb Verifies Identities}.'' \url{https://www.airbnb.com/help/article/1237}.
\newblock Accessed: 2024-02-29.

\bibitem{uberkyc}
``{Uber Delivery Agents: Identity Verification}.'' \url{https://www.uber.com/en-GB/blog/identity-verification/}.
\newblock Accessed: 2024-02-29.

\bibitem{idme}
``{ID.me for KYC by US Government Agencies}.'' \url{https://network.id.me/platform/identity-verification/}.
\newblock Accessed: 2024-02-29.

\bibitem{ets_gre}
``{ETS for At-Home Testing}.'' \url{https://www.ets.org/gre/test-takers/general-test/register/at-home-testing.html}.
\newblock Accessed: 2024-02-29.

\bibitem{koreanfreelancer}
T.~Register, ``{‘How not to hire a North Korean plant posing as a techie’ guide updated by US and South Korean authorities}.'' \url{https://www.theregister.com/2023/10/19/north_korea_fake_freelance_avoidance/}.
\newblock [Accessed: 24-Oct-2023].

\bibitem{shiohara2022detecting}
K.~Shiohara and T.~Yamasaki, ``Detecting deepfakes with self-blended images,'' in {\em Proceedings of the IEEE/CVF Conference on Computer Vision and Pattern Recognition}, pp.~18720--18729, 2022.

\bibitem{zheng2021exploring}
Y.~Zheng, J.~Bao, D.~Chen, M.~Zeng, and F.~Wen, ``Exploring temporal coherence for more general video face forgery detection,'' in {\em Proceedings of the IEEE/CVF International Conference on Computer Vision}, pp.~15044--15054, 2021.

\bibitem{celebdf}
Y.~Li, X.~Yang, P.~Sun, H.~Qi, and S.~Lyu, ``Celeb-df: A large-scale challenging dataset for deepfake forensics,'' in {\em Proceedings of the IEEE/CVF conference on computer vision and pattern recognition}, pp.~3207--3216, 2020.

\bibitem{dfdc}
B.~Dolhansky, J.~Bitton, B.~Pflaum, J.~Lu, R.~Howes, M.~Wang, and C.~C. Ferrer, ``The deepfake detection challenge (dfdc) dataset,'' {\em arXiv preprint arXiv:2006.07397}, 2020.

\bibitem{zhou2021face}
T.~Zhou, W.~Wang, Z.~Liang, and J.~Shen, ``Face forensics in the wild,'' in {\em Proceedings of the IEEE/CVF conference on computer vision and pattern recognition}, pp.~5778--5788, 2021.

\bibitem{hara2017learning}
K.~Hara, H.~Kataoka, and Y.~Satoh, ``Learning spatio-temporal features with 3d residual networks for action recognition,'' in {\em Proceedings of the IEEE international conference on computer vision workshops}, pp.~3154--3160, 2017.

\bibitem{kopuklu2021driver}
O.~Kopuklu, J.~Zheng, H.~Xu, and G.~Rigoll, ``Driver anomaly detection: A dataset and contrastive learning approach,'' in {\em Proceedings of the IEEE/CVF Winter Conference on Applications of Computer Vision}, pp.~91--100, 2021.

\bibitem{lugaresi2019mediapipe}
C.~Lugaresi, J.~Tang, H.~Nash, C.~McClanahan, E.~Uboweja, M.~Hays, F.~Zhang, C.-L. Chang, M.~G. Yong, J.~Lee, {\em et~al.}, ``Mediapipe: A framework for building perception pipelines,'' {\em arXiv preprint arXiv:1906.08172}, 2019.

\bibitem{yawpitch_compliance}
T.~Hempel, A.~A. Abdelrahman, and A.~Al-Hamadi, ``6d rotation representation for unconstrained head pose estimation,'' in {\em 2022 IEEE International Conference on Image Processing (ICIP)}, pp.~2496--2500, IEEE, 2022.

\bibitem{fer_compliance}
A.~V. Savchenko, ``Facial expression and attributes recognition based on multi-task learning of lightweight neural networks,'' in {\em 2021 IEEE 19th International Symposium on Intelligent Systems and Informatics (SISY)}, pp.~119--124, 2021.

\bibitem{liSeeingLivingRethinking2022a}
C.~Li, L.~Wang, S.~Ji, X.~Zhang, Z.~Xi, S.~Guo, and T.~Wang, ``Seeing is living? rethinking the security of facial liveness verification in the deepfake era,'' in {\em 31st USENIX Security Symposium (USENIX Security 22)}, pp.~2673--2690, 2022.

\bibitem{karras2019style}
T.~Karras, S.~Laine, and T.~Aila, ``A style-based generator architecture for generative adversarial networks,'' in {\em Proceedings of the IEEE/CVF conference on computer vision and pattern recognition}, pp.~4401--4410, 2019.

\bibitem{selfie_conscious}
``{Why iProov Genuine Presence Assurance Doesn’t Use Selfies}.'' \url{https://www.iproov.com/blog/genuine-presence-assurance-selfie-anxiety}.
\newblock Accessed: 2024-02-29.

\bibitem{drexelcvprw}
D.~S. Vahdati, T.~Duc~Nguyen, and M.~C. Stamm, ``Defending low-bandwidth talking head videoconferencing systems from real-time puppeteering attacks,'' in {\em 2023 IEEE/CVF Conference on Computer Vision and Pattern Recognition Workshops (CVPRW)}, pp.~983--992, 2023.

\bibitem{detectorsfail}
B.~Le, S.~Tariq, A.~Abuadbba, K.~Moore, and S.~Woo, ``Why do facial deepfake detectors fail?,'' in {\em Proceedings of the 2nd Workshop on Security Implications of Deepfakes and Cheapfakes}, WDC '23, (New York, NY, USA), p.~24–28, Association for Computing Machinery, 2023.

\bibitem{chandrasegaran2021closer}
K.~Chandrasegaran, N.-T. Tran, and N.-M. Cheung, ``A closer look at fourier spectrum discrepancies for cnn-generated images detection,'' in {\em Proceedings of the IEEE/CVF conference on computer vision and pattern recognition}, pp.~7200--7209, 2021.

\bibitem{roessler2019faceforensicspp}
A.~R\"ossler, D.~Cozzolino, L.~Verdoliva, C.~Riess, J.~Thies, and M.~Nie{\ss}ner, ``Face{F}orensics++: Learning to detect manipulated facial images,'' in {\em International Conference on Computer Vision (ICCV)}, 2019.

\bibitem{grohDeepfakeDetectionHuman2022}
M.~Groh, Z.~Epstein, C.~Firestone, and R.~Picard, ``Deepfake detection by human crowds, machines, and machine-informed crowds,'' {\em Proceedings of the National Academy of Sciences}, vol.~119, p.~e2110013119, Jan. 2022.

\bibitem{josephs2023artifact}
E.~Josephs, C.~Fosco, and A.~Oliva, ``Artifact magnification on deepfake videos increases human detection and subjective confidence,'' {\em arXiv preprint arXiv:2304.04733}, 2023.

\bibitem{uzunRtCaptchareal-timeCAPTCHA2018}
E.~Uzun, S.~P.~H. Chung, I.~Essa, and W.~Lee, ``{rtCaptcha}: {A} {Real}-{Time} {CAPTCHA} {Based} {Liveness} {Detection} {System},'' in {\em Proceedings 2018 {Network} and {Distributed} {System} {Security} {Symposium}}, (San Diego, CA), Internet Society, 2018.

\bibitem{sluganovic2016using}
I.~Sluganovic, M.~Roeschlin, K.~B. Rasmussen, and I.~Martinovic, ``Using reflexive eye movements for fast challenge-response authentication,'' in {\em Proceedings of the 2016 ACM SIGSAC Conference on Computer and Communications Security}, pp.~1056--1067, 2016.

\bibitem{yasur2023deepfake}
L.~Yasur, G.~Frankovits, F.~M. Grabovski, and Y.~Mirsky, ``Deepfake captcha: A method for preventing fake calls,'' in {\em Proceedings of the 2023 ACM Asia Conference on Computer and Communications Security}, ASIA CCS '23, (New York, NY, USA), p.~608–622, Association for Computing Machinery, 2023.

\bibitem{yamagishi2021asvspoof}
J.~Yamagishi, X.~Wang, M.~Todisco, M.~Sahidullah, J.~Patino, A.~Nautsch, X.~Liu, K.~A. Lee, T.~Kinnunen, N.~Evans, {\em et~al.}, ``Asvspoof 2021: accelerating progress in spoofed and deepfake speech detection,'' {\em arXiv preprint arXiv:2109.00537}, 2021.

\bibitem{muller2022does}
N.~M. M{\"u}ller, P.~Czempin, F.~Dieckmann, A.~Froghyar, and K.~B{\"o}ttinger, ``Does audio deepfake detection generalize?,'' {\em arXiv preprint arXiv:2203.16263}, 2022.

\bibitem{gerstnerDetectingreal-timeDeepFake}
C.~R. Gerstner and H.~Farid, ``Detecting real-time deep-fake videos using active illumination,'' in {\em Proceedings of the IEEE/CVF Conference on Computer Vision and Pattern Recognition}, pp.~53--60, 2022.

\bibitem{guo2023detection}
H.~Guo, X.~Wang, and S.~Lyu, ``Detection of real-time deepfakes in video conferencing with active probing and corneal reflection,'' in {\em ICASSP 2023-2023 IEEE International Conference on Acoustics, Speech and Signal Processing (ICASSP)}, pp.~1--5, IEEE, 2023.

\bibitem{hinton2015distilling}
G.~Hinton, O.~Vinyals, and J.~Dean, ``Distilling the knowledge in a neural network,'' {\em arXiv preprint arXiv:1503.02531}, 2015.

\bibitem{pmlr-v9-gutmann10a}
M.~Gutmann and A.~Hyvärinen, ``Noise-contrastive estimation: A new estimation principle for unnormalized statistical models,'' in {\em Proceedings of the Thirteenth International Conference on Artificial Intelligence and Statistics} (Y.~W. Teh and M.~Titterington, eds.), vol.~9 of {\em Proceedings of Machine Learning Research}, (Chia Laguna Resort, Sardinia, Italy), pp.~297--304, PMLR, 13--15 May 2010.

\end{thebibliography}
\renewcommand\thefigure{\thesection.\arabic{figure}}
\setcounter{figure}
{0}

\vfill
\pagebreak
\pagebreak

\renewcommand{\thesection}{A}

\section*{Supplementary Material for \textsc{Gotcha}}
\label{sec:appendix}

\subsection*{Outline}
The supplementary material contains further details and figures, along with their backlinks, as follows:
\begin{enumerate}[nosep,parsep=2pt,leftmargin=*]
  \item Dataset collection, referred in \S\ref{subsec:chal_dataset}. 
  \item Compute Environments used for training and/or inference of RTDFs (\S\ref{subsec:chal_dataset}).
  \item In-house fidelity score function, referred in \S\ref{subsec:chal_eval}.
  \item Human Evaluation experiments, referred in \S\ref{subsec:human_eval}.
  \item Explanation of Usability Benefits, referred in \S\ref{subsec:chal_design}.
\end{enumerate}

\subsection{Data Collection}
 \label{subsec:data_instructions}
     \noindent \textbf{Instructions for Participants.} We gave participants the following instructions and recorded a video for each task:
    \begin{enumerate}[nosep,parsep=2pt,leftmargin=*]
        \item While the camera moves around you for about two minutes, capturing the face from different angles, please sit still.
        \item Rotate your head from side to side, then look up and down. Please spend about 5 seconds on each side, rotating as comfortably as possible, totaling around 25 seconds.
        \item  Cover your eyes with a hand, followed by covering the left half, the right, and finally, the lower half of the face.
        \item  Put on the provided sunglasses, and then take them off. 
        \item Wear the provided clear glasses, ensuring they reflect a lamp light shining. After setting up the reflection, we start recording. Finally, the participant should remove the glasses.
        \item  Put on a face mask and count from 1 to 10 out loud. Then, remove the facemask.
        \item Press a finger against a cheek. 
        \item Stick out a small portion of the tongue.
        \item Laugh for 10 seconds, then frown, as if angry, for another 10 seconds.
        \item Slowly stand up and then sit back down.
        \item Dim the room light and keep the flashing light on the face for 10 seconds.
    \end{enumerate}
    
    \vspace{0.5em}
    Although we only evaluated eight challenges in this work, we collected more to develop insights into what challenges work and why. 
    
    It is important to note that we intentionally refrained from introducing artificial randomness into the dataset. While randomness could help prevent pre-computed solutions, it does not significantly contribute to creating more degrading artifacts. This approach ensured the dataset had natural variability, capturing a broad distribution of challenge performances across participants. Unless otherwise noted, all source images used in this paper are samples from the dataset collected.

\begin{figure}[t!]
        \subfloat[Race]{%
            \includegraphics[width=0.9\columnwidth]{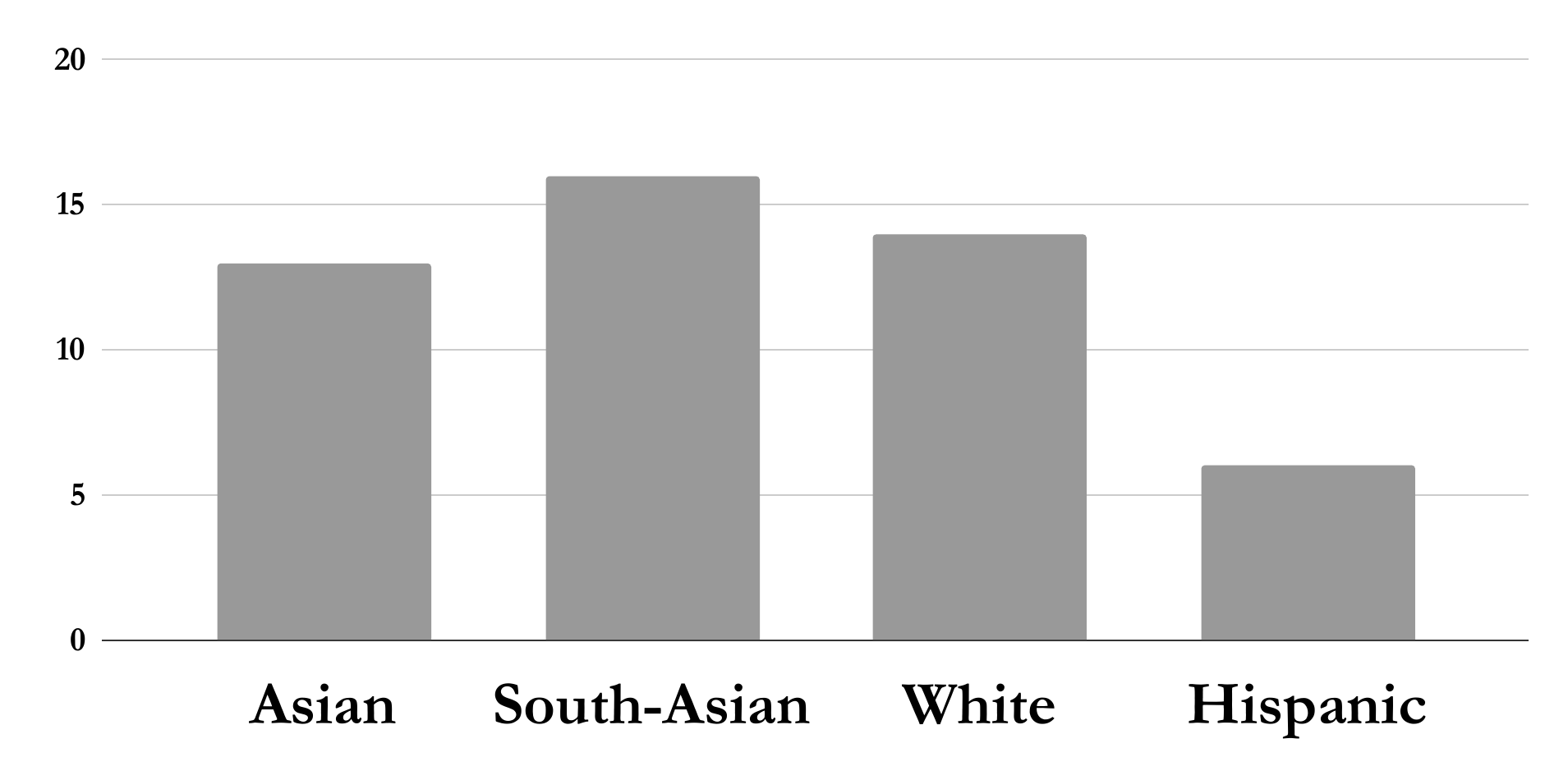}\hspace{-0.15em}%
        }
        
        \centering
        \subfloat[Gender]{%
            \includegraphics[width=0.7\columnwidth]{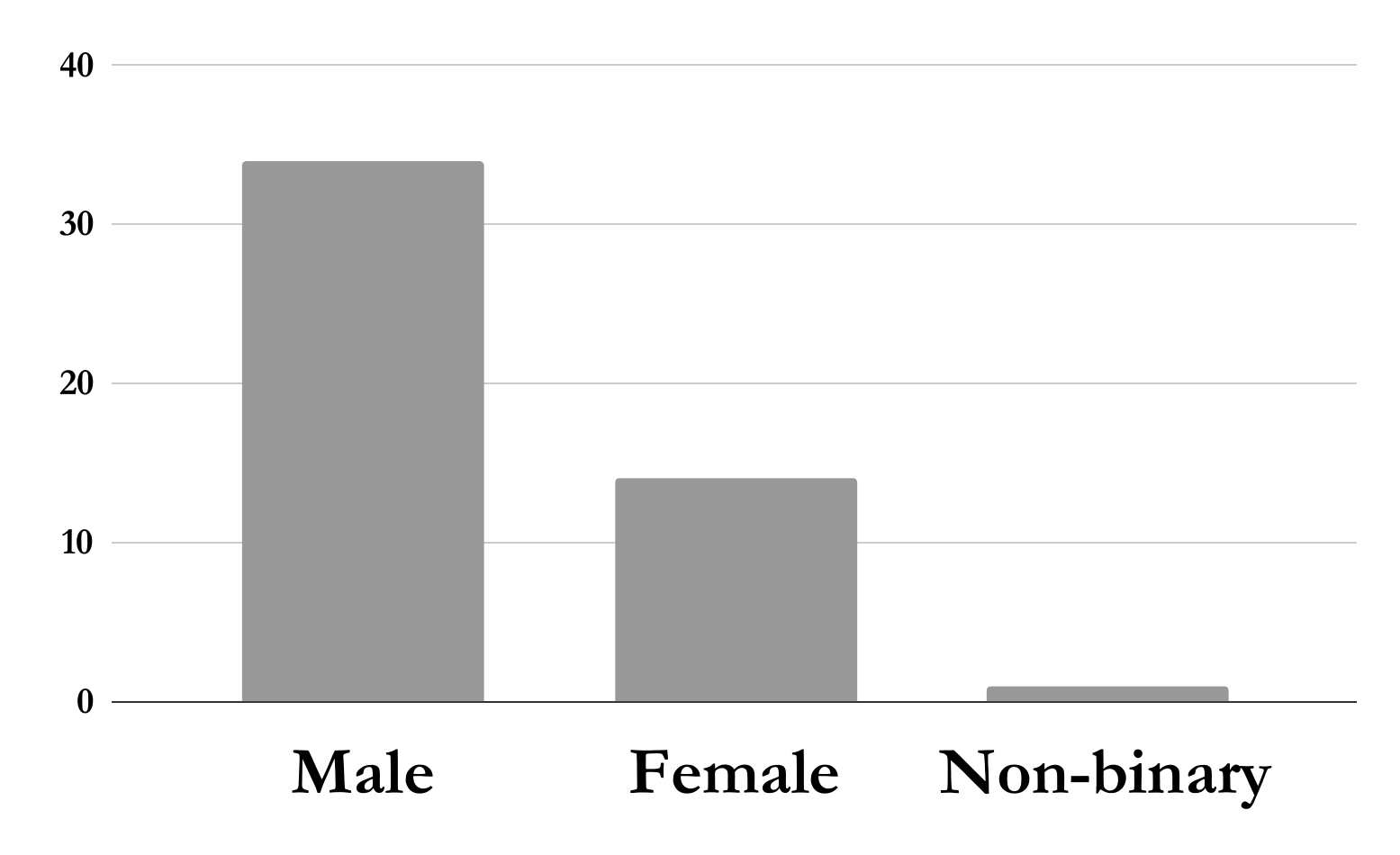}\hspace{-0.15em}
        }
        
        \centering
        \subfloat[Age]{%
            \includegraphics[width=0.5\columnwidth]{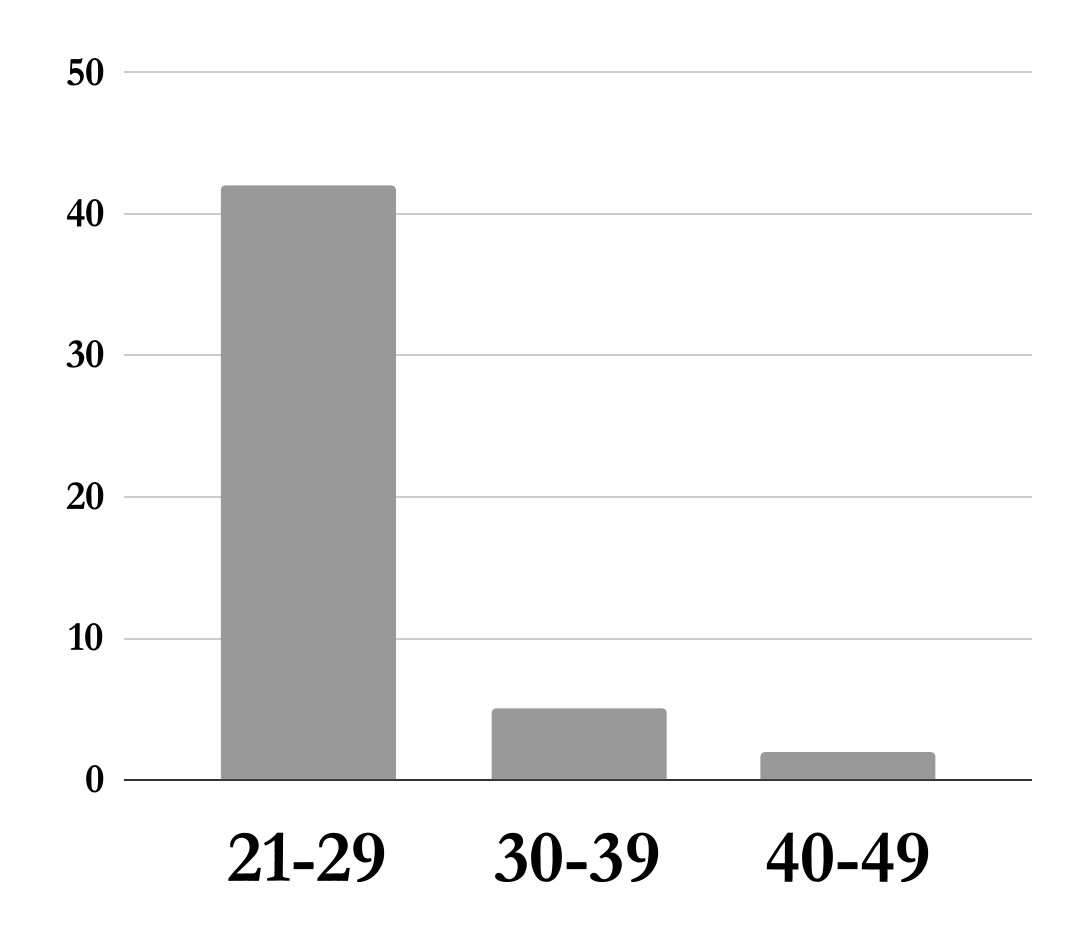}
        }
    \caption{Bar plots of Demographic distribution of participants (n = 47) recorded to create the original part of our dataset, with Y-axes indicating count per sub-category -- (a) Race, (b) Gender, and (c) Age.}
\label{fig:demographics}
\end{figure}

    \vspace{0.5em}
    \noindent \textbf{Demographics.} The demographics of the collected dataset is as follows (also illustrated in Fig.~\ref{fig:demographics}): 
    \begin{itemize}[nosep,parsep=2pt,leftmargin=*]
     \item \emph{Race} -- Asian: 27.7\%, Hispanic: 12.8\%, South Asian: 29.8\%, White: 29.8\%
     \item \emph{Age} --  21-29: 87.5\%, 30-39: 9.4\%, 40-49: 3.1\%
     \item \emph{Gender} -- Male: 59.4\%, Female: 37.5\%, Non-Binary: 3.1\%. 
     \end{itemize}

\subsection{Compute Environments}
\label{sec:compute}
  We trained DFL~\cite{iperov_deepfacelab_2021} models using a 16-core Intel(R) Xeon(R) Platinum 8358 CPU @ 2.60GHz with 80 GB of RAM and one NVIDIA A100 GPU equipped with 80 GB VRAM. We performed all RTDF pipeline inference tasks (e.g., during video calls) on an octa-core Intel(R) Core(TM) i7-9700K CPU @ 3.60GHz machine with 32 GB RAM and an NVIDIA RTX 3070 GPU with 8 GB VRAM.
  
\subsection{Fidelity Score Function}
\label{subsec:detector}
    \begin{figure*}[ht!]
        \centering
        \includegraphics[width=\textwidth]{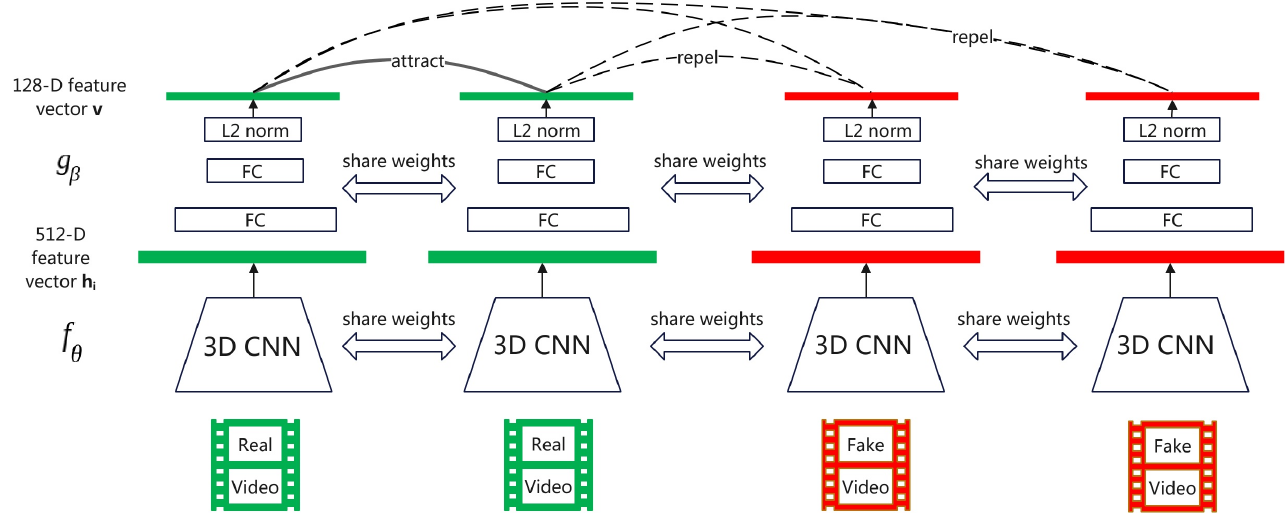}
        \caption{Training workflow of the fidelity score function. A 3D CNN processes a batch containing original and fake videos. The CNN along with the fully connected head optimize a loss function which maximizes the similarity of original videos (solid line) while minimizing the similarity in original-fake video pairs further apart (dotted lines).}
        \label{fig:fidelity_score}
    \end{figure*}
Given the unreliability of existing state-of-the-art deepfake detectors with our challenge videos, we felt the need to develop a custom deepfake detector for our automated evaluation. Drawing inspiration from the pioneering work by Kopuklu et al.~\cite{kopuklu2021driver}, who employed a contrastive learning paradigm for detecting anomalies during car-driving, we designed a similar approach.

\vspace{0.5em}
\noindent\textbf{Dataset Composition.}
In line with our earlier discussion in \S\ref{subsec:chal_eval}, the dataset selection involved:
    \begin{itemize}[nosep,parsep=2pt,leftmargin=*]
        \item Choosing both original and deepfake videos spanning four out of eight challenge types, including the 'no challenge' category.
        \item Ensuring 32 frames per sample, with each frame resized to $224 \times 224$.
        \item Incorporating 35 of the 47 target identities.
        \item Restricting to deepfake videos generated using the DFL approach.
    \end{itemize}
This curated selection resulted in 175 original sample and 8,050 deepfake sample for the training phase. All samples were normalized with a global mean and standard deviation. 

\vspace{0.5em}
\noindent\textbf{Framework Overview.}
Our primary objective was to learn representations that distinctly differentiate between authentic videos and deepfakes. By maximizing the similarity among genuine samples while minimizing the similarity between authentic and fake samples in a latent space, we aim to detect anomalies. The overall structure, visually represented in Fig.~\ref{fig:fidelity_score}, comprises:

\begin{itemize}[nosep,parsep=2pt,leftmargin=*]
    \item We use a \textbf{base encoder $enc_\theta(.)$} to extract vector representations of input video clips. $enc_\theta(.)$ is a 3D-CNN with parameters $\theta$. Specifically, We use a ResNet-18 as the candidate function to map an input clip $\textbf{x} \in \mathbb{R}^{N\times H\times W\times 3}$ into $\textbf{h}~=~\{h_i \in \mathbb{R}^{512}, s.t., \textbf{h}_i= enc_\theta(\textbf{x}_i), \forall i\}$.

    \item After we use a \textbf{projection head} $g_\beta (.)$ to map $\textbf{h}$ to a latent vector embedding $\textbf{v}$. Specifically, we use an MLP with one hidden layer and a ReLU activation, parameterized by $\beta$, to transform $\textbf{v} = \{\textbf{v}_i \in \mathbb{R}^{128}, s.t., \textbf{v}_i = g_\beta (\textbf{h}_i), \forall i \}$. After MLP, we normalize the embedding $\textbf{v}$ using L2 norm.

    \item Finally, a \textbf{contrastive loss} is used to impose that normalized embeddings from the original videos (positive class) are closer together than embeddings from deepfake videos (negative class). 
\end{itemize}
 Within a mini-batch, we have K original videos and M deepfake videos with $i \in \{1, \ldots, K+M\}$. Final embedding of the $i^{th}$ original and deepfake videos are denote as $v_{oi}$ and $v_{di}$, respectively. There are in total $K(K-1)$ positive pairs and $KM$ negative pairs in every mini-batch. Then the loss takes the final form:
\begin{gather*}
    \mathcal{L}_{ij} = -\text{log} \frac{exp(v_{oi}^T v_{oj}/\tau)}{exp(v_{oi}^T v_{oj}/\tau) + \sum_{m=1}^M exp(v_{oi}^T v_{dm}/\tau)} \\
    \mathcal{L} = \frac{1}{K(K-1)} \sum_{i=1}^K\sum_{j=1}^K \mathbf{1}_{j \neq i} \mathcal{L}_{ij}.
\end{gather*}
where $\tau \in (0, \infty)$ is a scalar temperature variable that can controls the spread of the learned distribution~\cite{hinton2015distilling}. Typically, $\tau$ is chosen between 0 and 1 to amplify the similarity between samples. The inner product of embedding vectors measures the cosine similarity, as they are all L2 normalized. By optimizing the above loss term, the encoder learns to maximize the similarity between the original feature vectors and minimizing the dissimilarity between the original feature vectors and the deepfake videos. 

We used Noise Constrastive estimation~\cite{pmlr-v9-gutmann10a} to estimate the full softmax distribution, to speed up the training process because it avoids the expensive normalization step. Instead NCE differentiates between the true data and artificially generated noise and over time, the model learns to assign higher probabilities to true data samples than to noise.

\vspace{0.5em}
\noindent\textbf{Training Hyperparameter Details.} We trained the whole framework from scratch for 120 epochs using Adam optimizer. The learning rate was 1e-3 with weight decay of 1e-5, temperature $\tau = 0.5$ and window size of 16 frames. Each mini-batch had $K = 4$ original and $M = 190$ deepfake samples, keeping $1:47$ ratio of target : imposters. We attempted to use a pre-trained 3D-CNN, however, in literature they have been predominantly used for action recognition tasks, which are devoid of facial features. 

\vspace{0.5em}
\noindent\textbf{Scoring an individual video.} For the test time scoring of individual videos, Kopuklu et al.~\cite{kopuklu2021driver} proposed a new evaluation protocol. After the training phase, only the 3D-CNN model is retained. This model encodes all N original training set videos into a set of L2 normalized 512-dimensional representations. Hence, the original template vector $v_o$ is given by:
\begin{gather*}
v_o = \frac{1}{N} \sum_{i=1}^{N} \frac{enc_\theta(x_i)}{||enc_\theta(x_i)||_2}.
\end{gather*}

To get a similarity score of a test video $x_{test}$, we simply compute the fidelity score $f$ to be cosine similarity between the encoded clip and $v_o$ by:
\begin{gather*}
f(x_{test}) = v_o^T \frac{enc_\theta(x_i)}{||enc_\theta(x_i)||_2}.
\end{gather*}
We used this score to calculate the difference between deepfake videos and their original counterparts. Hence, a larger deviation from the ground truth indicates lower fidelity of the deepfake video, serving as the fidelity score for deepfake evaluation. 

The self-supervised training paradigm adopted above gives itself to new video datasets and newer challenges with minimal data pre-processing and without the need for providing labels.

\begin{figure*}[ht!]
    \centering
        \subfloat[No Artifact]{%
            \includegraphics[width=0.40\columnwidth]{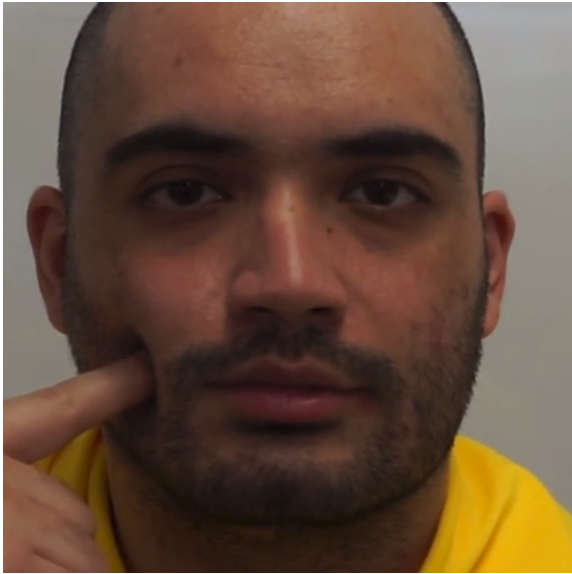}\hspace{0.1em}
        }
        \subfloat[Facial Boundary Artifacts]{%
            \includegraphics[width=0.40\columnwidth]{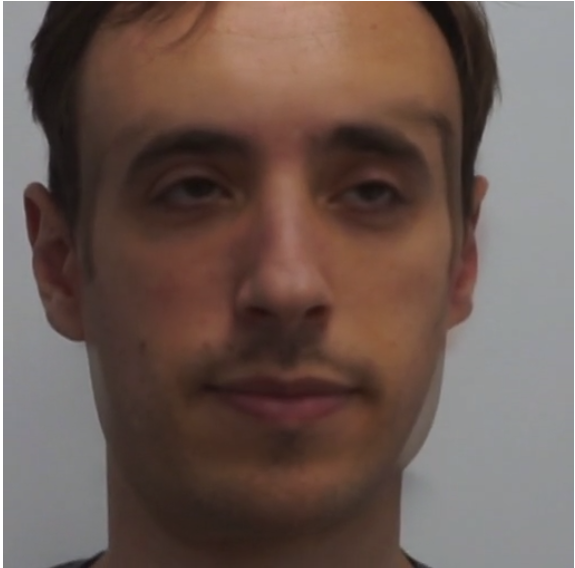}\hspace{0.1em}
        }
        \subfloat[Object Vanishes]{%
            \includegraphics[width=0.40\columnwidth]{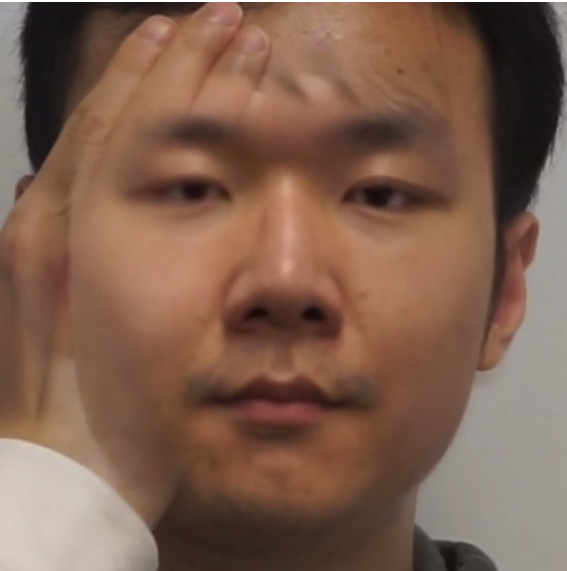}\hspace{0.1em}
        }
        \subfloat[Skin Texture Inconsistency]{%
            \includegraphics[width=0.41\columnwidth]{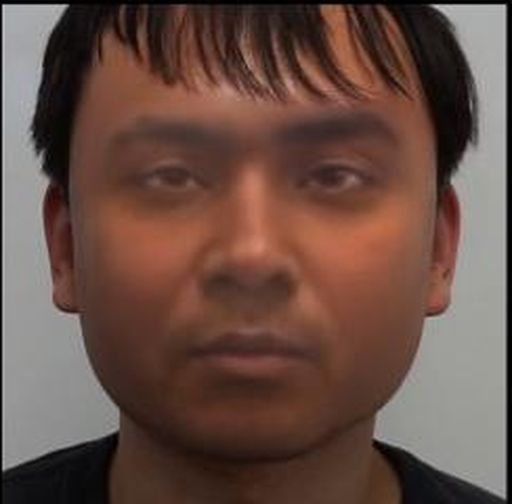}\hspace{0.1em}
        }
        \subfloat[Temporal Inconsistency]{%
            \includegraphics[width=0.40\columnwidth]{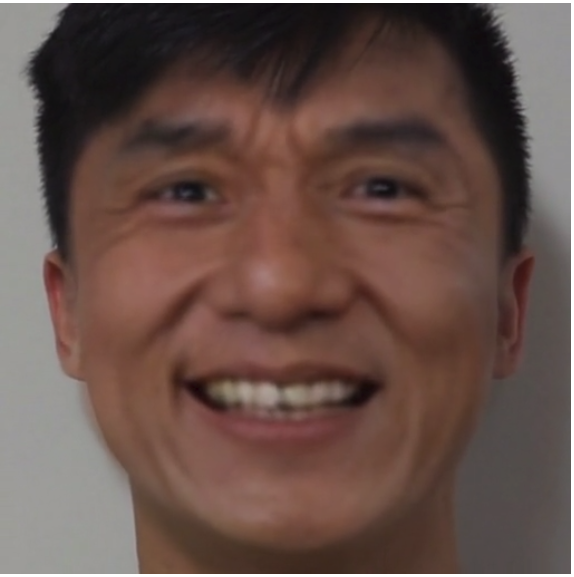}
        }
       
    \caption{Video samples that were shown to each participant along with definitions.}
\label{fig:he_artifacts}
\end{figure*}

\begin{figure*}[ht!]
    \centering
        \subfloat[Compliance]{%
            \includegraphics[width=0.8\columnwidth]{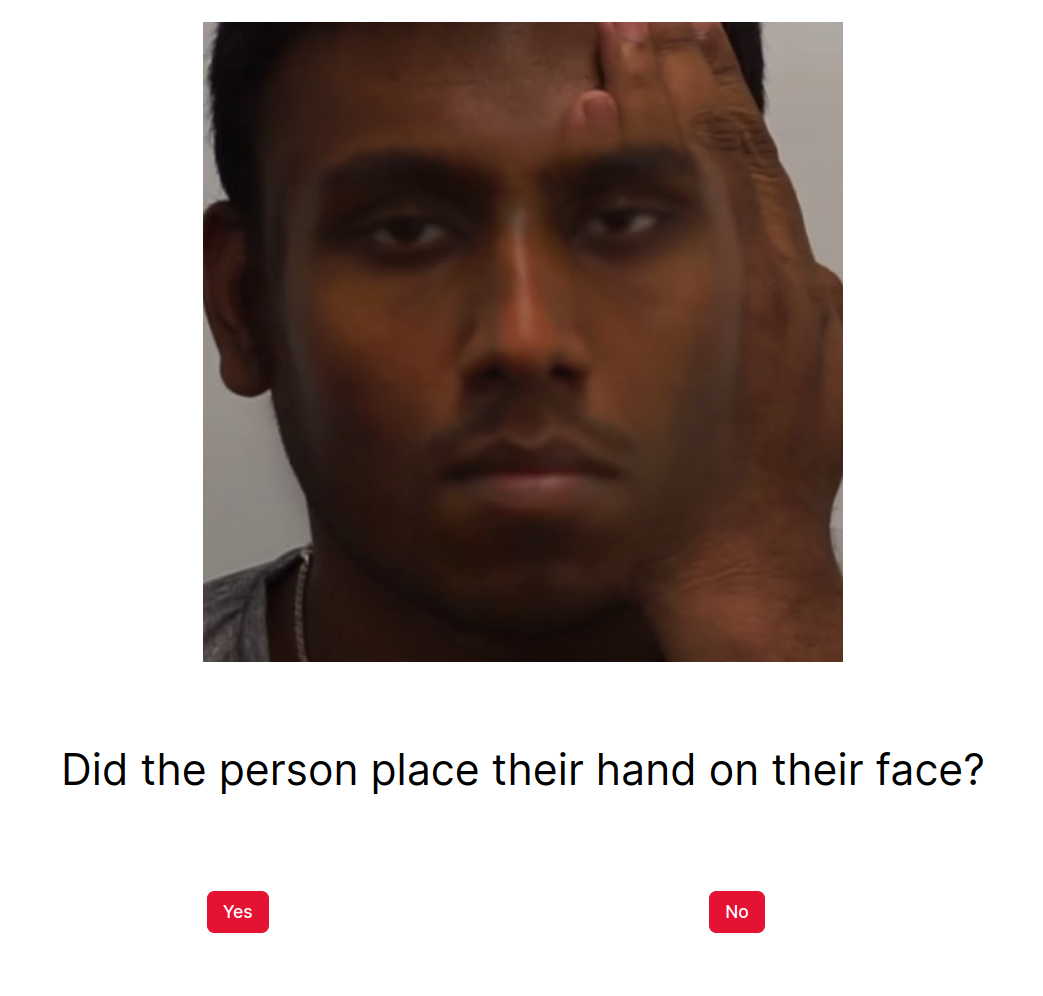}
       
        }%
        \centering
        \subfloat[Localization]{%
            \includegraphics[width=0.8\columnwidth]{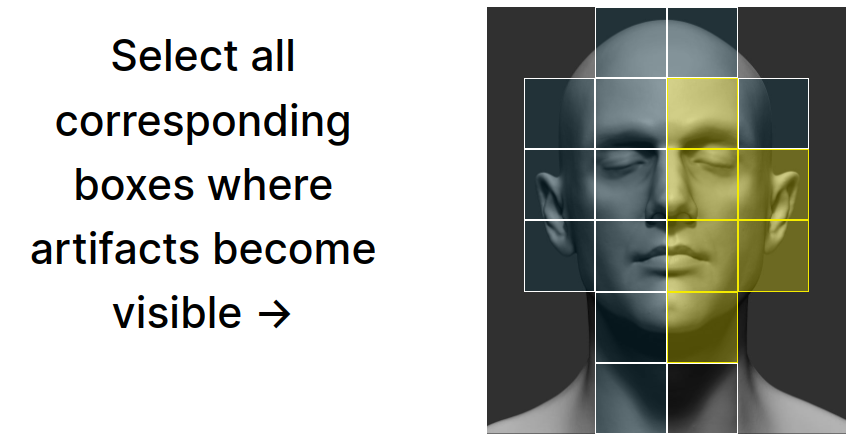}
        }

        \centering
        \subfloat[Realism]{%
             \includegraphics[width=0.8\columnwidth]{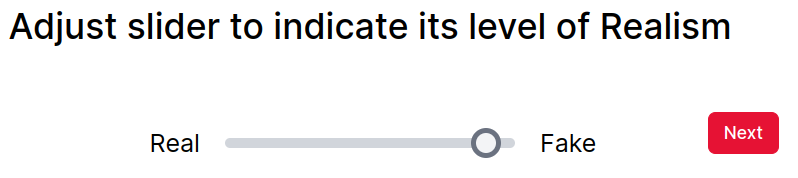}
        }%
        \centering
        \subfloat[Justification]{%
            \includegraphics[width=0.8\columnwidth]{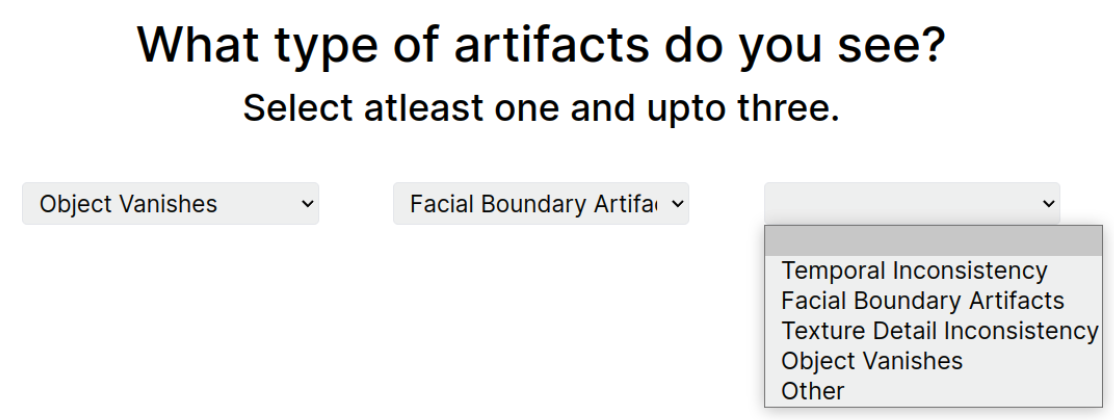}
        }
    \caption{Human evaluation of a deepfake video included four sub-responses -- (a) Compliance (b) Localization (c) Realism, and (d) Justification. Each sub-response was recorded in a separate window with the video available at all times. }
\label{fig:sdfl_human_eval}
\end{figure*}

\begin{figure}[ht!]
    \centering
    \includegraphics[width=\columnwidth]{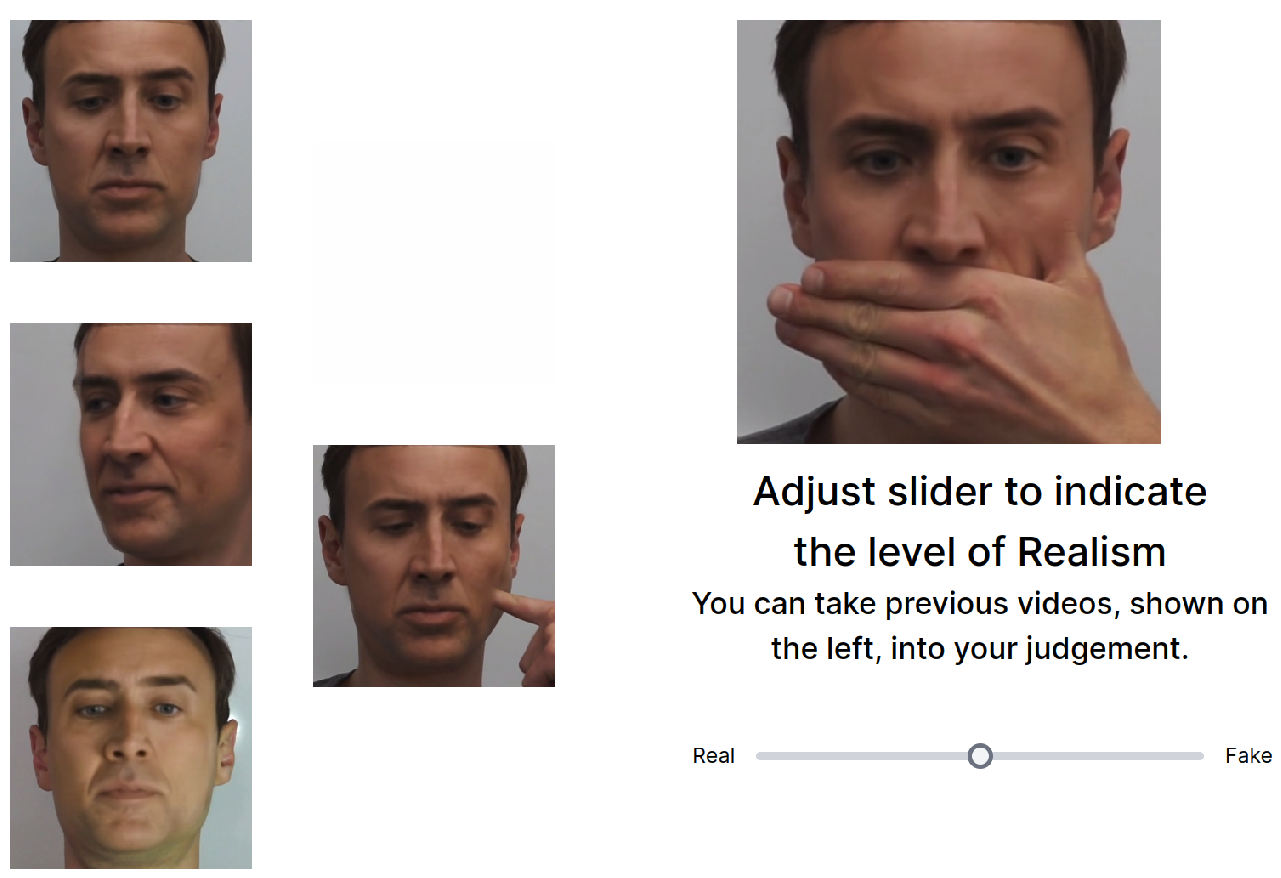}
    \caption{Human evaluation of a sequence of deepfake challenge videos. The videos were created using the adaptive RTDF pipeline. Each human response is supported with previous challenges in progression.}
    \label{fig:hdfl_human_eval}
\end{figure}

\subsection{Human Evaluation details}
\label{sec:human_eval_details}

We used Gorilla.sc for creating our human evaluation experiments and Prolific to recruit participants. The workflow for a consenting participant who passed the quiz was: Artifact~Definitions $\rightarrow$ Quiz $\rightarrow$ Video Evaluations.

\vspace{0.5em}
\noindent\textbf{Definitions of Artifacts.} All participants were provided with the following definitions and two sample videos of artifacts. Fig~\ref{fig:he_artifacts} lists one of the two samples.

\begin{itemize}[nosep,parsep=2pt,leftmargin=*]
\item \emph{No Artifact:} The video exhibits no inconsistencies or visible artifacts, appearing both natural and genuine. This category may also contain harmless artifacts or noise unrelated to deepfake manipulation. Note that the quivering of the whole video (observed on the right) is an artifact of headshot extraction and not a deepfake artifact.

\item \emph{Facial Boundary Artifacts:} These are mismatches or inconsistencies visible when the digitally manipulated face meets the original face, manifesting as unnatural transitions, blurring, or skin tone inconsistencies between facial regions.

\item \emph{Object Vanishes:} This artifact features abrupt disappearances of specific objects or features within the video frame (e.g., sunglasses or hands), creating an apparent visual inconsistency.

\item \emph{Skin Texture Inconsistency:} This artifact results from uneven or inconsistent skin texture or detail. For example, one facial region may appear unusually smooth, overly sharp, or lighted up, contrasting sharply with other areas. Note that the movement of the black boundary frame is an artifact of headshot extraction and not a deepfake artifact. On the left, the person looks side-to-side, and on the right, they stand and sit down quickly.

\item \emph{Temporal Inconsistency:} This artifact is characterized by inconsistencies in facial movements or expressions over time. Such inconsistencies are most noticeable when observing a sequence of frames. They can manifest as flickering skin or rigid facial movements that deviate from natural human behavior.
\end{itemize}

\vspace{0.5em}
\noindent\textbf{Quiz.} We administered a multiple-choice quiz, asking participants to pick the most apparent artifact in ten unseen videos. We have kept the videos close to the original videos shown with artifact definitions, and failure was granted if they scored below $6/10$. We used this process as a proxy to filter un-attentive participants and teach them about artifacts. 43 out of 60 recruited participants passed the quiz. We paid the disqualified participants partially for their time. 

\vspace{0.5em}
\noindent\textbf{Video Evaluations.} Fig.~\ref{fig:sdfl_human_eval} illustrates how we collected responses for each deepfake and original video from each consenting and passing human evaluator.

\vspace{0.5em}
\noindent\textbf{Video Evaluations of the adaptive adversary.} Fig.~\ref{fig:hdfl_human_eval} illustrates how we collected responses for each deepfake and original video from each consenting and passing human evaluator for adaptive adversaries. The main difference was showing videos from past challenges with the new challenge. 

\subsection{Usability Benefits}
\label{subsec:usability}
  This category contains participant-facing benefits, which indicate how practical a challenge is in a video-call setting. As this category only includes participant-facing benefits, a particular benefit could either awarded an \textit{Offered}, \textit{Not-offered} or \textit{Quasi} (partially offered) status. The following challenges are granted an \textit{Offered} status:

  \begin{itemize}[nosep,parsep=2pt,leftmargin=*]
    \item \textbf{Easy-to-Comprehend}: If the challenge requires action (i.e., is active), it is also easily understood, and the task comes naturally to a human participant. \textit{Quasi} is granted if it is difficult to understand, even if done automatically by the camera.
    
    \item \textbf{Appropriate-to-Request}: If the task must be completed, it can be completed without hesitation or embarrassment. It is still permitted if done automatically without the participant's awareness during the call.
    
    \item \textbf{Physically-Effortless}:   If the participant is not required to do anything except press a button. \textit{Quasi} is granted if performed challenges do not require more than a simple task, which a participant might even undertake naturally during a video call interaction. Passive challenges are deemed physically low-effort by default.

    \item \textbf{No-Equipment-Needed}: Suppose the participant does not require extra equipment to complete the task. They would join the call as usual, with no expectations.

    \item \textbf{Detected-by-Humans}: If the challenge introduces artifacts or inconsistencies that humans could perceive distinctly. \textit{Quasi} is granted if the artifacts could be seen with the keen eye of a human evaluator who might be specifically looking for them. If the challenge would not introduce any particular observable artifacts, \emph{Not-Offered} is granted. 
  \end{itemize}

\end{document}